\documentclass[longauth]{aaEC}

\usepackage{euclid}
\usepackage{graphicx}
\usepackage{natbib}
\usepackage{scalerel}
\usepackage{verbatim}
\usepackage[table]{xcolor}
\usepackage{multirow}  
\usepackage{lastpage}
\usepackage{amsmath,amsfonts,amssymb}
\usepackage{txfonts}
\usepackage{yfonts}
\usepackage{longtable}

\usepackage{soul,color,lscape}

\usepackage[pdfencoding=auto,psdextra]{hyperref}
\hypersetup{
    colorlinks=true,
    linkcolor=blue,
    filecolor=magenta,      
    urlcolor=blue,
    citecolor=blue
}
\makeatletter
\renewcommand*\aa@pageof{, page \thepage{} of \pageref*{LastPage}}
\makeatother

\usepackage[utf8]{inputenc}

\usepackage{physics}

\newcommand{\done}[1]{}

\definecolor{brilliantlavender}{rgb}{0.96, 0.73, 1.0}
\definecolor{cerulean}{rgb}{0.2,0.3,0.8}

\definecolor{prune}{rgb}{0.44, 0.11, 0.11}
\definecolor{dgold}{rgb}{.75, .72, .1}

\definecolor{kindofgreen}{RGB}{0, 120, 130}

\defcitealias{Ajani-EP29}{Paper I}

\newcommand{\inlinecomment}[1]{}
\usepackage{cancel}

\usepackage[normalem]{ulem}
\usepackage[switch, modulo]{lineno}

\begin{document}

\title{\Euclid preparation}
\subtitle{LXXXV. Toward a DR1 application of higher-order weak lensing statistics}

\newcommand{\orcid}[1]{}	   
\author{Euclid Collaboration: S.~Vinciguerra\orcid{0009-0005-4018-3184}\thanks{\email{simone.vinciguerra@lam.fr}}\inst{\ref{aff1}}
\and F.~Bouch\`e\orcid{0000-0002-4663-1786}\inst{\ref{aff2},\ref{aff3}}
\and N.~Martinet\orcid{0000-0003-2786-7790}\inst{\ref{aff1}}
\and L.~Castiblanco\orcid{0000-0002-2324-7335}\inst{\ref{aff4},\ref{aff5}}
\and C.~Uhlemann\orcid{0000-0001-7831-1579}\inst{\ref{aff5},\ref{aff4}}
\and S.~Pires\orcid{0000-0002-0249-2104}\inst{\ref{aff6}}
\and J.~Harnois-D\'eraps\orcid{0000-0002-4864-1240}\inst{\ref{aff4}}
\and C.~Giocoli\orcid{0000-0002-9590-7961}\inst{\ref{aff7},\ref{aff8}}
\and M.~Baldi\orcid{0000-0003-4145-1943}\inst{\ref{aff9},\ref{aff7},\ref{aff8}}
\and V.~F.~Cardone\inst{\ref{aff10},\ref{aff11}}
\and A.~Vadal\`a\orcid{0009-0004-3863-0179}\inst{\ref{aff10},\ref{aff11},\ref{aff12},\ref{aff13}}
\and N.~Dagoneau\orcid{0000-0002-1361-2562}\inst{\ref{aff14}}
\and L.~Linke\orcid{0000-0002-2622-8113}\inst{\ref{aff15}}
\and E.~Sellentin\inst{\ref{aff16},\ref{aff17}}
\and P.~L.~Taylor\orcid{0000-0001-6999-4718}\inst{\ref{aff18},\ref{aff19}}
\and J.~C.~Broxterman\inst{\ref{aff20},\ref{aff17}}
\and S.~Heydenreich\inst{\ref{aff21},\ref{aff22}}
\and V.~Tinnaneri~Sreekanth\orcid{0009-0008-3885-5131}\inst{\ref{aff6}}
\and N.~Porqueres\orcid{0000-0002-7599-966X}\inst{\ref{aff23}}
\and L.~Porth\orcid{0000-0003-1176-6346}\inst{\ref{aff22}}
\and M.~Gatti\inst{\ref{aff24}}
\and D.~Grand\'on\orcid{0000-0003-4953-7864}\inst{\ref{aff16},\ref{aff25}}
\and A.~Barthelemy\orcid{0000-0003-1060-3959}\inst{\ref{aff23},\ref{aff26}}
\and F.~Bernardeau\orcid{0009-0007-3015-2581}\inst{\ref{aff27},\ref{aff28}}
\and A.~Tersenov\orcid{0009-0007-5348-6701}\inst{\ref{aff23},\ref{aff29},\ref{aff30}}
\and H.~Hoekstra\orcid{0000-0002-0641-3231}\inst{\ref{aff17}}
\and J.-L.~Starck\orcid{0000-0003-2177-7794}\inst{\ref{aff6}}
\and S.~Cheng\orcid{0000-0002-9156-7461}\inst{\ref{aff31},\ref{aff32}}
\and P.~A.~Burger\orcid{0000-0001-8637-6305}\inst{\ref{aff33}}
\and I.~Tereno\orcid{0000-0002-4537-6218}\inst{\ref{aff34},\ref{aff35}}
\and R.~Scaramella\orcid{0000-0003-2229-193X}\inst{\ref{aff10}}
\and B.~Altieri\orcid{0000-0003-3936-0284}\inst{\ref{aff36}}
\and S.~Andreon\orcid{0000-0002-2041-8784}\inst{\ref{aff37}}
\and N.~Auricchio\orcid{0000-0003-4444-8651}\inst{\ref{aff7}}
\and C.~Baccigalupi\orcid{0000-0002-8211-1630}\inst{\ref{aff38},\ref{aff39},\ref{aff40},\ref{aff41}}
\and S.~Bardelli\orcid{0000-0002-8900-0298}\inst{\ref{aff7}}
\and A.~Biviano\orcid{0000-0002-0857-0732}\inst{\ref{aff39},\ref{aff38}}
\and E.~Branchini\orcid{0000-0002-0808-6908}\inst{\ref{aff42},\ref{aff43},\ref{aff37}}
\and M.~Brescia\orcid{0000-0001-9506-5680}\inst{\ref{aff44},\ref{aff45}}
\and S.~Camera\orcid{0000-0003-3399-3574}\inst{\ref{aff46},\ref{aff47},\ref{aff48}}
\and G.~Ca\~nas-Herrera\orcid{0000-0003-2796-2149}\inst{\ref{aff49},\ref{aff17}}
\and V.~Capobianco\orcid{0000-0002-3309-7692}\inst{\ref{aff48}}
\and C.~Carbone\orcid{0000-0003-0125-3563}\inst{\ref{aff50}}
\and J.~Carretero\orcid{0000-0002-3130-0204}\inst{\ref{aff51},\ref{aff52}}
\and M.~Castellano\orcid{0000-0001-9875-8263}\inst{\ref{aff10}}
\and G.~Castignani\orcid{0000-0001-6831-0687}\inst{\ref{aff7}}
\and S.~Cavuoti\orcid{0000-0002-3787-4196}\inst{\ref{aff45},\ref{aff3}}
\and K.~C.~Chambers\orcid{0000-0001-6965-7789}\inst{\ref{aff53}}
\and A.~Cimatti\inst{\ref{aff54}}
\and C.~Colodro-Conde\inst{\ref{aff55}}
\and G.~Congedo\orcid{0000-0003-2508-0046}\inst{\ref{aff56}}
\and L.~Conversi\orcid{0000-0002-6710-8476}\inst{\ref{aff57},\ref{aff36}}
\and Y.~Copin\orcid{0000-0002-5317-7518}\inst{\ref{aff58}}
\and F.~Courbin\orcid{0000-0003-0758-6510}\inst{\ref{aff59},\ref{aff60},\ref{aff61}}
\and H.~M.~Courtois\orcid{0000-0003-0509-1776}\inst{\ref{aff62}}
\and M.~Cropper\orcid{0000-0003-4571-9468}\inst{\ref{aff63}}
\and A.~Da~Silva\orcid{0000-0002-6385-1609}\inst{\ref{aff34},\ref{aff64}}
\and H.~Degaudenzi\orcid{0000-0002-5887-6799}\inst{\ref{aff65}}
\and S.~de~la~Torre\inst{\ref{aff1}}
\and G.~De~Lucia\orcid{0000-0002-6220-9104}\inst{\ref{aff39}}
\and H.~Dole\orcid{0000-0002-9767-3839}\inst{\ref{aff66}}
\and F.~Dubath\orcid{0000-0002-6533-2810}\inst{\ref{aff65}}
\and X.~Dupac\inst{\ref{aff36}}
\and S.~Dusini\orcid{0000-0002-1128-0664}\inst{\ref{aff67}}
\and S.~Escoffier\orcid{0000-0002-2847-7498}\inst{\ref{aff68}}
\and M.~Farina\orcid{0000-0002-3089-7846}\inst{\ref{aff69}}
\and R.~Farinelli\inst{\ref{aff7}}
\and S.~Farrens\orcid{0000-0002-9594-9387}\inst{\ref{aff6}}
\and F.~Faustini\orcid{0000-0001-6274-5145}\inst{\ref{aff10},\ref{aff70}}
\and S.~Ferriol\inst{\ref{aff58}}
\and F.~Finelli\orcid{0000-0002-6694-3269}\inst{\ref{aff7},\ref{aff71}}
\and M.~Frailis\orcid{0000-0002-7400-2135}\inst{\ref{aff39}}
\and E.~Franceschi\orcid{0000-0002-0585-6591}\inst{\ref{aff7}}
\and M.~Fumana\orcid{0000-0001-6787-5950}\inst{\ref{aff50}}
\and S.~Galeotta\orcid{0000-0002-3748-5115}\inst{\ref{aff39}}
\and K.~George\orcid{0000-0002-1734-8455}\inst{\ref{aff72}}
\and B.~Gillis\orcid{0000-0002-4478-1270}\inst{\ref{aff56}}
\and J.~Gracia-Carpio\inst{\ref{aff73}}
\and A.~Grazian\orcid{0000-0002-5688-0663}\inst{\ref{aff74}}
\and F.~Grupp\inst{\ref{aff73},\ref{aff26}}
\and S.~V.~H.~Haugan\orcid{0000-0001-9648-7260}\inst{\ref{aff75}}
\and W.~Holmes\inst{\ref{aff76}}
\and F.~Hormuth\inst{\ref{aff77}}
\and A.~Hornstrup\orcid{0000-0002-3363-0936}\inst{\ref{aff78},\ref{aff79}}
\and P.~Hudelot\inst{\ref{aff28}}
\and K.~Jahnke\orcid{0000-0003-3804-2137}\inst{\ref{aff80}}
\and M.~Jhabvala\inst{\ref{aff81}}
\and B.~Joachimi\orcid{0000-0001-7494-1303}\inst{\ref{aff82}}
\and E.~Keih\"anen\orcid{0000-0003-1804-7715}\inst{\ref{aff83}}
\and S.~Kermiche\orcid{0000-0002-0302-5735}\inst{\ref{aff68}}
\and A.~Kiessling\orcid{0000-0002-2590-1273}\inst{\ref{aff76}}
\and M.~Kilbinger\orcid{0000-0001-9513-7138}\inst{\ref{aff6}}
\and B.~Kubik\orcid{0009-0006-5823-4880}\inst{\ref{aff58}}
\and M.~Kunz\orcid{0000-0002-3052-7394}\inst{\ref{aff84}}
\and H.~Kurki-Suonio\orcid{0000-0002-4618-3063}\inst{\ref{aff85},\ref{aff86}}
\and A.~M.~C.~Le~Brun\orcid{0000-0002-0936-4594}\inst{\ref{aff87}}
\and S.~Ligori\orcid{0000-0003-4172-4606}\inst{\ref{aff48}}
\and P.~B.~Lilje\orcid{0000-0003-4324-7794}\inst{\ref{aff75}}
\and V.~Lindholm\orcid{0000-0003-2317-5471}\inst{\ref{aff85},\ref{aff86}}
\and I.~Lloro\orcid{0000-0001-5966-1434}\inst{\ref{aff88}}
\and G.~Mainetti\orcid{0000-0003-2384-2377}\inst{\ref{aff89}}
\and D.~Maino\inst{\ref{aff90},\ref{aff50},\ref{aff91}}
\and O.~Mansutti\orcid{0000-0001-5758-4658}\inst{\ref{aff39}}
\and O.~Marggraf\orcid{0000-0001-7242-3852}\inst{\ref{aff22}}
\and M.~Martinelli\orcid{0000-0002-6943-7732}\inst{\ref{aff10},\ref{aff11}}
\and F.~Marulli\orcid{0000-0002-8850-0303}\inst{\ref{aff92},\ref{aff7},\ref{aff8}}
\and R.~J.~Massey\orcid{0000-0002-6085-3780}\inst{\ref{aff93}}
\and E.~Medinaceli\orcid{0000-0002-4040-7783}\inst{\ref{aff7}}
\and S.~Mei\orcid{0000-0002-2849-559X}\inst{\ref{aff94},\ref{aff95}}
\and M.~Melchior\inst{\ref{aff96}}
\and Y.~Mellier\inst{\ref{aff97},\ref{aff28}}
\and M.~Meneghetti\orcid{0000-0003-1225-7084}\inst{\ref{aff7},\ref{aff8}}
\and G.~Meylan\inst{\ref{aff98}}
\and A.~Mora\orcid{0000-0002-1922-8529}\inst{\ref{aff99}}
\and M.~Moresco\orcid{0000-0002-7616-7136}\inst{\ref{aff92},\ref{aff7}}
\and L.~Moscardini\orcid{0000-0002-3473-6716}\inst{\ref{aff92},\ref{aff7},\ref{aff8}}
\and C.~Neissner\orcid{0000-0001-8524-4968}\inst{\ref{aff100},\ref{aff52}}
\and S.-M.~Niemi\orcid{0009-0005-0247-0086}\inst{\ref{aff49}}
\and C.~Padilla\orcid{0000-0001-7951-0166}\inst{\ref{aff100}}
\and S.~Paltani\orcid{0000-0002-8108-9179}\inst{\ref{aff65}}
\and F.~Pasian\orcid{0000-0002-4869-3227}\inst{\ref{aff39}}
\and K.~Pedersen\inst{\ref{aff101}}
\and V.~Pettorino\orcid{0000-0002-4203-9320}\inst{\ref{aff49}}
\and G.~Polenta\orcid{0000-0003-4067-9196}\inst{\ref{aff70}}
\and M.~Poncet\inst{\ref{aff102}}
\and L.~A.~Popa\inst{\ref{aff103}}
\and F.~Raison\orcid{0000-0002-7819-6918}\inst{\ref{aff73}}
\and A.~Renzi\orcid{0000-0001-9856-1970}\inst{\ref{aff104},\ref{aff67}}
\and J.~Rhodes\orcid{0000-0002-4485-8549}\inst{\ref{aff76}}
\and G.~Riccio\inst{\ref{aff45}}
\and E.~Romelli\orcid{0000-0003-3069-9222}\inst{\ref{aff39}}
\and M.~Roncarelli\orcid{0000-0001-9587-7822}\inst{\ref{aff7}}
\and R.~Saglia\orcid{0000-0003-0378-7032}\inst{\ref{aff26},\ref{aff73}}
\and Z.~Sakr\orcid{0000-0002-4823-3757}\inst{\ref{aff105},\ref{aff106},\ref{aff107}}
\and A.~G.~S\'anchez\orcid{0000-0003-1198-831X}\inst{\ref{aff73}}
\and D.~Sapone\orcid{0000-0001-7089-4503}\inst{\ref{aff108}}
\and B.~Sartoris\orcid{0000-0003-1337-5269}\inst{\ref{aff26},\ref{aff39}}
\and P.~Schneider\orcid{0000-0001-8561-2679}\inst{\ref{aff22}}
\and T.~Schrabback\orcid{0000-0002-6987-7834}\inst{\ref{aff15}}
\and A.~Secroun\orcid{0000-0003-0505-3710}\inst{\ref{aff68}}
\and G.~Seidel\orcid{0000-0003-2907-353X}\inst{\ref{aff80}}
\and S.~Serrano\orcid{0000-0002-0211-2861}\inst{\ref{aff109},\ref{aff110},\ref{aff111}}
\and C.~Sirignano\orcid{0000-0002-0995-7146}\inst{\ref{aff104},\ref{aff67}}
\and G.~Sirri\orcid{0000-0003-2626-2853}\inst{\ref{aff8}}
\and A.~Spurio~Mancini\orcid{0000-0001-5698-0990}\inst{\ref{aff112}}
\and L.~Stanco\orcid{0000-0002-9706-5104}\inst{\ref{aff67}}
\and J.~Steinwagner\orcid{0000-0001-7443-1047}\inst{\ref{aff73}}
\and P.~Tallada-Cresp\'{i}\orcid{0000-0002-1336-8328}\inst{\ref{aff51},\ref{aff52}}
\and A.~N.~Taylor\inst{\ref{aff56}}
\and N.~Tessore\orcid{0000-0002-9696-7931}\inst{\ref{aff63}}
\and S.~Toft\orcid{0000-0003-3631-7176}\inst{\ref{aff113},\ref{aff114}}
\and R.~Toledo-Moreo\orcid{0000-0002-2997-4859}\inst{\ref{aff115}}
\and F.~Torradeflot\orcid{0000-0003-1160-1517}\inst{\ref{aff52},\ref{aff51}}
\and I.~Tutusaus\orcid{0000-0002-3199-0399}\inst{\ref{aff111},\ref{aff109},\ref{aff106}}
\and J.~Valiviita\orcid{0000-0001-6225-3693}\inst{\ref{aff85},\ref{aff86}}
\and T.~Vassallo\orcid{0000-0001-6512-6358}\inst{\ref{aff39},\ref{aff72}}
\and Y.~Wang\orcid{0000-0002-4749-2984}\inst{\ref{aff116}}
\and J.~Weller\orcid{0000-0002-8282-2010}\inst{\ref{aff26},\ref{aff73}}
\and A.~Zacchei\orcid{0000-0003-0396-1192}\inst{\ref{aff39},\ref{aff38}}
\and G.~Zamorani\orcid{0000-0002-2318-301X}\inst{\ref{aff7}}
\and F.~M.~Zerbi\inst{\ref{aff37}}
\and E.~Zucca\orcid{0000-0002-5845-8132}\inst{\ref{aff7}}
\and M.~Ballardini\orcid{0000-0003-4481-3559}\inst{\ref{aff117},\ref{aff118},\ref{aff7}}
\and M.~Bolzonella\orcid{0000-0003-3278-4607}\inst{\ref{aff7}}
\and A.~Boucaud\orcid{0000-0001-7387-2633}\inst{\ref{aff94}}
\and E.~Bozzo\orcid{0000-0002-8201-1525}\inst{\ref{aff65}}
\and C.~Burigana\orcid{0000-0002-3005-5796}\inst{\ref{aff119},\ref{aff71}}
\and R.~Cabanac\orcid{0000-0001-6679-2600}\inst{\ref{aff106}}
\and M.~Calabrese\orcid{0000-0002-2637-2422}\inst{\ref{aff120},\ref{aff50}}
\and A.~Cappi\inst{\ref{aff121},\ref{aff7}}
\and J.~A.~Escartin~Vigo\inst{\ref{aff73}}
\and L.~Gabarra\orcid{0000-0002-8486-8856}\inst{\ref{aff122}}
\and W.~G.~Hartley\inst{\ref{aff65}}
\and R.~Maoli\orcid{0000-0002-6065-3025}\inst{\ref{aff13},\ref{aff10}}
\and J.~Mart\'{i}n-Fleitas\orcid{0000-0002-8594-569X}\inst{\ref{aff123}}
\and S.~Matthew\orcid{0000-0001-8448-1697}\inst{\ref{aff56}}
\and N.~Mauri\orcid{0000-0001-8196-1548}\inst{\ref{aff54},\ref{aff8}}
\and R.~B.~Metcalf\orcid{0000-0003-3167-2574}\inst{\ref{aff92},\ref{aff7}}
\and A.~Pezzotta\orcid{0000-0003-0726-2268}\inst{\ref{aff37}}
\and M.~P\"ontinen\orcid{0000-0001-5442-2530}\inst{\ref{aff85}}
\and I.~Risso\orcid{0000-0003-2525-7761}\inst{\ref{aff37},\ref{aff43}}
\and V.~Scottez\orcid{0009-0008-3864-940X}\inst{\ref{aff97},\ref{aff124}}
\and M.~Sereno\orcid{0000-0003-0302-0325}\inst{\ref{aff7},\ref{aff8}}
\and M.~Tenti\orcid{0000-0002-4254-5901}\inst{\ref{aff8}}
\and M.~Viel\orcid{0000-0002-2642-5707}\inst{\ref{aff38},\ref{aff39},\ref{aff41},\ref{aff40},\ref{aff125}}
\and M.~Wiesmann\orcid{0009-0000-8199-5860}\inst{\ref{aff75}}
\and Y.~Akrami\orcid{0000-0002-2407-7956}\inst{\ref{aff126},\ref{aff127}}
\and I.~T.~Andika\orcid{0000-0001-6102-9526}\inst{\ref{aff128},\ref{aff129}}
\and R.~E.~Angulo\orcid{0000-0003-2953-3970}\inst{\ref{aff130},\ref{aff131}}
\and S.~Anselmi\orcid{0000-0002-3579-9583}\inst{\ref{aff67},\ref{aff104},\ref{aff132}}
\and M.~Archidiacono\orcid{0000-0003-4952-9012}\inst{\ref{aff90},\ref{aff91}}
\and F.~Atrio-Barandela\orcid{0000-0002-2130-2513}\inst{\ref{aff133}}
\and E.~Aubourg\orcid{0000-0002-5592-023X}\inst{\ref{aff94},\ref{aff134}}
\and D.~Bertacca\orcid{0000-0002-2490-7139}\inst{\ref{aff104},\ref{aff74},\ref{aff67}}
\and M.~Bethermin\orcid{0000-0002-3915-2015}\inst{\ref{aff135}}
\and A.~Blanchard\orcid{0000-0001-8555-9003}\inst{\ref{aff106}}
\and L.~Blot\orcid{0000-0002-9622-7167}\inst{\ref{aff136},\ref{aff87}}
\and M.~Bonici\orcid{0000-0002-8430-126X}\inst{\ref{aff33},\ref{aff50}}
\and S.~Borgani\orcid{0000-0001-6151-6439}\inst{\ref{aff137},\ref{aff38},\ref{aff39},\ref{aff40},\ref{aff125}}
\and M.~L.~Brown\orcid{0000-0002-0370-8077}\inst{\ref{aff138}}
\and S.~Bruton\orcid{0000-0002-6503-5218}\inst{\ref{aff139}}
\and A.~Calabro\orcid{0000-0003-2536-1614}\inst{\ref{aff10}}
\and B.~Camacho~Quevedo\orcid{0000-0002-8789-4232}\inst{\ref{aff38},\ref{aff41},\ref{aff39}}
\and F.~Caro\inst{\ref{aff10}}
\and C.~S.~Carvalho\inst{\ref{aff35}}
\and T.~Castro\orcid{0000-0002-6292-3228}\inst{\ref{aff39},\ref{aff40},\ref{aff38},\ref{aff125}}
\and F.~Cogato\orcid{0000-0003-4632-6113}\inst{\ref{aff92},\ref{aff7}}
\and S.~Conseil\orcid{0000-0002-3657-4191}\inst{\ref{aff58}}
\and A.~R.~Cooray\orcid{0000-0002-3892-0190}\inst{\ref{aff140}}
\and G.~Desprez\orcid{0000-0001-8325-1742}\inst{\ref{aff141}}
\and A.~D\'iaz-S\'anchez\orcid{0000-0003-0748-4768}\inst{\ref{aff142}}
\and J.~J.~Diaz\orcid{0000-0003-2101-1078}\inst{\ref{aff55}}
\and S.~Di~Domizio\orcid{0000-0003-2863-5895}\inst{\ref{aff42},\ref{aff43}}
\and J.~M.~Diego\orcid{0000-0001-9065-3926}\inst{\ref{aff143}}
\and M.~Y.~Elkhashab\orcid{0000-0001-9306-2603}\inst{\ref{aff39},\ref{aff40},\ref{aff137},\ref{aff38}}
\and Y.~Fang\orcid{0000-0002-0334-6950}\inst{\ref{aff26}}
\and P.~G.~Ferreira\orcid{0000-0002-3021-2851}\inst{\ref{aff122}}
\and A.~Finoguenov\orcid{0000-0002-4606-5403}\inst{\ref{aff85}}
\and A.~Franco\orcid{0000-0002-4761-366X}\inst{\ref{aff144},\ref{aff145},\ref{aff146}}
\and K.~Ganga\orcid{0000-0001-8159-8208}\inst{\ref{aff94}}
\and J.~Garc\'ia-Bellido\orcid{0000-0002-9370-8360}\inst{\ref{aff126}}
\and T.~Gasparetto\orcid{0000-0002-7913-4866}\inst{\ref{aff10}}
\and V.~Gautard\inst{\ref{aff23}}
\and R.~Gavazzi\orcid{0000-0002-5540-6935}\inst{\ref{aff1},\ref{aff28}}
\and E.~Gaztanaga\orcid{0000-0001-9632-0815}\inst{\ref{aff111},\ref{aff109},\ref{aff147}}
\and F.~Giacomini\orcid{0000-0002-3129-2814}\inst{\ref{aff8}}
\and F.~Gianotti\orcid{0000-0003-4666-119X}\inst{\ref{aff7}}
\and G.~Gozaliasl\orcid{0000-0002-0236-919X}\inst{\ref{aff148},\ref{aff85}}
\and M.~Guidi\orcid{0000-0001-9408-1101}\inst{\ref{aff9},\ref{aff7}}
\and C.~M.~Gutierrez\orcid{0000-0001-7854-783X}\inst{\ref{aff149}}
\and A.~Hall\orcid{0000-0002-3139-8651}\inst{\ref{aff56}}
\and S.~Hemmati\orcid{0000-0003-2226-5395}\inst{\ref{aff150}}
\and H.~Hildebrandt\orcid{0000-0002-9814-3338}\inst{\ref{aff151}}
\and J.~Hjorth\orcid{0000-0002-4571-2306}\inst{\ref{aff101}}
\and J.~J.~E.~Kajava\orcid{0000-0002-3010-8333}\inst{\ref{aff152},\ref{aff153}}
\and Y.~Kang\orcid{0009-0000-8588-7250}\inst{\ref{aff65}}
\and D.~Karagiannis\orcid{0000-0002-4927-0816}\inst{\ref{aff117},\ref{aff154}}
\and K.~Kiiveri\inst{\ref{aff83}}
\and J.~Kim\orcid{0000-0003-2776-2761}\inst{\ref{aff122}}
\and C.~C.~Kirkpatrick\inst{\ref{aff83}}
\and S.~Kruk\orcid{0000-0001-8010-8879}\inst{\ref{aff36}}
\and L.~Legrand\orcid{0000-0003-0610-5252}\inst{\ref{aff155},\ref{aff156}}
\and M.~Lembo\orcid{0000-0002-5271-5070}\inst{\ref{aff28},\ref{aff117},\ref{aff118}}
\and F.~Lepori\orcid{0009-0000-5061-7138}\inst{\ref{aff157}}
\and G.~Leroy\orcid{0009-0004-2523-4425}\inst{\ref{aff158},\ref{aff93}}
\and G.~F.~Lesci\orcid{0000-0002-4607-2830}\inst{\ref{aff92},\ref{aff7}}
\and J.~Lesgourgues\orcid{0000-0001-7627-353X}\inst{\ref{aff159}}
\and T.~I.~Liaudat\orcid{0000-0002-9104-314X}\inst{\ref{aff134}}
\and J.~Macias-Perez\orcid{0000-0002-5385-2763}\inst{\ref{aff160}}
\and M.~Magliocchetti\orcid{0000-0001-9158-4838}\inst{\ref{aff69}}
\and F.~Mannucci\orcid{0000-0002-4803-2381}\inst{\ref{aff161}}
\and C.~J.~A.~P.~Martins\orcid{0000-0002-4886-9261}\inst{\ref{aff162},\ref{aff163}}
\and L.~Maurin\orcid{0000-0002-8406-0857}\inst{\ref{aff66}}
\and M.~Miluzio\inst{\ref{aff36},\ref{aff164}}
\and P.~Monaco\orcid{0000-0003-2083-7564}\inst{\ref{aff137},\ref{aff39},\ref{aff40},\ref{aff38}}
\and C.~Moretti\orcid{0000-0003-3314-8936}\inst{\ref{aff39},\ref{aff38},\ref{aff40},\ref{aff41}}
\and G.~Morgante\inst{\ref{aff7}}
\and S.~Nadathur\orcid{0000-0001-9070-3102}\inst{\ref{aff147}}
\and K.~Naidoo\orcid{0000-0002-9182-1802}\inst{\ref{aff147},\ref{aff82}}
\and A.~Navarro-Alsina\orcid{0000-0002-3173-2592}\inst{\ref{aff22}}
\and S.~Nesseris\orcid{0000-0002-0567-0324}\inst{\ref{aff126}}
\and D.~Paoletti\orcid{0000-0003-4761-6147}\inst{\ref{aff7},\ref{aff71}}
\and F.~Passalacqua\orcid{0000-0002-8606-4093}\inst{\ref{aff104},\ref{aff67}}
\and K.~Paterson\orcid{0000-0001-8340-3486}\inst{\ref{aff80}}
\and L.~Patrizii\inst{\ref{aff8}}
\and A.~Pisani\orcid{0000-0002-6146-4437}\inst{\ref{aff68}}
\and D.~Potter\orcid{0000-0002-0757-5195}\inst{\ref{aff157}}
\and S.~Quai\orcid{0000-0002-0449-8163}\inst{\ref{aff92},\ref{aff7}}
\and M.~Radovich\orcid{0000-0002-3585-866X}\inst{\ref{aff74}}
\and S.~Sacquegna\orcid{0000-0002-8433-6630}\inst{\ref{aff165}}
\and M.~Sahl\'en\orcid{0000-0003-0973-4804}\inst{\ref{aff166}}
\and D.~B.~Sanders\orcid{0000-0002-1233-9998}\inst{\ref{aff53}}
\and E.~Sarpa\orcid{0000-0002-1256-655X}\inst{\ref{aff41},\ref{aff125},\ref{aff40}}
\and A.~Schneider\orcid{0000-0001-7055-8104}\inst{\ref{aff157}}
\and D.~Sciotti\orcid{0009-0008-4519-2620}\inst{\ref{aff10},\ref{aff11}}
\and L.~C.~Smith\orcid{0000-0002-3259-2771}\inst{\ref{aff167}}
\and K.~Tanidis\orcid{0000-0001-9843-5130}\inst{\ref{aff122}}
\and C.~Tao\orcid{0000-0001-7961-8177}\inst{\ref{aff68}}
\and G.~Testera\inst{\ref{aff43}}
\and R.~Teyssier\orcid{0000-0001-7689-0933}\inst{\ref{aff168}}
\and S.~Tosi\orcid{0000-0002-7275-9193}\inst{\ref{aff42},\ref{aff43},\ref{aff37}}
\and A.~Troja\orcid{0000-0003-0239-4595}\inst{\ref{aff104},\ref{aff67}}
\and M.~Tucci\inst{\ref{aff65}}
\and D.~Vergani\orcid{0000-0003-0898-2216}\inst{\ref{aff7}}
\and G.~Verza\orcid{0000-0002-1886-8348}\inst{\ref{aff169}}
\and N.~A.~Walton\orcid{0000-0003-3983-8778}\inst{\ref{aff167}}}
										   
\institute{Aix-Marseille Universit\'e, CNRS, CNES, LAM, Marseille, France\label{aff1}
\and
Scuola Superiore Meridionale, Via Mezzocannone 4, 80138, Napoli, Italy\label{aff2}
\and
INFN section of Naples, Via Cinthia 6, 80126, Napoli, Italy\label{aff3}
\and
School of Mathematics, Statistics and Physics, Newcastle University, Herschel Building, Newcastle-upon-Tyne, NE1 7RU, UK\label{aff4}
\and
Fakult\"at f\"ur Physik, Universit\"at Bielefeld, Postfach 100131, 33501 Bielefeld, Germany\label{aff5}
\and
Universit\'e Paris-Saclay, Universit\'e Paris Cit\'e, CEA, CNRS, AIM, 91191, Gif-sur-Yvette, France\label{aff6}
\and
INAF-Osservatorio di Astrofisica e Scienza dello Spazio di Bologna, Via Piero Gobetti 93/3, 40129 Bologna, Italy\label{aff7}
\and
INFN-Sezione di Bologna, Viale Berti Pichat 6/2, 40127 Bologna, Italy\label{aff8}
\and
Dipartimento di Fisica e Astronomia, Universit\`a di Bologna, Via Gobetti 93/2, 40129 Bologna, Italy\label{aff9}
\and
INAF-Osservatorio Astronomico di Roma, Via Frascati 33, 00078 Monteporzio Catone, Italy\label{aff10}
\and
INFN-Sezione di Roma, Piazzale Aldo Moro, 2 - c/o Dipartimento di Fisica, Edificio G. Marconi, 00185 Roma, Italy\label{aff11}
\and
Dipartimento di Fisica, Universit\`a di Roma Tor Vergata, Via della Ricerca Scientifica 1, Roma, Italy\label{aff12}
\and
Dipartimento di Fisica, Sapienza Universit\`a di Roma, Piazzale Aldo Moro 2, 00185 Roma, Italy\label{aff13}
\and
Universit\'e Paris-Saclay, CEA, D\'epartement d'\'Electronique des D\'etecteurs et d'Informatique pour la Physique, 91191, Gif-sur-Yvette, France\label{aff14}
\and
Universit\"at Innsbruck, Institut f\"ur Astro- und Teilchenphysik, Technikerstr. 25/8, 6020 Innsbruck, Austria\label{aff15}
\and
Mathematical Institute, University of Leiden, Einsteinweg 55, 2333 CA Leiden, The Netherlands\label{aff16}
\and
Leiden Observatory, Leiden University, Einsteinweg 55, 2333 CC Leiden, The Netherlands\label{aff17}
\and
Center for Cosmology and AstroParticle Physics, The Ohio State University, 191 West Woodruff Avenue, Columbus, OH 43210, USA\label{aff18}
\and
Department of Physics, The Ohio State University, Columbus, OH 43210, USA\label{aff19}
\and
Institute Lorentz, Leiden University, Niels Bohrweg 2, 2333 CA Leiden, The Netherlands\label{aff20}
\and
Department of Astronomy and Astrophysics, University of California, Santa Cruz, 1156 High Street, Santa Cruz, CA 95064, USA\label{aff21}
\and
Universit\"at Bonn, Argelander-Institut f\"ur Astronomie, Auf dem H\"ugel 71, 53121 Bonn, Germany\label{aff22}
\and
CEA Saclay, DFR/IRFU, Service d'Astrophysique, Bat. 709, 91191 Gif-sur-Yvette, France\label{aff23}
\and
Kavli Institute for Cosmological Physics, University of Chicago, Chicago, IL 60637, USA\label{aff24}
\and
Instituto de F\'isica y Astronom\'ia, Facultad de Ciencias, Universidad de Valpara\'iso, Avenida Gran Bretana 1111, Valpara\'iso, Chile\label{aff25}
\and
Universit\"ats-Sternwarte M\"unchen, Fakult\"at f\"ur Physik, Ludwig-Maximilians-Universit\"at M\"unchen, Scheinerstrasse 1, 81679 M\"unchen, Germany\label{aff26}
\and
Institut de Physique Th\'eorique, CEA, CNRS, Universit\'e Paris-Saclay 91191 Gif-sur-Yvette Cedex, France\label{aff27}
\and
Institut d'Astrophysique de Paris, UMR 7095, CNRS, and Sorbonne Universit\'e, 98 bis boulevard Arago, 75014 Paris, France\label{aff28}
\and
Institutes of Computer Science and Astrophysics, Foundation for Research and Technology - Hellas (FORTH), N. Plastira 100, Voutes GR-70013 Heraklion, Greece\label{aff29}
\and
University of Crete, Department of Physics, GR-70013 Heraklion, Greece\label{aff30}
\and
Perimeter Institute for Theoretical Physics, Waterloo, Ontario N2L 2Y5, Canada\label{aff31}
\and
Institute for Advanced Study, 1 Einstein Dr., NJ 08540, USA\label{aff32}
\and
Waterloo Centre for Astrophysics, University of Waterloo, Waterloo, Ontario N2L 3G1, Canada\label{aff33}
\and
Departamento de F\'isica, Faculdade de Ci\^encias, Universidade de Lisboa, Edif\'icio C8, Campo Grande, PT1749-016 Lisboa, Portugal\label{aff34}
\and
Instituto de Astrof\'isica e Ci\^encias do Espa\c{c}o, Faculdade de Ci\^encias, Universidade de Lisboa, Tapada da Ajuda, 1349-018 Lisboa, Portugal\label{aff35}
\and
ESAC/ESA, Camino Bajo del Castillo, s/n., Urb. Villafranca del Castillo, 28692 Villanueva de la Ca\~nada, Madrid, Spain\label{aff36}
\and
INAF-Osservatorio Astronomico di Brera, Via Brera 28, 20122 Milano, Italy\label{aff37}
\and
IFPU, Institute for Fundamental Physics of the Universe, via Beirut 2, 34151 Trieste, Italy\label{aff38}
\and
INAF-Osservatorio Astronomico di Trieste, Via G. B. Tiepolo 11, 34143 Trieste, Italy\label{aff39}
\and
INFN, Sezione di Trieste, Via Valerio 2, 34127 Trieste TS, Italy\label{aff40}
\and
SISSA, International School for Advanced Studies, Via Bonomea 265, 34136 Trieste TS, Italy\label{aff41}
\and
Dipartimento di Fisica, Universit\`a di Genova, Via Dodecaneso 33, 16146, Genova, Italy\label{aff42}
\and
INFN-Sezione di Genova, Via Dodecaneso 33, 16146, Genova, Italy\label{aff43}
\and
Department of Physics "E. Pancini", University Federico II, Via Cinthia 6, 80126, Napoli, Italy\label{aff44}
\and
INAF-Osservatorio Astronomico di Capodimonte, Via Moiariello 16, 80131 Napoli, Italy\label{aff45}
\and
Dipartimento di Fisica, Universit\`a degli Studi di Torino, Via P. Giuria 1, 10125 Torino, Italy\label{aff46}
\and
INFN-Sezione di Torino, Via P. Giuria 1, 10125 Torino, Italy\label{aff47}
\and
INAF-Osservatorio Astrofisico di Torino, Via Osservatorio 20, 10025 Pino Torinese (TO), Italy\label{aff48}
\and
European Space Agency/ESTEC, Keplerlaan 1, 2201 AZ Noordwijk, The Netherlands\label{aff49}
\and
INAF-IASF Milano, Via Alfonso Corti 12, 20133 Milano, Italy\label{aff50}
\and
Centro de Investigaciones Energ\'eticas, Medioambientales y Tecnol\'ogicas (CIEMAT), Avenida Complutense 40, 28040 Madrid, Spain\label{aff51}
\and
Port d'Informaci\'{o} Cient\'{i}fica, Campus UAB, C. Albareda s/n, 08193 Bellaterra (Barcelona), Spain\label{aff52}
\and
Institute for Astronomy, University of Hawaii, 2680 Woodlawn Drive, Honolulu, HI 96822, USA\label{aff53}
\and
Dipartimento di Fisica e Astronomia "Augusto Righi" - Alma Mater Studiorum Universit\`a di Bologna, Viale Berti Pichat 6/2, 40127 Bologna, Italy\label{aff54}
\and
Instituto de Astrof\'{\i}sica de Canarias, V\'{\i}a L\'actea, 38205 La Laguna, Tenerife, Spain\label{aff55}
\and
Institute for Astronomy, University of Edinburgh, Royal Observatory, Blackford Hill, Edinburgh EH9 3HJ, UK\label{aff56}
\and
European Space Agency/ESRIN, Largo Galileo Galilei 1, 00044 Frascati, Roma, Italy\label{aff57}
\and
Universit\'e Claude Bernard Lyon 1, CNRS/IN2P3, IP2I Lyon, UMR 5822, Villeurbanne, F-69100, France\label{aff58}
\and
Institut de Ci\`{e}ncies del Cosmos (ICCUB), Universitat de Barcelona (IEEC-UB), Mart\'{i} i Franqu\`{e}s 1, 08028 Barcelona, Spain\label{aff59}
\and
Instituci\'o Catalana de Recerca i Estudis Avan\c{c}ats (ICREA), Passeig de Llu\'{\i}s Companys 23, 08010 Barcelona, Spain\label{aff60}
\and
Institut de Ciencies de l'Espai (IEEC-CSIC), Campus UAB, Carrer de Can Magrans, s/n Cerdanyola del Vall\'es, 08193 Barcelona, Spain\label{aff61}
\and
UCB Lyon 1, CNRS/IN2P3, IUF, IP2I Lyon, 4 rue Enrico Fermi, 69622 Villeurbanne, France\label{aff62}
\and
Mullard Space Science Laboratory, University College London, Holmbury St Mary, Dorking, Surrey RH5 6NT, UK\label{aff63}
\and
Instituto de Astrof\'isica e Ci\^encias do Espa\c{c}o, Faculdade de Ci\^encias, Universidade de Lisboa, Campo Grande, 1749-016 Lisboa, Portugal\label{aff64}
\and
Department of Astronomy, University of Geneva, ch. d'Ecogia 16, 1290 Versoix, Switzerland\label{aff65}
\and
Universit\'e Paris-Saclay, CNRS, Institut d'astrophysique spatiale, 91405, Orsay, France\label{aff66}
\and
INFN-Padova, Via Marzolo 8, 35131 Padova, Italy\label{aff67}
\and
Aix-Marseille Universit\'e, CNRS/IN2P3, CPPM, Marseille, France\label{aff68}
\and
INAF-Istituto di Astrofisica e Planetologia Spaziali, via del Fosso del Cavaliere, 100, 00100 Roma, Italy\label{aff69}
\and
Space Science Data Center, Italian Space Agency, via del Politecnico snc, 00133 Roma, Italy\label{aff70}
\and
INFN-Bologna, Via Irnerio 46, 40126 Bologna, Italy\label{aff71}
\and
University Observatory, LMU Faculty of Physics, Scheinerstrasse 1, 81679 Munich, Germany\label{aff72}
\and
Max Planck Institute for Extraterrestrial Physics, Giessenbachstr. 1, 85748 Garching, Germany\label{aff73}
\and
INAF-Osservatorio Astronomico di Padova, Via dell'Osservatorio 5, 35122 Padova, Italy\label{aff74}
\and
Institute of Theoretical Astrophysics, University of Oslo, P.O. Box 1029 Blindern, 0315 Oslo, Norway\label{aff75}
\and
Jet Propulsion Laboratory, California Institute of Technology, 4800 Oak Grove Drive, Pasadena, CA, 91109, USA\label{aff76}
\and
Felix Hormuth Engineering, Goethestr. 17, 69181 Leimen, Germany\label{aff77}
\and
Technical University of Denmark, Elektrovej 327, 2800 Kgs. Lyngby, Denmark\label{aff78}
\and
Cosmic Dawn Center (DAWN), Denmark\label{aff79}
\and
Max-Planck-Institut f\"ur Astronomie, K\"onigstuhl 17, 69117 Heidelberg, Germany\label{aff80}
\and
NASA Goddard Space Flight Center, Greenbelt, MD 20771, USA\label{aff81}
\and
Department of Physics and Astronomy, University College London, Gower Street, London WC1E 6BT, UK\label{aff82}
\and
Department of Physics and Helsinki Institute of Physics, Gustaf H\"allstr\"omin katu 2, University of Helsinki, 00014 Helsinki, Finland\label{aff83}
\and
Universit\'e de Gen\`eve, D\'epartement de Physique Th\'eorique and Centre for Astroparticle Physics, 24 quai Ernest-Ansermet, CH-1211 Gen\`eve 4, Switzerland\label{aff84}
\and
Department of Physics, P.O. Box 64, University of Helsinki, 00014 Helsinki, Finland\label{aff85}
\and
Helsinki Institute of Physics, Gustaf H{\"a}llstr{\"o}min katu 2, University of Helsinki, 00014 Helsinki, Finland\label{aff86}
\and
Laboratoire d'etude de l'Univers et des phenomenes eXtremes, Observatoire de Paris, Universit\'e PSL, Sorbonne Universit\'e, CNRS, 92190 Meudon, France\label{aff87}
\and
SKAO, Jodrell Bank, Lower Withington, Macclesfield SK11 9FT, UK\label{aff88}
\and
Centre de Calcul de l'IN2P3/CNRS, 21 avenue Pierre de Coubertin 69627 Villeurbanne Cedex, France\label{aff89}
\and
Dipartimento di Fisica "Aldo Pontremoli", Universit\`a degli Studi di Milano, Via Celoria 16, 20133 Milano, Italy\label{aff90}
\and
INFN-Sezione di Milano, Via Celoria 16, 20133 Milano, Italy\label{aff91}
\and
Dipartimento di Fisica e Astronomia "Augusto Righi" - Alma Mater Studiorum Universit\`a di Bologna, via Piero Gobetti 93/2, 40129 Bologna, Italy\label{aff92}
\and
Department of Physics, Institute for Computational Cosmology, Durham University, South Road, Durham, DH1 3LE, UK\label{aff93}
\and
Universit\'e Paris Cit\'e, CNRS, Astroparticule et Cosmologie, 75013 Paris, France\label{aff94}
\and
CNRS-UCB International Research Laboratory, Centre Pierre Bin\'etruy, IRL2007, CPB-IN2P3, Berkeley, USA\label{aff95}
\and
University of Applied Sciences and Arts of Northwestern Switzerland, School of Engineering, 5210 Windisch, Switzerland\label{aff96}
\and
Institut d'Astrophysique de Paris, 98bis Boulevard Arago, 75014, Paris, France\label{aff97}
\and
Institute of Physics, Laboratory of Astrophysics, Ecole Polytechnique F\'ed\'erale de Lausanne (EPFL), Observatoire de Sauverny, 1290 Versoix, Switzerland\label{aff98}
\and
Telespazio UK S.L. for European Space Agency (ESA), Camino bajo del Castillo, s/n, Urbanizacion Villafranca del Castillo, Villanueva de la Ca\~nada, 28692 Madrid, Spain\label{aff99}
\and
Institut de F\'{i}sica d'Altes Energies (IFAE), The Barcelona Institute of Science and Technology, Campus UAB, 08193 Bellaterra (Barcelona), Spain\label{aff100}
\and
DARK, Niels Bohr Institute, University of Copenhagen, Jagtvej 155, 2200 Copenhagen, Denmark\label{aff101}
\and
Centre National d'Etudes Spatiales -- Centre spatial de Toulouse, 18 avenue Edouard Belin, 31401 Toulouse Cedex 9, France\label{aff102}
\and
Institute of Space Science, Str. Atomistilor, nr. 409 M\u{a}gurele, Ilfov, 077125, Romania\label{aff103}
\and
Dipartimento di Fisica e Astronomia "G. Galilei", Universit\`a di Padova, Via Marzolo 8, 35131 Padova, Italy\label{aff104}
\and
Institut f\"ur Theoretische Physik, University of Heidelberg, Philosophenweg 16, 69120 Heidelberg, Germany\label{aff105}
\and
Institut de Recherche en Astrophysique et Plan\'etologie (IRAP), Universit\'e de Toulouse, CNRS, UPS, CNES, 14 Av. Edouard Belin, 31400 Toulouse, France\label{aff106}
\and
Universit\'e St Joseph; Faculty of Sciences, Beirut, Lebanon\label{aff107}
\and
Departamento de F\'isica, FCFM, Universidad de Chile, Blanco Encalada 2008, Santiago, Chile\label{aff108}
\and
Institut d'Estudis Espacials de Catalunya (IEEC),  Edifici RDIT, Campus UPC, 08860 Castelldefels, Barcelona, Spain\label{aff109}
\and
Satlantis, University Science Park, Sede Bld 48940, Leioa-Bilbao, Spain\label{aff110}
\and
Institute of Space Sciences (ICE, CSIC), Campus UAB, Carrer de Can Magrans, s/n, 08193 Barcelona, Spain\label{aff111}
\and
Department of Physics, Royal Holloway, University of London, Surrey TW20 0EX, UK\label{aff112}
\and
Cosmic Dawn Center (DAWN)\label{aff113}
\and
Niels Bohr Institute, University of Copenhagen, Jagtvej 128, 2200 Copenhagen, Denmark\label{aff114}
\and
Universidad Polit\'ecnica de Cartagena, Departamento de Electr\'onica y Tecnolog\'ia de Computadoras,  Plaza del Hospital 1, 30202 Cartagena, Spain\label{aff115}
\and
Infrared Processing and Analysis Center, California Institute of Technology, Pasadena, CA 91125, USA\label{aff116}
\and
Dipartimento di Fisica e Scienze della Terra, Universit\`a degli Studi di Ferrara, Via Giuseppe Saragat 1, 44122 Ferrara, Italy\label{aff117}
\and
Istituto Nazionale di Fisica Nucleare, Sezione di Ferrara, Via Giuseppe Saragat 1, 44122 Ferrara, Italy\label{aff118}
\and
INAF, Istituto di Radioastronomia, Via Piero Gobetti 101, 40129 Bologna, Italy\label{aff119}
\and
Astronomical Observatory of the Autonomous Region of the Aosta Valley (OAVdA), Loc. Lignan 39, I-11020, Nus (Aosta Valley), Italy\label{aff120}
\and
Universit\'e C\^{o}te d'Azur, Observatoire de la C\^{o}te d'Azur, CNRS, Laboratoire Lagrange, Bd de l'Observatoire, CS 34229, 06304 Nice cedex 4, France\label{aff121}
\and
Department of Physics, Oxford University, Keble Road, Oxford OX1 3RH, UK\label{aff122}
\and
Aurora Technology for European Space Agency (ESA), Camino bajo del Castillo, s/n, Urbanizacion Villafranca del Castillo, Villanueva de la Ca\~nada, 28692 Madrid, Spain\label{aff123}
\and
ICL, Junia, Universit\'e Catholique de Lille, LITL, 59000 Lille, France\label{aff124}
\and
ICSC - Centro Nazionale di Ricerca in High Performance Computing, Big Data e Quantum Computing, Via Magnanelli 2, Bologna, Italy\label{aff125}
\and
Instituto de F\'isica Te\'orica UAM-CSIC, Campus de Cantoblanco, 28049 Madrid, Spain\label{aff126}
\and
CERCA/ISO, Department of Physics, Case Western Reserve University, 10900 Euclid Avenue, Cleveland, OH 44106, USA\label{aff127}
\and
Technical University of Munich, TUM School of Natural Sciences, Physics Department, James-Franck-Str.~1, 85748 Garching, Germany\label{aff128}
\and
Max-Planck-Institut f\"ur Astrophysik, Karl-Schwarzschild-Str.~1, 85748 Garching, Germany\label{aff129}
\and
Donostia International Physics Center (DIPC), Paseo Manuel de Lardizabal, 4, 20018, Donostia-San Sebasti\'an, Guipuzkoa, Spain\label{aff130}
\and
IKERBASQUE, Basque Foundation for Science, 48013, Bilbao, Spain\label{aff131}
\and
Laboratoire Univers et Th\'eorie, Observatoire de Paris, Universit\'e PSL, Universit\'e Paris Cit\'e, CNRS, 92190 Meudon, France\label{aff132}
\and
Departamento de F{\'\i}sica Fundamental. Universidad de Salamanca. Plaza de la Merced s/n. 37008 Salamanca, Spain\label{aff133}
\and
IRFU, CEA, Universit\'e Paris-Saclay 91191 Gif-sur-Yvette Cedex, France\label{aff134}
\and
Universit\'e de Strasbourg, CNRS, Observatoire astronomique de Strasbourg, UMR 7550, 67000 Strasbourg, France\label{aff135}
\and
Center for Data-Driven Discovery, Kavli IPMU (WPI), UTIAS, The University of Tokyo, Kashiwa, Chiba 277-8583, Japan\label{aff136}
\and
Dipartimento di Fisica - Sezione di Astronomia, Universit\`a di Trieste, Via Tiepolo 11, 34131 Trieste, Italy\label{aff137}
\and
Jodrell Bank Centre for Astrophysics, Department of Physics and Astronomy, University of Manchester, Oxford Road, Manchester M13 9PL, UK\label{aff138}
\and
California Institute of Technology, 1200 E California Blvd, Pasadena, CA 91125, USA\label{aff139}
\and
Department of Physics \& Astronomy, University of California Irvine, Irvine CA 92697, USA\label{aff140}
\and
Kapteyn Astronomical Institute, University of Groningen, PO Box 800, 9700 AV Groningen, The Netherlands\label{aff141}
\and
Departamento F\'isica Aplicada, Universidad Polit\'ecnica de Cartagena, Campus Muralla del Mar, 30202 Cartagena, Murcia, Spain\label{aff142}
\and
Instituto de F\'isica de Cantabria, Edificio Juan Jord\'a, Avenida de los Castros, 39005 Santander, Spain\label{aff143}
\and
INFN, Sezione di Lecce, Via per Arnesano, CP-193, 73100, Lecce, Italy\label{aff144}
\and
Department of Mathematics and Physics E. De Giorgi, University of Salento, Via per Arnesano, CP-I93, 73100, Lecce, Italy\label{aff145}
\and
INAF-Sezione di Lecce, c/o Dipartimento Matematica e Fisica, Via per Arnesano, 73100, Lecce, Italy\label{aff146}
\and
Institute of Cosmology and Gravitation, University of Portsmouth, Portsmouth PO1 3FX, UK\label{aff147}
\and
Department of Computer Science, Aalto University, PO Box 15400, Espoo, FI-00 076, Finland\label{aff148}
\and
Instituto de Astrof\'\i sica de Canarias, c/ Via Lactea s/n, La Laguna 38200, Spain. Departamento de Astrof\'\i sica de la Universidad de La Laguna, Avda. Francisco Sanchez, La Laguna, 38200, Spain\label{aff149}
\and
Caltech/IPAC, 1200 E. California Blvd., Pasadena, CA 91125, USA\label{aff150}
\and
Ruhr University Bochum, Faculty of Physics and Astronomy, Astronomical Institute (AIRUB), German Centre for Cosmological Lensing (GCCL), 44780 Bochum, Germany\label{aff151}
\and
Department of Physics and Astronomy, Vesilinnantie 5, University of Turku, 20014 Turku, Finland\label{aff152}
\and
Serco for European Space Agency (ESA), Camino bajo del Castillo, s/n, Urbanizacion Villafranca del Castillo, Villanueva de la Ca\~nada, 28692 Madrid, Spain\label{aff153}
\and
Department of Physics and Astronomy, University of the Western Cape, Bellville, Cape Town, 7535, South Africa\label{aff154}
\and
DAMTP, Centre for Mathematical Sciences, Wilberforce Road, Cambridge CB3 0WA, UK\label{aff155}
\and
Kavli Institute for Cosmology Cambridge, Madingley Road, Cambridge, CB3 0HA, UK\label{aff156}
\and
Department of Astrophysics, University of Zurich, Winterthurerstrasse 190, 8057 Zurich, Switzerland\label{aff157}
\and
Department of Physics, Centre for Extragalactic Astronomy, Durham University, South Road, Durham, DH1 3LE, UK\label{aff158}
\and
Institute for Theoretical Particle Physics and Cosmology (TTK), RWTH Aachen University, 52056 Aachen, Germany\label{aff159}
\and
Univ. Grenoble Alpes, CNRS, Grenoble INP, LPSC-IN2P3, 53, Avenue des Martyrs, 38000, Grenoble, France\label{aff160}
\and
INAF-Osservatorio Astrofisico di Arcetri, Largo E. Fermi 5, 50125, Firenze, Italy\label{aff161}
\and
Centro de Astrof\'{\i}sica da Universidade do Porto, Rua das Estrelas, 4150-762 Porto, Portugal\label{aff162}
\and
Instituto de Astrof\'isica e Ci\^encias do Espa\c{c}o, Universidade do Porto, CAUP, Rua das Estrelas, PT4150-762 Porto, Portugal\label{aff163}
\and
HE Space for European Space Agency (ESA), Camino bajo del Castillo, s/n, Urbanizacion Villafranca del Castillo, Villanueva de la Ca\~nada, 28692 Madrid, Spain\label{aff164}
\and
INAF - Osservatorio Astronomico d'Abruzzo, Via Maggini, 64100, Teramo, Italy\label{aff165}
\and
Theoretical astrophysics, Department of Physics and Astronomy, Uppsala University, Box 516, 751 37 Uppsala, Sweden\label{aff166}
\and
Institute of Astronomy, University of Cambridge, Madingley Road, Cambridge CB3 0HA, UK\label{aff167}
\and
Department of Astrophysical Sciences, Peyton Hall, Princeton University, Princeton, NJ 08544, USA\label{aff168}
\and
Center for Computational Astrophysics, Flatiron Institute, 162 5th Avenue, 10010, New York, NY, USA\label{aff169}}

\date{}

\setcounter{page}{1}

\abstract{

    This is the second paper in the HOWLS (higher-order weak lensing statistics) series exploring the usage of non-Gaussian statistics for cosmology inference within \textit{Euclid}. With respect to our first paper, we develop a full tomographic analysis based on realistic photometric redshifts that allows us to derive Fisher forecasts in the ($\sigma_8$, $w_0$) plane for a \textit{Euclid}-like data release 1 (DR1) setup. We find that the five higher-order statistics (HOS) that satisfy the Gaussian likelihood assumption of the Fisher formalism (one-point probability distribution function, $\ell$1-norm, peak counts, Minkowski functionals, and Betti numbers) each outperform the shear two-point correlation functions by a factor of $2.5$ on the $w_0$ forecasts, with only marginal improvement when used in combination with two-point estimators, suggesting that every HOS is able to retrieve both the non-Gaussian and Gaussian information of the matter density field. The similar performance of the different estimators is explained by a homogeneous use of multi-scale and tomographic information, optimized to lower computational costs. These results hold for the three mass mapping techniques of the \textit{Euclid} pipeline, aperture mass, Kaiser--Squires, and Kaiser--Squires plus, and they are unaffected by the application of realistic star masks. Finally, we explored the use of HOS with the Bernardeau--Nishimichi--Taruya (BNT) nulling scheme approach, finding promising results toward applying physical scale cuts to HOS.
    
}


   \titlerunning{HOWLS Paper II - Toward DR1}
   \authorrunning{Euclid Collaboration: Vinciguerra et al.}

   \maketitle

\nolinenumbers

\section{Introduction}
\label{sec:intro}

The accelerated expansion of the Universe (\citealp{1998AJ....116.1009R}, \citealp{Perlmutter+99}), together with recent tensions measured from late and early Universe probes \citep{Riess+16, Hildebrandt+17} are some of the most profound puzzles in modern cosmology. Tracing the large-scale structure (LSS) of the Universe allows us to tackle the questions related to the expansion and tensions by robustly measuring key cosmological parameters (CPs) such as the matter density, $\Omm$; the matter fluctuations, $\sigma_8$; and the dark energy equation of state parameter, $w$. The \textit{Euclid} survey (\citealp{EuclidOverview}) offers an extraordinary opportunity to push these measurements beyond the current state of the art \citep[e.g.,][]{wright2025kidslegacycosmologicalconstraintscosmic, desicollaboration2025desidr2resultsii} by using weak lensing (WL) cosmic shear: the subtle distortion of the shapes of distant galaxies caused by the gravitational influence of LSS.

In WL surveys, classical analyses have traditionally relied on two-point statistics, such as the cosmic shear two-point correlation functions ($\gamma$-2PCFs, see, e.g., \citealp{schneider_twopoint}; \citealp{KilbingerReview}), measuring how the ellipticities of galaxy pairs are correlated as a function of their separation. 
By construction, these probes solely capture the Gaussian properties of the lensing field, thus up to the second-order moment of its distribution.
However, the non-linear evolution of matter -- driven by gravitational interactions -- introduces non-Gaussian features in the small-scale regime. 
This physical effect makes the lensing fields statistically non-Gaussian, therefore encoding information in higher-order moments beyond the variance (second), for example the skewness (third) and kurtosis (fourth), which cannot be captured by the standard two-point statistics. 
An opportunity to better exploit the potential of the data is to use higher-order statistics (HOS). 
In fact, by capturing information about the shape and connectivity of LSS, such probes are sensitive to both the Gaussian and non-Gaussian regimes of information of the lensing maps.
The goal of the higher-order weak lensing statistics (HOWLS) activities in \textit{Euclid} is to exploit the non-Gaussian information that the standard two-point analysis discards, thus improving the standard cosmological inference.

In preparation for the future observational \textit{Euclid} data analyses with HOS, the HOWLS project released a first analysis (\citealp{Ajani-EP29}, \citetalias{Ajani-EP29} hereafter) in which non-Gaussian statistics have been benchmarked against the traditional two-point estimators in a non-tomographic Fisher analysis based on $N$-body simulations mimicking the final and third data release of \textit{Euclid} (DR3, hereafter).
Taken individually, each HOS outperforms two-point correlation functions by a factor of 2 in terms of the forecasted precision within the ($\Omm$, $\sigma_8$) plane.
These results reinforce the possibility to enhance the classical WL analysis, thus encouraging for new studies.

Building on this foundation, we conduct a follow-up analysis that takes essential steps toward applying HOS to the upcoming \textit{Euclid} DR1.
The main effort is to systematically address the challenges of realistic data not tackled in the past work.
First, while in \citetalias{Ajani-EP29} we limited our inference to the ($\Omm$, $\sigma_8$) plane, we now present forecasts in the ($\sigma_{8}$, $w_{0}$) configuration, thereby directly responding to one of \Euclid{}’s primary science goals. 
Since HOS mostly lack theoretical models, dedicated mock catalogs are developed to estimate their covariance and predict their signal in the cosmological parameter space.
We constructed new realistic galaxy mock catalogs relying on the same $N$-body simulations as in \citetalias{Ajani-EP29}, the Scinet LIght-Cones (SLICS; \citealp{SLICS}) for the numerical covariance and DUSTGRAIN-\textit{pathfinder} (\citealp{giocoli18b}) for derivatives estimation, plus two extra simulations with a large variation of $w_0$ to reduce the impact of numerical noise, which prevented us from constraining this parameter in \citetalias{Ajani-EP29}. 

The novel suite of mocks increases the number of realizations and now mimics realistic redshift properties sampled from the Flagship (FS, hereafter) $N$-body simulation (\citealp{castander_fs2}) retrieved from the \texttt{CosmoHub} portal (\citealp{Carretero_2017}; \citealp{TALLADA2020100391}).
In detail, we focus on reproducing a DR1-like redshift distribution $n(z)$ with realistic properties of true redshifts and photometric redshifts. 
In contrast to \citetalias{Ajani-EP29}, this enables 
a tomographic analysis that incorporates both single and combined bin correlations (\citealp{Martinet+21a}), as a way to enhance the sensitivity to the CPs.

Our HOS analysis relies on mass maps that are reconstructed from the (reduced) shear. 
In this work, we implement the three mass mapping algorithms of the official \textit{Euclid} processing pipelines: Kaiser--Squires (KS; \citealp{Kaiser+93}), Kaiser--Squires+ (KS+; \citealp{Pires+20}), and aperture mass (APM; \citealp{Schneider1998}) with external codes.

We also include realistic DR1-like masks based on \textit{Gaia} stars (\citealp{gaiamission_2016}) to add the realism expected with the observational data.
In particular, we compare the HOS performance among the mass mapping algorithms in a mask-free and masked configuration.

We also explore additional aspects that could improve future observational analysis.
We investigate solutions to de-bias the small-scales from known systematics such as the baryonic feedback (e.g. \citealp{Semboloni+13}; \citealp{Martinet+21b}; \citealp{Lee_2022}; \citealp{Ferlito_2022}; \citealp{Broxterman_2024}), 
which represent a potential bottleneck in the extraction of small-scale information, the key feature of HOS. 
To this end, we tested the Bernardeau--Nishimichi--Taruya (BNT; \citealp{Bernardeau:2013rda}) transformation and its alternative approach, BNT smoothing (Taylor et al., in prep.), paving the way for future extensions. 
Finally, we explore the combination of HOS measured from multi-scale filtered mass maps, which was shown to be useful in breaking some parameter degeneracies (e.g. \citealp{Ajani+20}, \citeyear{Ajani_2023}).

We consider a selection of HOS divided in two categories: the one-point statistics including the  probability density function (PDF hereafter, e.g. \citealp{Liu2019}; \citealp{Barthelemy20a}; \citealp{Boyle2020}; \citealp{Thiele2020}, \citeyear{Thiele_2023}; \citealp{Castiblanco:2024xnd}) and the $\ell1$-norm (e.g. \citealp{Ajani_l1norm}; \citealp{sreekanth2024theoreticalwaveletell1normonepoint}), and topological descriptors gathering peak counts (peaks hereafter, e.g. \citealp{Marian+09}; \citealp{Dietrich+10}; \citealp{Kacprzak+16}; \citealp{Martinet+18}, \citeyear{Martinet+21b}; \citealp{Harnois-Deraps+21}, \citeyear{harnoisderaps2024kids1000desy1combinedcosmology}; \citealp{2024MNRAS.528.4513M}; \citealp{2025MNRAS.536.2064G}), Minkowski functionals (MFs hereafter, e.g. \citealp{Kratochvil+12}; \citealp{Petri+15}; \citealp{Vicinanza_2019}; \citealp{Parroni_2020}; \citealp{2025MNRAS.537.3553A}), and Betti numbers (BNs hereafter, e.g. \citealp{Feldbrugge_2019}; \citealp{Parroni_2021}). Compared to \citetalias{Ajani-EP29}, we had to discard our implementation of higher order moments (e.g. \citealp{Porth+21}; \citealp{Gatti2021}), persistent homology (e.g. \citealp{Heydenreich:2021}, \citeyear{Heydenreich:2022}), and scattering transforms (e.g. \citealp{Cheng2020}, \citeyear{Cheng2021b}), as they do not satisfy the Gaussian likelihood approximation assumed by the Fisher formalism.
These probes will be considered again when shifting to a more robust likelihood approach based on a Markov chain Monte Carlo (MCMC) analysis.

On top of that, and differently to our first paper, we selected the same filter and smoothing settings across all HOS. 
This is to decouple the effect of the filtering from the one due to the intrinsic nature of the non-Gaussian probes, enabling a fair assessment of statistics efficiency in extracting the cosmological information. 

The structure of this paper is as follows.
In Sect.~\ref{sec:mocks_production}, we start by introducing the dataset composed of $N$-body simulations used for covariance and derivatives estimation. 
Section~\ref{sec:dv_extraction} details the selected statistics that fulfill the Fisher requirements as well as their treatment in the presence of masked pixels. 
The Fisher formalism is introduced in Sect.~\ref{sec:fisher_forecasts} and includes choice of parameter constraints and the test of the Gaussian likelihood approximation. 
An iterative scheme for optimizing tomographic bin combinations is proposed in Sect.~\ref{sec:tomography}. 
Section~\ref{sec:BNT} explores the BNT transformation for a fixed smoothing scale.
We extend our tomographic analysis to multiple smoothing scales as described in Sect.~\ref{sec:multiscale}, and we present a cross validation of our simulations with theory in Sect.~\ref{sec:theory_validation}.
Section~\ref{sec:mass_mapping_and_masks} examines mass mapping and masking effects, concluding the results with a ranking across Gaussian and non-Gaussian probes in Sect.~\ref{sec:comparing_probes}. 
Finally, Sect.~\ref{sec:conclusions} presents a summary of our findings and outlines future steps on the road to the upcoming analyses of real \textit{Euclid} data. 
%


\section{Mocks}
\label{sec:mocks_production}

\subsection{Simulations}

\begin{table}
\caption{Cosmological parameters used in our $N$-body simulations.}
\begin{tabular}{lccc}
\hline
\hline
  & $\Omm$ & $\sigma_8$ & $w_0$ \\
 SLICS &  0.2905 & 0.826& $-1.0$ \\ 
 DUSTGRAIN FID & 0.31345 & 0.841917 & $-1.0$\\
 DUSTGRAIN $\Omm-$ & 0.2  & 0.841917 & $-1.0$ \\
 DUSTGRAIN $\Omm+$ & 0.4  & 0.841917 & $-1.0$ \\
 DUSTGRAIN $\sigma_8-$  & 0.31345 & 0.707210 & $-1.0$ \\
 DUSTGRAIN $\sigma_8+$ & 0.31345 & 0.976624 & $-1.0$ \\
 DUSTGRAIN $w_0-$ & 0.31345 & 0.841917 & $-0.40$\\
 DUSTGRAIN $w_0+$ & 0.31345 & 0.841917 & $-1.60$\\
   \hline
   \hline
\end{tabular}
\label{table:cosmo}
\end{table}

This work relies on two suites of $N$-body simulations introduced in \citetalias{Ajani-EP29} and that have been expanded since. We will therefore briefly summarize their properties here, focusing only on the novelties, with more detailed discussions available in the references provided below.

The Fisher formalism requires computing derivatives of given estimators, as well as their associated data covariance matrix. Following \citetalias{Ajani-EP29}, these two tasks are achieved with the DUSTGRAIN-{\it pathfinder} \citep{giocoli18a} and the SLICS simulations \citep{SLICS_V1, 
SLICS}, respectively. The DUSTGRAIN-{\it pathfinder} series, or DUSTGRAIN for short, are constructed from $N$-body simulations that evolve $768^3$ particles in cubic volumes of $750 \, h^{-1}$Mpc on the side with MG-Gadget \citep{Puchwein_etal_2013}, in which the CPs ($\Omm$, $\sigma_8$, $w_0$) are varied one at a time, allowing for an accurate estimation of two-point stencil numerical derivatives. The values adopted for the fiducial CPs are consistent with the {\it Planck} 2015 results \citep{planck1_15}, and detailed in Table \ref{table:cosmo}. The parameters not listed in this table are fixed to their fiducial values. For DUSTGRAIN and SLICS, these are set to $(\Omega_{\rm b},h,n_{\rm s}) = (0.0491, 0.6731, 0.9658)$ and $(0.045,0.6898,0.96)$, respectively. Neutrinos are assumed to be massless in both cases. Compared to \citetalias{Ajani-EP29}, the series we use here has new $w_0$ nodes, $-0.40$ ($-60\%$) and $-1.60$ ($+60\%$), to improve the precision of the derivative about this parameter. We also used the DUSTGRAIN fiducial simulations to validate some of our theoretical predictions of HOS. The SLICS simulations are a series of 924 fully independent $N$-body runs that each evolve $1536^3$ particles in a box that is $505 h^{-1}$ Mpc on the side. These are all produced at the same cosmology, also listed in Table \ref{table:cosmo}, but the initial conditions are varied in each case, making this an ideal tool to estimate covariance matrices numerically. About $75\%$ of the super-sample covariance term on two-point statistics is captured by these simulations \citep{cosmo-SLICS}. 

Both sets of simulations provide projected flat-sky mass sheets, $\delta_\mathrm{2D}(\boldsymbol{\theta},z_{\sfont{L}})$, at a series of predetermined lens redshifts, $z_{\sfont{L}}$, from which we constructed past light-cones with at least 18 mass sheets up to $z_{\sfont{S}}=3$. The DUSTGRAIN light-cones have an opening angle of 25 deg$^2$ and an angular resolution of \ang{;;8.8}, while the SLICS are 100 deg$^2$ wide with a resolution of \ang{;;4.8}. These maps are obtained by interpolating the projected particle distributions onto our pixels, after choosing an origin and a direction for our light-cone. Multiple cones are carved out from a single $N$-body run by randomly shifting these two attributes. A total of 256 cones were extracted from the DUSTGRAIN simulations, allowing us to suppress the sample variance associated with the choice of sub-volume used in the light-cone, and therefore achieving higher accuracy in our derivatives. The requirement for the SLICS is the exact opposite: We want to accurately estimate the sample variance with a large number of realizations. Each of the SLICS light-cones was sampled only twice; over-sampling the simulations with multiple light-cones would down-weight the variance due to the initial conditions themselves and provide a potentially biased data covariance matrix. It is shown in \citet{cosmo-SLICS} that this bias is relatively small, even with hundreds of resamplings; however we decided to be conservative and limited this number to a small amount, reaching a total of 1910 light-cones. Each of these were further split into four sub-patches of size $5 \times 5$ deg$^2$, matching the area of the DUSTGRAIN simulations, for a total of 7616 SLICS patches. 

Convergence maps are produced under the Born approximations, an integral over the mass sheets, weighted by the lensing kernel.
Source planes $z_{\sfont{S}}$ are placed at the far end of every projected mass shell,  yielding a set of convergence maps that are next transformed into shear maps using the KS inversion method. These sets of convergence and shear maps are then used to assign lensing quantities to galaxies populating the light-cones, as described in the next section. 

We have established in \citetalias{Ajani-EP29} that these simulations meet the angular resolution requirement for our forecasts, and hence we do not reproduce these tests here. It should be noted, however, that a number of effects not included here will be required for future data analysis, including source clustering, intrinsic alignments of galaxies (e.g. \citealp{2024OJAp....7E..14L}), and baryonic feedback (e.g. \citealp{2024PhRvD.110j3539G}). Furthermore, the number of mass sheets is quite low and will need to be optimized in the future to protect our inference analyses against choices of light-cone hyper-parameters \citep{ZorillaMatilla}.

\subsection{Mock properties}

From the simulations described above, we build \textit{Euclid}-like mocks following the prescriptions of \citetalias{Ajani-EP29}. Briefly, shape-noise is sampled from a Gaussian distribution of zero-mean and standard deviation $\sigma_\epsilon = 0.26$ per component.
This value, as reported in \citet{EuclidIV}, was derived from a sample of galaxies observed by the \textit{Hubble} Space Telescope with photometric characteristics comparable to those predicted for the Visible Camera (VIS, hereafter) dataset, with typical $I_{814}$ magnitudes around 24.5 (\citealp{Schrabback+18}).
Galaxy positions and intrinsic ellipticity amplitudes are identical across cosmologies of the same line-of-sight (LOS) and vary LOS by LOS.

\begin{figure}
    \centering
    \includegraphics[scale=0.45]{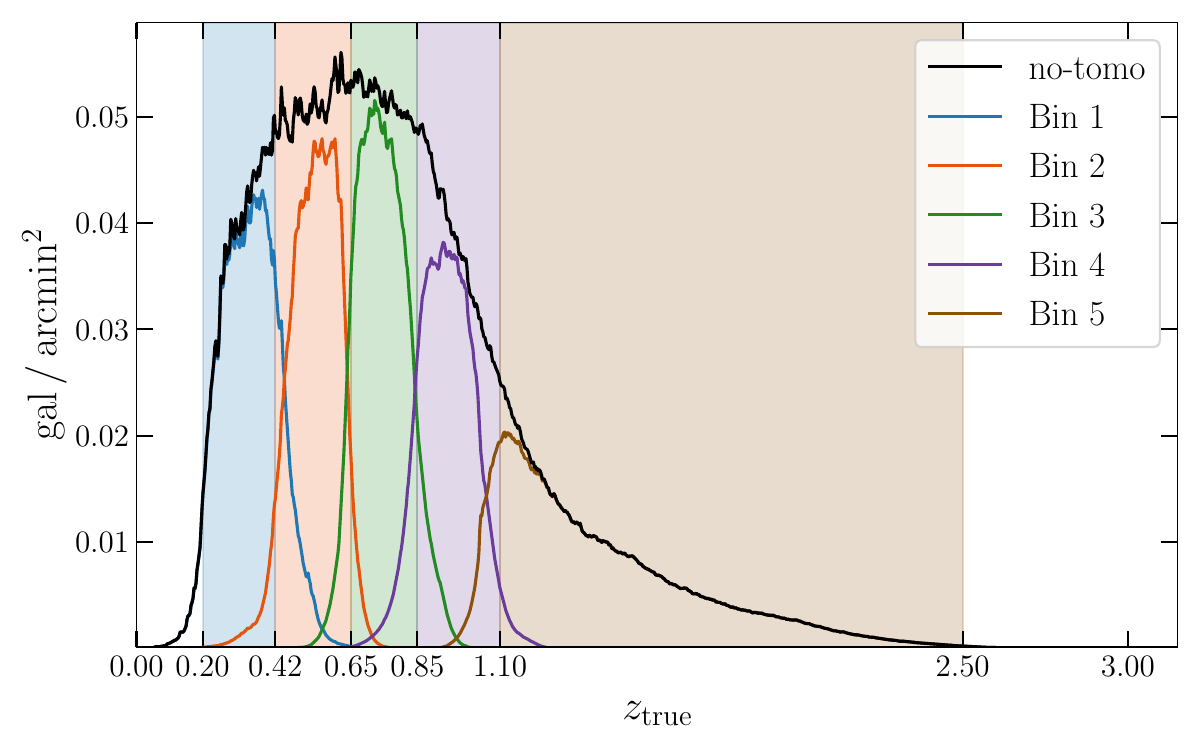}
    \caption{Mean DR1-opt galaxy density distribution as function of true-redshift shown in units of galaxy density for the non-tomographic and five tomographic bins. Shaded areas represent the photometric-redshift edges used to build the five equi-populated slices.}
    \label{fig:DR1_opt_nofz}
\end{figure}

The main difference, however, is that we now include realistic photometric redshifts and galaxy densities. For each mock we pick a random area in the FS catalog, retrieve the true redshift and photometric redshift of every object and assign shear from the corresponding DUSTGRAIN or SLICS simulation at the FS true redshift and using random (RA, Dec) positions uniformly distributed. This ensures the ability to derive tomographic bins based on realistic photometric redshifts with proper overlap of the true redshift distributions across the bins, as shown in Fig.~\ref{fig:DR1_opt_nofz}.
This approach neglects source clustering, which can play an important role \citep{Gatti24_source_clusteringDESY3} and which will need to be modeled in a full data analysis.

\begin{table*}[h!]
    \centering
    \caption{Summary of the statistics investigated in this work, their abbreviations, and the implemented code.}
    \renewcommand{\arraystretch}{1.2}
    {
    \begin{tabular}{c c c c c c c}
        \hline
        \multicolumn{1}{c}{Statistics} & Abbreviation & Algorithm \\
        \hline
        shear two-point correlation functions & $\gamma$-2PCFs ($\xi_{+}/\xi_{-}$) & \texttt{TreeCorr} \\
        convergence two-point correlation function & $\kappa$-2PCF ($\xi_{\kappa}$) & \texttt{TreeCorr} \\
        one-point probability distribution function & PDF ($\mathcal{P}$) & \citetalias{Ajani-EP29},  \citet{Castiblanco:2024xnd} \\
        $\rm \ell1 \! - \! norm$ & $\rm \ell1 \! - \! norm$ & \citet{sreekanth2024theoreticalwaveletell1normonepoint} \\
        peak counts & peaks ($N$) & \texttt{lenspack} \\
        Minkowski functionals & MFs ($V_0, V_1, V_2$) & \citet{Vicinanza_2019} \\
        Betti numbers & BNs ($\beta_0, \beta_1$) & \citet{Parroni_2021} \\
        \hline
    \end{tabular}%
    }
    \label{tab:statistics_summary}
    \tablefoot{All estimators are measured from $\kappa$ (low-pass top-hat filter) and $M_{\mathrm{ap}, \gamma}$ (high-pass Starlet wavelet filter, \citealp{Leonard+12}). Finally, three smoothing scales [$\ang{;2.34;}$, $\ang{;4.68;}$, $\ang{;9.38;}$] are applied to the mass maps prior to the statistic measurement.}

\end{table*}

The FS photometric redshifts are computed from the \Euclid VIS and Near Infrared (NIR, hereafter) bands as well as the Legacy Survey of Space and Time (LSST, hereafter) of the Vera C. Rubin Observatory (\citealp{LSSTSRD}) ground-based simulated photometry, and therefore represent a DR3-like configuration. We nonetheless use the same $n(z)$ for our DR1 mocks. More specifically, our DR3 mocks have a total galaxy density of $24.3$ arcmin$^{-2}$ corresponding to a cut at $\IE \leq24.5$ in the FS mocks while for DR1 we derive two sets of mocks depending on ground-based  photometry cuts: the DR1 optimistic (DR1-opt, hereafter) and pessimistic (DR1-pes, hereafter) cases where we remove objects below the Dark Energy Survey (DES, \citealp{DES}) $10 \, \sigma$ detection limit of the $i$ band and of the four $griz$ DES bands respectively giving non-tomographic galaxy density of $16.2$ arcmin$^{-2}$ and  $8.3$ arcmin$^{-2}$.
Unless for some specific cases where we use the DR3 mocks (BNT in Sect. \ref{sec:BNT}, mass mapping in Sect. \ref{sec:mass_mapping_and_masks}), all our results focus on the DR1-opt configuration.
In the following sections, we refer to the DR1-opt tomographic setup of KS-reconstructed mass maps as our reference configuration.

We explore various tomographic configurations: equi-distant and equi-populated binning and several tomographic slices. For this analysis, we choose to use only the equi-populated framework, the equi-distant option creating galaxy densities at high redshift which are too low for accurate mass reconstruction, and focus on five tomographic bins in the range $0.2 \leq 
\zph \leq 2.5$ as a DR1 realistic option based on the past choices of the KiDS and DES analyses. We have also tested the configuration based on 6 equi-populated redshift bins, finding consistent results.

\section{Data vector extraction}
\label{sec:dv_extraction}

This section briefly describes the data vector (DV) measurements from the mocks introduced in the preceding section. A full description of their implementation can be found in Sect. 3 of \citetalias{Ajani-EP29}.
The probes implemented in this work span two different varieties: one-point (PDF, $\ell$1-norm) tracking the information encoded into the cumulants of the field distribution, and topological (peaks, MFs, BNs) describing the density of stationary points and their connectivity, the main features of the field manifold.
A list of the estimators is presented in Table \ref{tab:statistics_summary}.

In contrast to \citetalias{Ajani-EP29}, all statistics are now compared at the same mass maps and filtering properties, allowing us to decouple the contribution of the mass mapping from that of the estimator itself.

Finally, we describe below how to account for masked pixels in the DV extraction with some new developments in the case of topological descriptors. 
We define a generic 2D Cartesian mask-map $\mathcal{M}$ containing masked (unmasked) pixel as $\mathcal{M}_{ij}$ = \texttt{NaN} (1). 
We account for the mask in each input mask-free map by performing an element-by-element product between the two. \

\subsection{Two-point correlation functions}
\label{two-point_extraction}
The WL $\gamma$-2PCFs ($\kappa$-2PCF) probe the large-scale distribution of matter by measuring the statistical correlation of the shear (convergence) field at two different points on the sky.
We relied on the public code \texttt{TreeCorr}\footnote{\url{https://github.com/rmjarvis/TreeCorr}.} (\citealp{Jarvis+04}) to perform measurements from maps in ten log-bins ranging from the pixel side (\ang{;1.17;}) up to the map diagonal (\ang{;424.26;}).
While in \citetalias{Ajani-EP29} we applied the algorithm at catalog level, relying on a grid reduces the computational time and enables for a fair comparison with the other map-based estimators.
In the presence of a mask, e.g. due to bright star regions, it is sufficient to neglect all pixel pairs which include at least one masked pixel.

\subsection{One-point probability distribution and $\rm \ell1 \! - \! norm$}
\label{pdf_and_l1norm_extraction}

The PDF counts the occurrence of each value of a random variable, in our case either the $\kappa$ or APM field \citep{Barthelemy:2020yva}. As such it provides access to different density environments as well as higher-order information that complement the shear correlation functions. As an example, \citet{Castiblanco:2024xnd} showed that the $\kappa$-PDF brings information about the skewness of the $\kappa$ distribution in different redshift bins enabling an improved complementarity in a tomographic approach. Similarly, the $\rm \ell1 \! - \! norm$, which focuses on the absolute values of the probed field, has been shown to also contain rich cosmological information \citep{Ajani_l1norm, sreekanth2024theoreticalwaveletell1normonepoint}.
It is worth noting that the PDF and the $\rm \ell1 \! - \! norm$ target similar features of the maps: the former counts the number of pixels within specified convergence intervals, while the latter takes the sum of the absolute values of those pixels.
Unlike HOS such as peak and minimum counts, the $\rm \ell1 \! - \! norm$ and PDF efficiently capture information from all values in the map.

As with any statistics involving a filtered field, missing or masked pixels will mix into and thus dilute the signal.
The impact of masked pixels can be reduced by discarding mostly masked regions from the histogram samples and/or modeled through a mode mixing matrix \citep{Gatti19,Barthelemy2024}. 

\subsection{Peak counts}
\label{peakcounts_extraction}
Positive- and negative-valued WL peaks of mass maps represent local-maxima found in overdense (e.g. due to single or superposed massive LSS objects along the LOS) and underdense regions, respectively. Peaks may also be caused by (shape) noise which are not expected to carry any cosmological information (\citealp{2011PhRvD..84d3529Y}).
Consistent with \citetalias{Ajani-EP29}, we probe the distribution of peak heights in signal-to-noise ratio (S/N) bins, where a pixel is detected as a peak if it has a larger value than its eight neighbors. It is worth noting that, alternative features of peaks, such as their steepness, are also explored in the literature (e.g. \citealp{Li_2023}).
The backbone of the algorithm is extracted from the \cite{Peel+18} analysis, whose code is publicly released in the \texttt{lenspack}\footnote{\url{https://github.com/CosmoStat/lenspack/blob/master/lenspack/peaks.py}} repository. 

Detections in masked maps are performed by assigning the map's minimum value to the masked pixels so that none of them can represent a peak.
We remove any detection found in a masked region from our analysis, but peaks can still arise near the masked pixels. 
This effect is not corrected for in the present analysis, but it should be cosmology-independent as the implemented masks do not correlate with LSS.

\subsection{Minkowski functionals}
\label{mfs_extraction}
MFs represent morphological descriptors that quantify the geometry and topology of a scalar field. 
When applied to a mass map, according to Hadwiger’s theorem, they consist of three functionals that characterize the area, boundary length, and Euler characteristic of its excursion sets.
This probe has been measured with the same algorithm considered in Sect. 3.6 of \citetalias{Ajani-EP29} based on the implementation of \cite{Vicinanza_2019}. 
In order to reduce the CPU time, we adopted a coarser initial binning.

We refined the code to take into account masked pixels by propagating the input mask $\mathcal{M}$ to the first- and second-order derivatives needed in Eqs. (27--28) of \citetalias{Ajani-EP29}. 
Specifically, each gradient pixel is well defined as long as the surrounding neighbor-values are not masked. 
As at least one masked pixel is involved in the derivative computation, the gradient in such a coordinate is masked. 
Consequently, the first- and second-order gradients respectively define two wider masks $\mathcal{M}_{,i}$, $\mathcal{M}_{,ij}$ satisfying $\mathcal{M} \subseteq  \mathcal{M}_{,i} \subseteq  \mathcal{M}_{,ij}$. 
Following this approach, we can easily measure the MFs neglecting the masked pixels of the input map, first- and second-order gradient map. 

\subsection{Betti numbers}
\label{bns_extraction}

As described in Sect. 3.7 of \citetalias{Ajani-EP29}, other topological statistics can be used to extract cosmological information from mass maps. 
Among them, we consider the BNs which describe the connectivity of minima, maxima, and saddle points in a field.
We rely on the implementation of \cite{Parroni_2021}.
All the mass map-based HOS introduced so far are measured in masked configurations by explicitly neglecting the masked pixels prior to their measurement.
This is a conservative approach which allows us to avoid any eventual bias arising from compromised observations.
On the other hand, while those algorithms can deal with masked regions, others, such as BNs, would rather give highly unreliable measurements because of the presence of a significant amount of masked pixels.
However, this strategy can be relaxed by probing the mass map reconstructed pixels even if originally masked in the shear maps.
We discuss this in more detail in Appendix \ref{app:noremasking_vs_remasking} where we also show that, in our configuration, not reapplying the mask footprint after the mass reconstruction has a negligible impact on the Fisher forecasts. 
We therefore follow this approach for BNs.

\section{Fisher forecasts}
\label{sec:fisher_forecasts}
We use the Fisher formalism to quantify the amount of cosmological information encoded in our different estimators.
However, with respect to \citetalias{Ajani-EP29}, the increase of the DV-length due to the tomographic approach introduces some new caveats.
We first recall the main equations of the Fisher formalism in Sect. \ref{subsec:fisher_formalism}.
We then explain in Sect. \ref{subsec:fisher_numerical_stability} how numerical instabilities propagate into the Fisher matrix and get amplified in its inverse (ill-conditioning matrix), which justifies our choice to focus on the ($\sigma_{8}$, $w_{0}$) plane for the inference to minimize the impact of numerical noise in our Fisher framework.
Finally, Sect. \ref{subsec:fisher_SMAPE_test} introduces the methodology used to assure the Gaussian likelihood approximation, a fundamental assumption imposed by the Fisher formalism.

\subsection{Fisher formalism}
\label{subsec:fisher_formalism}
The Fisher matrix $\mathsf{F}$, based on a Gaussian-likelihood assumption, quantifies the information content of the data described, in our case, by the CPs. 
It is defined as the expectation value of the second derivative of the log-likelihood (see Eqs. 37--38 in Sect. 4.1 of \citetalias{Ajani-EP29}).

Key ingredients are the data covariance matrix (and its inverse) $\mathsf{COV}(N_{\rm f}, N_{\rm d})$ of the DV ($\mathbf{D}$) measured on $N_{\rm f}=7616$ realizations of the SLICS fiducial cosmology containing $N_{\rm d}$ entries (this number varies across the statistics and configurations), estimated as
\begin{equation} 
\mathsf{COV} = \left\langle \, (\mathbf{D} - \langle \mathbf{D} \rangle) \, (\mathbf{D} - \langle \mathbf{D} \rangle)^T \, \right\rangle \, ,
\end{equation}
as well as DV model derivatives $\partial \mathbf{D} / \partial p_{\alpha}$ with respect to $N_{\rm p} = (2, 3)$ model parameters $p_{\alpha} = (\Omm, \sigma_{8}, w_{0})$\footnote{Note that we consider the full set of CPs $(\Omm, \sigma_{8}, w_{0})$ only to investigate the numerical noise stability.}. 
Under the assumption of a cosmology-independent covariance, the Fisher matrix elements $\mathsf{F}_{\alpha \beta}(N_{\rm f}, N_{\rm d}, N_{\rm c}, N_{\rm p})$ simplify to
\begin{equation} \mathsf{F}_{\alpha \beta} = \frac{\partial \mathbf{D}}{\partial p_{\alpha}} \, \mathsf{COV}^{-1} \, \frac{\partial \mathbf{D}}{\partial p_{\beta}}^{\rm T} \, , \end{equation}
where the set of DV model derivatives are numerically computed via finite differences across $N_{\rm c}$ realizations.
These are computed both from simulations and theory when possible.
Because of the Cramer–Rao inequality, the inverse of the Fisher matrix $C$ provides a lower limit to the marginalized errors, which include all degeneracies with respect to other parameters.
These errors can be estimated from its diagonal elements as
\begin{equation}
\sigma(p_{\alpha}) = C_{\alpha \alpha}^{1/2} = \left( [\mathsf{F}^{-1}]_{\alpha \alpha} \right)^{1/2} \, ,
\end{equation}
and we report them as percentage values with respect to the fiducial CPs [i.e., $\delta p_{\alpha} (\%) = 100 \, \sigma(p_{\alpha}) \, / \, p_{\alpha}^{\rm fid} \,$].
In the Fisher formalism, the ability to constrain parameters is also quantified by the figure of merit (FoM), defined as
\begin{equation}\label{eq:fom}
    \mathrm{FoM}_{\alpha \beta} = \sqrt{\det \left( \Tilde{\mathrm{F}}_{\alpha\beta} \right)} \, ,
\end{equation}
where $\Tilde{\mathrm{F}}_{\alpha\beta}$ is the marginalized Fisher submatrix for the parameters $p_{\alpha}$ and $p_{\beta}$. The FoM is inversely proportional to the area of the Fisher ellipse in the $(p_{\alpha}, p_{\beta})$ plane. Additionally, the Fisher matrix is instrumental in determining the degeneracy directions in parameter space. 

In the rest of the analysis, consistent with \citetalias{Ajani-EP29}, we assume a linear rescaling of the data covariance matrix from the simulation area to the desired survey area (DR3\footnote{DR3 area is expected to be at most $14 \, 000 \, \rm deg^2$. However, lacking of a definitive coverage, we choose the same value consistently with \citetalias{Ajani-EP29} to ease any comparison in this setup.}: $15 \, 000 \, \rm deg^2$, DR1: $1900 \, \rm deg^2$).
Finally, we corrected for the noise due to the finite number of covariance realizations, $N_{\rm f}$, by multiplying the inverse Fisher matrix with the correction factor $s$ (see Eq. 28 of \citealp{Sellentin+17}):
\begin{equation}
s = \frac{(N_{\rm f} - 1)}{(N_{\rm f} - N_{\rm d} + N_{\rm p} - 1)} \, .
\end{equation}

Although the correction above accounts for the fact that for a low number of data realizations the likelihood follows a $t$-Student distribution rather than a Gaussian one, it does not include the \citet{Dodelson+13} effect which quantifies biases to the likelihood due to the use of a particular random realization of the covariance itself. Following \citet{Percival+22}, this second effect can be calculated as a multiplicative factor to $s$ of the form
\begin{equation}
    s' = 1+B \, (N_{\rm d}-N_{\rm p}) \, ,
\end{equation}
where $B$ is given in Eq. (55) of the mentioned paper
\begin{equation}
    B = \frac{N_{\rm f} - N_{\rm d} - 2}{(N_{\rm f} - N_{\rm d} - 1)(N_{\rm f} - N_{\rm d} - 4)} \, .
\end{equation}
Due to the large number of SLICS realizations for the covariance matrix, in our analysis this factor would result in inflating the 1D confidence intervals by less than 1\% for most of our probes (with multi-scale tomographic DV length of the order of a few hundreds elements) and up to 6\% for the longest DV (MFs with 890 elements). Given the uncertainty imposed by the ill-conditioning of the Fisher matrix, we neglect this correction in the present analysis. We however highlight that it should be accounted for in a future MCMC approach unless we increase further $N_{\rm f}$ or improve on a compression scheme for our DVs \citep[e.g.,][]{Homer+25}.

\subsection{Numerical stability: Ill-conditioning}
\label{subsec:fisher_numerical_stability}
Multiple studies have attempted to characterize the bias in the Fisher forecast estimation due to the numerical noise imprinted on the data covariance matrix and model derivatives (e.g. \citealp{Vallisneri_2008}).
Depending on the properties of the input DVs, different robustness tests can be implemented by relying on theoretical models (e.g. \citealp{wilson2024fishersmiragenoisetightening}) or on a more numerical approach (e.g. \citealp{Yahia_2021}).
However, numerical instabilities in a matrix can also be amplified by its inversion, a collateral effect which commonly takes the name of ill-conditioning.
Specifically, an ill-conditioned matrix refers to a matrix that is nearly singular, meaning that its determinant is close to zero or its condition number (i.e., the ratio between the largest eigenvalue to the smallest eigenvalue, \citealp{belsley2005regression}) is very large.
In the Fisher context, small perturbations of an ill-conditioned Fisher matrix result in large deviations in the error data covariance matrix, thus compromising the reliability of the inferred constraints. 

Lacking analytical models for most of the probes, we assess the stability of the data covariance matrix, Fisher and error covariance matrices following a pure numerical approach, inspired by the vibration technique reviewed in \citet{Yahia_2021}.
The numerical stability of a data covariance matrix $\mathsf{COV}(N_{\rm f}, N_{\rm d})$ is characterized by estimating the average residuals $\delta_{\mathsf{COV}}$ [where $\delta_{\mathsf{COV}} = \langle \, 1 - \mathsf{COV}(N_{ \rm f}) \, / \, \mathsf{COV}(N_{ \rm f}^{\rm max}) \, \rangle$] across its elements for an increasing number of data covariance realizations $N_{\rm f}$.
This test is implemented for the Fisher matrix $\mathsf{F}(N_{\rm f}, N_{\rm d}, N_{\rm c}, N_{\rm p})$ as well as its inverse $C(N_{\rm f}, N_{\rm d}, N_{\rm c}, N_{\rm p})$ estimating respectively $\delta_{\mathsf{F}}$ and $\delta_{C}$ for varying number of realizations per cosmology $N_{\rm c}$.
These residuals reflect the order of convergence hampered by the presence of numerical noise.

We explore the stability behavior of $ \delta_{\mathsf{COV}}, \delta_{\mathsf{F}}, \delta_{C}$ when using $N_{\rm p}$ = 3 (i.e., varying $\Omm$, $\sigma_{8}$, $w_{0}$), and $N_{\rm p}$ = 2 (i.e., varying 2 parameters and fixing the third one, for all the combinations).
In this step, we aim to retrieve consistent results when changing probe, smoothing scale and tomographic configuration.
A clear symptom of instability occurs when the residual $\delta$ shows no strong evidence of convergence as the number of realizations increases.
This is the case for the data covariance matrix when $N_{\rm d} \approx N_{\rm f}$ (i.e., $\delta_{\mathsf{COV}} \approx 30\%$).
Instead, a sign of ill-conditioning can be noticed whenever $\delta_{\mathsf{COV}^{-1}} \gg \delta_{\mathsf{COV}}$ or $\delta_{C} \gg \delta_{\mathsf{F}}$.

In agreement with the literature cited above, we find that ill-conditioning depends on several aspects of the analysis. These include the dimensionality of both the data and the model, the correlation between data and model parameters, the sensitivity to model parameters, and the numerical estimation of covariance and derivatives, including the step $\Delta p_{\alpha}$ used for the latter.
Overall, we found evidence of ill-conditioning on both the data covariance and Fisher matrices.
However, we checked that the impact of the ill-conditioned data covariance matrices is negligible in the forecasts computation. 
On the other hand, the ill-conditioning of the Fisher matrices strongly affects the parameter constraints, leading to scatter when varying the derivative realizations $N_{\rm c}$. 
In detail, configurations with $N_{\rm p}=2$ generally result to be less ill-conditioned ($\delta_{C} \approx [1.1, \,1.3] \, \delta_{\mathsf{F}}$) than $N_{\rm p}=3$ ($\delta_{C} \approx [1.3, \,1.5] \, \delta_{\mathsf{F}}$), minimizing the effect when probing less degenerate CPs, i.e., ($\Omm$, $w_{0}$) and ($\sigma_{8}$, $w_{0}$) ($\delta_{C} \approx 1.1 \, \delta_{\mathsf{F}}$).
Motivated by these results, we set our fiducial Fisher analysis to the 2-dimensional model parameter plane defined by ($\sigma_{8}$, $w_{0}$).
In addition to being less ill-conditioned, such a configuration allows us to constrain the most interesting CPs probed by HOS, now including the dark energy equation of state parameter $w_{0}$. 
We stress here that the size and orientation of parameter errors highly depend on the model parameter dimensionality. Hence, results obtained in this paper for ($\sigma_{8}$, $w_{0}$) cannot be fairly compared with a similar Fisher analysis in which a marginalization procedure is performed.

\subsection{SMAPE test}
\label{subsec:fisher_SMAPE_test}
The Fisher formalism relies on the assumption of Gaussianity for the likelihood near its maximum. However, this assumption is explicitly violated by both HOS (\citetalias{Ajani-EP29}) and two-point correlation functions \citep{sellentin2018,Lin:2019omj, Oehl:2024gbm}. To avoid introducing bias into our Fisher analysis, we iteratively exclude DV elements starting from the DV domain edges until deviations from Gaussianity are sufficiently suppressed. Before this step, a rebinning of the DVs is performed to increase the S/N, by combining groups of $N$ consecutive bins with $N$ kept constant throughout the entire DV. We checked that our results are independent of the rebinning, provided that we avoid regimes of strong under- and over-sampling of the DVs.

As in \citetalias{Ajani-EP29}, the metric used to quantify the distance between the SLICS dataset and a Gaussian distributed one is the weighted average SMAPE (symmetrized mean absolute percentage deviation, \citealp{Rizzato:2022hbu}). We computed the quantity
\begin{equation}
    y_i = \left(\textbf{D}_{i} - \langle \textbf{D} \rangle \right)^\mathrm{T}  \mathsf{COV}^{-1}  \left(\textbf{D}_{i} - \langle \textbf{D} \rangle \right) \, ,
\end{equation}
where $\textbf{D}_{i}$ is the DV obtained from the $i$-th SLICS map, with a length of $N_{\rm d}$; $\langle \textbf{D} \rangle$ denotes the mean computed across $N_{\rm f}$ SLICS realizations, and $\mathsf{COV}$ refers to the numerical data covariance matrix derived from the complete set of SLICS simulations. For Gaussian distributed DVs and in the limit of $N_{\rm d}\ll N_{\rm f}$, $y_i$ follows a $\chi^2$ distribution, the SMAPE between $y_i$ and $\chi^2$ therefore quantifies the departure from Gaussianity. However, in the tomographic setup, the number of degrees of freedom approaches the number of realizations, $N_{\rm f}$, and hence we cannot rely on the $\chi^2$ as our target distribution. We thus compute the $y_i$ vector for 500 realizations of a Gaussian dataset, generated from a multivariate Gaussian distribution with the same mean and covariance as SLICS. Our target distribution is the mean across the 500 Gaussian realizations of the probability distribution $P_{\text{gauss}}\left(y_i\right)$.
We checked that 500 realizations safely lead to convergent results. Following the same approach as \citetalias{Ajani-EP29}, we compute the observed SMAPE value, $\mathcal{S}_{\text{obs}}$, by comparing the $y_i$ computed from the SLICS dataset to the target distribution. The Gaussianity threshold, $\mathcal{S}_{\text{lim}}$, is then computed as 
\begin{equation}
    \mathcal{S}_{\text{lim}} = \langle \mathcal{S}_{\text{obs}}^{\text{gauss}} \rangle + 2 \, \sigma \left( \mathcal{S}_{\text{obs}}^{\text{gauss}} \right) \, ,
\end{equation}
where we take the mean and variance across the 500 Gaussian realizations.

We note that we iteratively remove the elements from each tomographic data vector, starting from the edges of the DV domain. The SMAPE Gaussianity test is then applied to the vector obtained by concatenating all the individual DVs from each tomographic bin. This results in slightly over-aggressive tail cuts, thus ensuring the Gaussianity of the dataset built from any subsample of the tomographic bins.

\section{Tomography}
\label{sec:tomography}

The constraining power of WL HOS can be enhanced through tomographic analysis, particularly when considering both auto- and cross-correlations among the tomographic bins (\citealp{Martinet+21a}).
We present in Sect. \ref{subsec:tomography_tomo-strategy} the strategy followed to build the whole ensemble of tomographic mass maps reconstructed from combined redshift-slices, establishing an iterative scheme to preserve the cosmological information while reducing the dimensionality of the tomographic DVs.
We discuss the results of this optimal tomographic setup and compare it with other configurations in Sect. \ref{subsec:tomography_tomo-results}.

\subsection{Tomographic strategy}
\label{subsec:tomography_tomo-strategy}

\begin{figure*}
    \centering
    \includegraphics[scale=0.52]{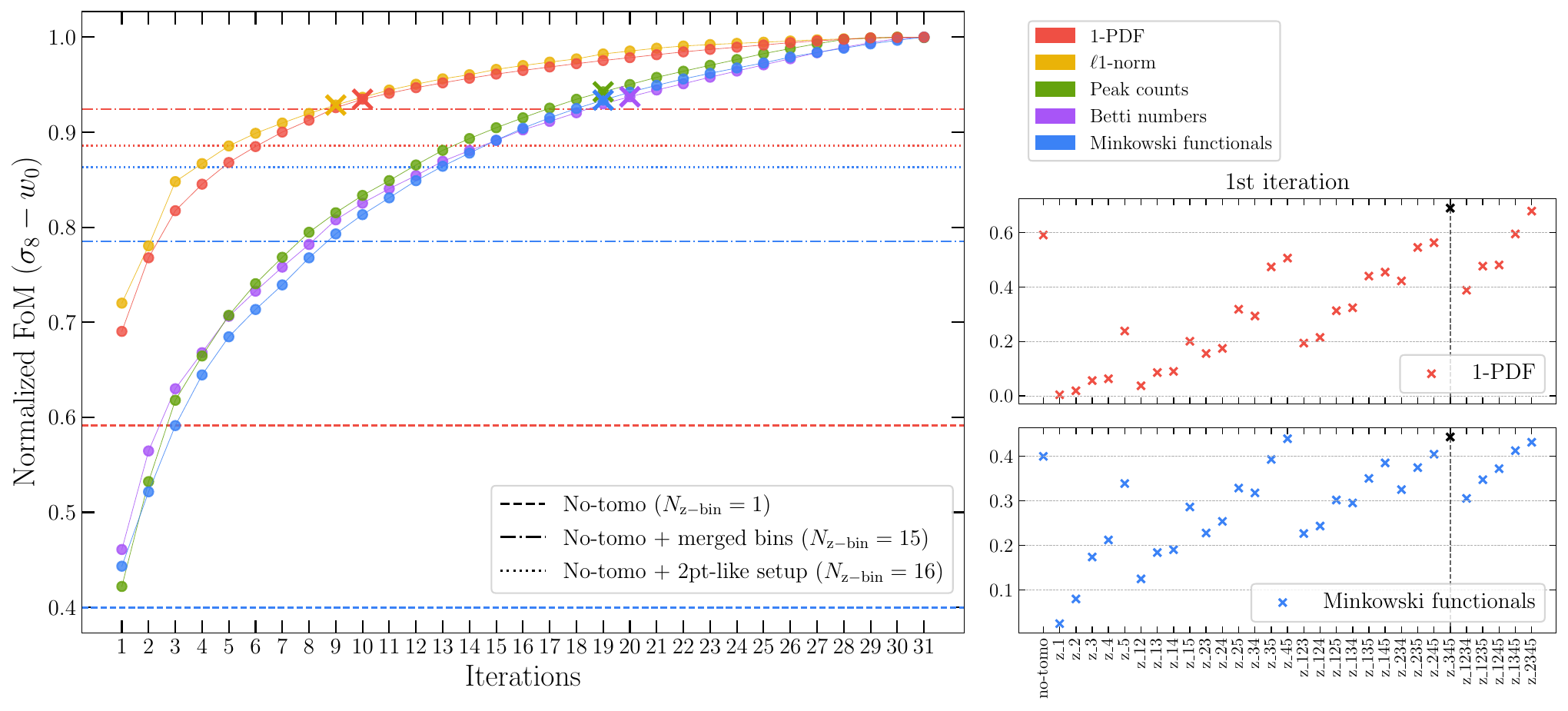}
    \caption{Optimal tomographic DV selection for the reference setup with a smoothing radius of $\ang{;2.34;}$. \textit{Main panel}: Colored dots show the normalized FoM for each HOS at successive iterations of the greedy algorithm, descibed in Sect. \ref{subsec:tomography_tomo-strategy}. The FoM for each statistic is normalized to its full-tomography value. The colored crosses indicate the iteration where the $1\%$ improvement threshold is reached; bins beyond this point contribute negligible information and were discarded. Horizontal lines mark reference configurations: non-tomographic (dashed), non-tomographic with merged bins (dash-dotted), and non-tomographic with two-point-like setup (dotted). Only $\kappa$-1PDF and $\kappa$-MFs are shown as representatives of one-point and topological statistics. \textit{Right panels}: FoM contribution from individual tomographic bins shown for $\kappa$-1PDF and $\kappa$-MFs, with the most constraining bin highlighted in black.}
    \label{fig:tomo_selection}
\end{figure*}

We optimize the extraction of the tomographic WL signal by investigating the correlation between redshift bins (\citealp{Martinet+21a, Castiblanco:2024xnd}). Starting from five equi-populated tomographic slices, we generate shear maps for all possible combinations of redshift bins, ranging from single bins ($z_i$, with $i \in [1,5]$) to quadruplets ($z_i \cup z_j \cup z_k \cup z_l$, with $i,j,k,l \in [1,5]$), resulting in a total of 30 tomographic shear maps.

When extending the tomographic approach from single to combined bins, some of the cosmological information contained in the resulting 30 shear maps becomes redundant (\citealp{Martinet+21a}), unnecessarily increasing the uncertainty in the covariance. We thus apply a greedy \citep{greedy_book} algorithm to determine the minimum number of redshift bins required to extract most of the cosmological information. At each iteration, the algorithm selects the locally optimal choice, ultimately yielding heuristic solutions that closely approximate the global optimum. In the initial step, we compute the Fisher forecast for each of the 31 redshift bins (non-tomographic, single bins, and combined bins) and select the bin that provides the highest ($\sigma_{8}$, $w_{0}$) FoM. We then iteratively selected the tomographic bin combinations that maximize the FoM of the DV formed by concatenating it with all previously selected z-bins. This approach is equivalent to iteratively excluding the tomographic bins that provide the least contribution to the eigenvectors of the inverse data covariance matrix. To reduce the size of the final DV and avoid the inclusion of redundant redshift slices, we set a $1\%$ threshold on the minimum FoM improvement required at each iteration for a tomographic bin to be included. In this work, we adopt the output of the greedy algorithm as our optimal tomographic DVs, discarding any redshift slices that improve the ($\sigma_{8}$, $w_{0}$) FoM by less than $1\%$ relative to the cumulative signal of the more informative bins.

\subsection{Tomography results}
\label{subsec:tomography_tomo-results}

The core result of our tomographic analysis is the substantial improvement enabled by redshift slicing compared to the non-tomographic configuration. For one-point estimators, the enhancement in the FoM ranges from approximately $70\%$ at $\ang{;2.34;}$, where shape noise is more significant, to around $150\%$ at $\ang{;9.38;}$. In the case of topological statistics, the improvement is consistently around $150\%$ across all scales, indicating a higher robustness to the effects of shape noise. 

In the context of optimal tomographic DV selection, we find that a significant reduction in dimensionality can be achieved with minimal information loss, as shown in Fig. \ref{fig:tomo_selection}. For one-point statistics, the optimal configuration retains approximately one-third of the tomographic bins at the $\ang{;2.34;}$ smoothing scale and about one-half at $\ang{;9.38;}$, while for topological estimators, around two-thirds of the bins are selected across all scales. Discarding the unselected DVs results in only a $\approx [7, 9]\%$ decrease in the FoM compared to the full tomographic DV, independent of the statistic or scale. This loss remains within the ill-conditioning uncertainty of our analysis.

In Fig. \ref{fig:tomo_selection}, we show the tomographic selection for the five HOS in our fiducial configuration (DR1-opt, KS reconstruction, $\ang{;2.34;}$ smoothing radius). The horizontal lines show the FoM associated to specific configurations: the merged bins (dash-dotted) comprise the non-tomographic slice, the single bins and the combined bins constructed from consecutive tomographic slices (\citealp{Castiblanco:2024xnd}). They are explicitly designed to mitigate shape noise. The two-point-like configuration (dotted) is built through the concatenation of the non-tomographic bin with the single bins and the pairs.

For one-point estimators, the optimal subset of DVs yield constraints comparable to those obtained from the merged bins and the two-point-like configuration, despite involving roughly $30\%$ fewer elements. However, we can see that the merged bins always slightly outperform the two-point-like setup. In contrast, for topological statistics, the optimal selection improves the FoM by approximately $15\%$ over the merged bins case and by $\approx [5, 10]\%$ with respect to the two-point-like set of tomographic bins, though at the cost of $\approx30\%$ more elements.

It is important to highlight that the most impactful selections occur within the first four iterations of the optimization procedure, where the merged bins involving the highest redshift slice, namely $z_5$, $z_{45}$, $z_{345}$, $z_{2345}$ and the non-tomographic bin, are preferentially selected. The redshift slice $z_5$ indeed carries most of the WL information, as photons from high-redshift sources travel through a larger volume of large-scale structures. Since bin $z_5$ is strongly affected by shape noise, merging this tomographic slice with adjacent bins effectively mitigates noise, enabling a more robust extraction of the WL signal. For one-point statistics, this subset of five bins alone captures approximately $85\%$ of the total information at small smoothing scales, and around $70\%$ at larger scales. For topological estimators, the corresponding fractions are about $70\%$ and $60\%$, respectively. Beyond this core set, subsequent iterations enter a regime where the Fisher ill-conditioning uncertainty dominates, thus rendering further selection steps less interpretable and statistically less robust. 

Throughout this paper, we use the optimal tomographic DV for our analysis, as it enables an almost lossless compression of information when combining multiple scales. Nevertheless, for the upcoming DR1 analysis, we recommend using the merged bins for one-point statistics and the two-point-like setup for topological estimators. As well-defined sets of tomographic bins, they can be selected a priori, thus avoiding a case-by-case optimal selection, which would be unfeasible within the MCMC framework. Moreover, they constitute the subset of redshift slices that most closely approximate the optimal DV, independently of the smoothing scale, and allow for a $50\%$ reduction (or even greater when considering a larger number of tomographic slices) in the number of mass maps to be computed relative to the full tomographic configuration. When combining multiple scales, an appropriate compression technique should be applied to prevent issues in the computation of the data covariance matrix, which may arise due to the limited number of available simulations. It is important to emphasize that we also tested the configuration with six equi-populated bins, finding consistent results across both realistic tomographic setups expected for DR1. However, these results are specific to the framework of our analysis, which includes only shape noise and neglects other sources of systematic error. Therefore, further validation within a fully realistic setup will be necessary for the DR1 analysis.

\section{BNT transform}
\label{sec:BNT}
The broad lensing efficiency kernels make the lensing signal of galaxies within a narrow tomographic bin sensitive to structure across a wide range of redshifts. This leads to scale-mixing and strong correlations when the tomographic approach is used. The BNT transformation (\citealp{Bernardeau:2013rda}) has been proposed as a method to effectively disentangle scales and separate the cosmological information content of the different redshift bins. This effect helps in better targeting the small physical scales, which are hard to model accurately, and that are often cut or smoothed (e.g. \citealp{DES:2021vln}). As our mocks do not include baryonic feedback or other small-scale physical effects, we focus here on validating the BNT transformation and the associated smoothing schemes -- specifically through HOS -- at the level of shear and convergence maps. We do not attempt to determine the specific scales that should be smoothed to mitigate small-scale physics biases.

In this section, we first describe how we implement the BNT transformation for single and combined bins as well as the two smoothing schemes that we tested within the BNT framework. Then, in Sect. \ref{subsec: BNT-results}, we discuss the results from these two approaches and compare them with that of the standard HOS analysis, where no BNT is applied. We finally discuss a possible way forward for the DR1 analysis in the BNT setup.


\subsection{BNT implementation}
\label{subsec:BNT-implementation}
The BNT technique is a nulling scheme that extends the initial proposition of \citet{Joachimi:2008ea} that aims at separating different contributions. The BNT transformation is based on a linear combination of tomographic bins, which results in a set of lensing efficiency kernels that can be localized in redshift. The standard kernels can be defined as
\begin{equation}
\label{standard_kernels}
    q_i(\chi) = \frac{3}{2} \Omm \left( \frac{H_0}{c} \right)^2 \frac{\chi}{a(\chi)} \int_\chi^{\chi_{\mathrm{H}}} \dd\chi' \, n_i(\chi') \, \frac{\chi' - \chi}{\chi'} \, ,
\end{equation}
where the subscript $i$ points to the $i$-th tomographic bin, with galaxy distribution $n_i(\chi)$, and $\chi_{\mathrm{H}}$ is the comoving distance to the horizon. The BNT transformation acts as a linear combination on the set of lensing kernels 
\begin{equation}\label{BNT_kernels}
    \Tilde{q}_i(\chi) = \sum_j M_{ij} \, q_j(\chi) \, ,
\end{equation}
where the elements of the BNT matrix, $M$, can be derived for continuous redshift distributions through the following constraint equations
\begin{equation}
\begin{aligned}
M_{ii} &= 1 \,, \quad
M_{ij} = 0 && \text{for } i<j \text{ or } j<i{-}2 \,, \\
\sum_{j=i-2}^{i} M_{ij} &= 0 \,, \quad
\sum_{j=i-2}^{i} M_{ij} \, B_{j} = 0 \,.
\end{aligned}
\end{equation}
where
\begin{equation}
    B_j = \int_0^{\chi_{\mathrm{max}}} \dd \chi \, \frac{n_j(\chi)}{\chi} \, ,
\end{equation}
and $\chi_{\mathrm{max}}$ is the maximum comoving distance observed by the survey.
In this paper, the BNT matrix is computed using the publicly available code from \citet{Taylor:2020zcg}\footnote{\href{https://github.com/pltaylor16/x-cut}{https://github.com/pltaylor16/x-cut}}, taking into account overlapping tomographic bins due to photometric redshift uncertainty. It is worth noticing that the BNT matrix carries negligible cosmological dependence, as shown in \citet{Bernardeau:2013rda} for discrete source planes and \citet{Taylor:2020zcg} for the continuous case. Therefore, we always compute the transformation matrix at the DUSTGRAIN fiducial cosmology throughout the analysis.

While the BNT transformation can be implemented in the theoretical treatment of the two-point correlation function (\citealp{Taylor:2020imc}), in the case of the HOS -- most of which lack theoretical modeling -- we have to directly measure the DVs on the BNT-transformed simulated shear fields. 
Due to the tomographic slicing, however, the WL maps may show empty pixels where no source galaxies were observed. 
Since the BNT transformation combines each redshift bin with the previous two, these blank pixels propagate through the tomographic maps, resulting in a reduced effective survey area. To mitigate this effect, the propagation of the missing data has been limited to the involved $z$-bin triplets rather than masking the whole line-of-sight. Moreover, contrary to the rest of the paper, which makes use of the DR1-opt configuration, we focus here on the DR3 configuration, where the presence of empty pixels is highly reduced.

Since the primary goal of the BNT transformation is to reduce the mixing of signals from different scales -- thereby preserving valuable cosmological information when cutting or smoothing small-scales -- we investigate two alternative smoothing schemes within the BNT setup. On the one hand, we apply the BNT transformation to the single bin shear maps, perform the mass mapping reconstruction to obtain the BNT-transformed convergence maps, and then apply the smoothing before extracting the summary statistics. In the second approach (BNT smoothing, hereafter), also presented by Taylor et al. (in prep.), we apply the BNT transformation to the single bin shear maps, smooth the transformed maps, and then apply the inverse BNT transformation before performing the mass mapping reconstruction and measuring the HOS. Both schemes are designed to better target the physical scales of interest at different redshifts.

In addition to the BNT-transformed single bins, we construct combinations,  $\left< \gamma_{i\,...\,j}^{\text{BNT}} \right>$, where each index points to the tomographic slices considered in the combined bin. Since it is not possible to directly compute the transformation matrix for the combined bins, we first apply the BNT transformation to each single bin, $\left< \gamma_{i} \right>$. Next, we de-normalize the shear maps by multiplying each pixel, $p$, by the number of galaxies it contains, $N_{\text{gal},i p}$. We then combine the single bins to form combined bins by summing the de-normalized shear maps pixel by pixel. Finally, we renormalized the combined bin shear map by dividing each pixel by the total galaxy count in that pixel, obtained by summing the galaxy counts from the constituent single bins,
\begin{equation}\label{BNT_combined_bins}
    \left< \gamma_{i\,...\,j}^{\text{BNT}} \right>_p = \frac{\sum_{k \in [i,\,...,\,j]} N_{\text{gal},k p} \, \left< \gamma_{k}^{\text{BNT}} \right>_p}{\sum_{k \in [i,\,...,\,j]} N_{\text{gal},k p}} \, .
\end{equation}

\begin{figure}
    \centering
    \includegraphics[scale=0.46]{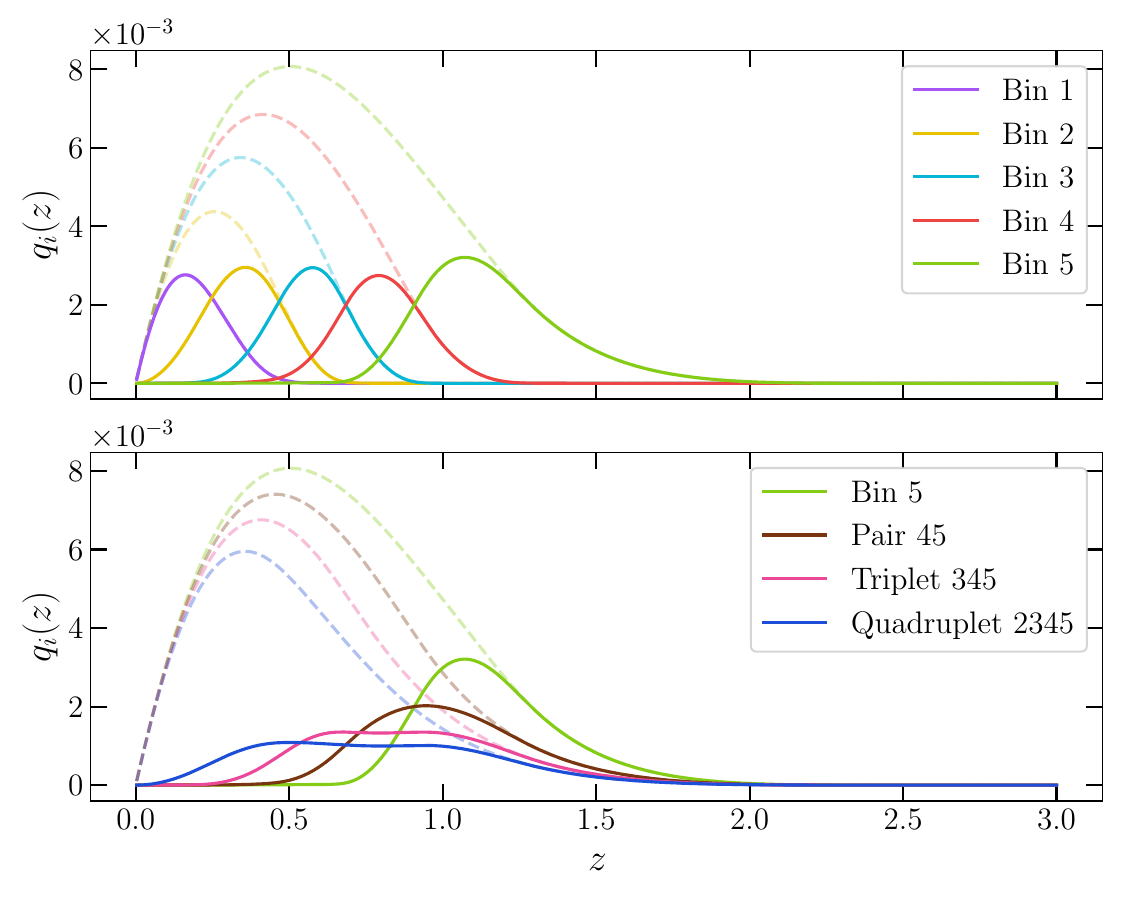}
    \caption{Lensing efficiency kernels for the DR3 configuration. The dashed lines represent the standard kernels (Eq. \ref{standard_kernels}), while the BNT-transformed kernels (Eq. \ref{BNT_kernels}) are shown in solid lines. Top-panel: Kernels for the single bins. Bottom panel: Kernels for the merged bins involving $z_5$.}
    \label{fig:lensing_kernels}
\end{figure}

In Fig. \ref{fig:lensing_kernels}, we compare the standard lensing kernels, described in Eq. (\ref{standard_kernels}), with those transformed using the BNT formalism, as defined in Eq. (\ref{BNT_kernels}). The upper panel shows the kernels corresponding to the single bins. The standard kernels display significant overlap, leading to scale mixing in the WL signal, whereas the BNT transformation produces almost disentangled single bin kernels. However, they are significantly suppressed, as they are now sensitive to a much narrower range of lenses. In the lower panel, we present the kernels associated with the merged bins involving the highest redshift slice ($z_{5}$, $z_{45}$, $z_{345}$, $z_{2345}$). Due to the overlap of the redshift distributions, the transformed kernels still exhibit a high degree of superposition. Nevertheless, they allow for the investigation of correlations between structures at different redshifts.

\begin{figure*}
    \centering
    \includegraphics[scale=0.4]{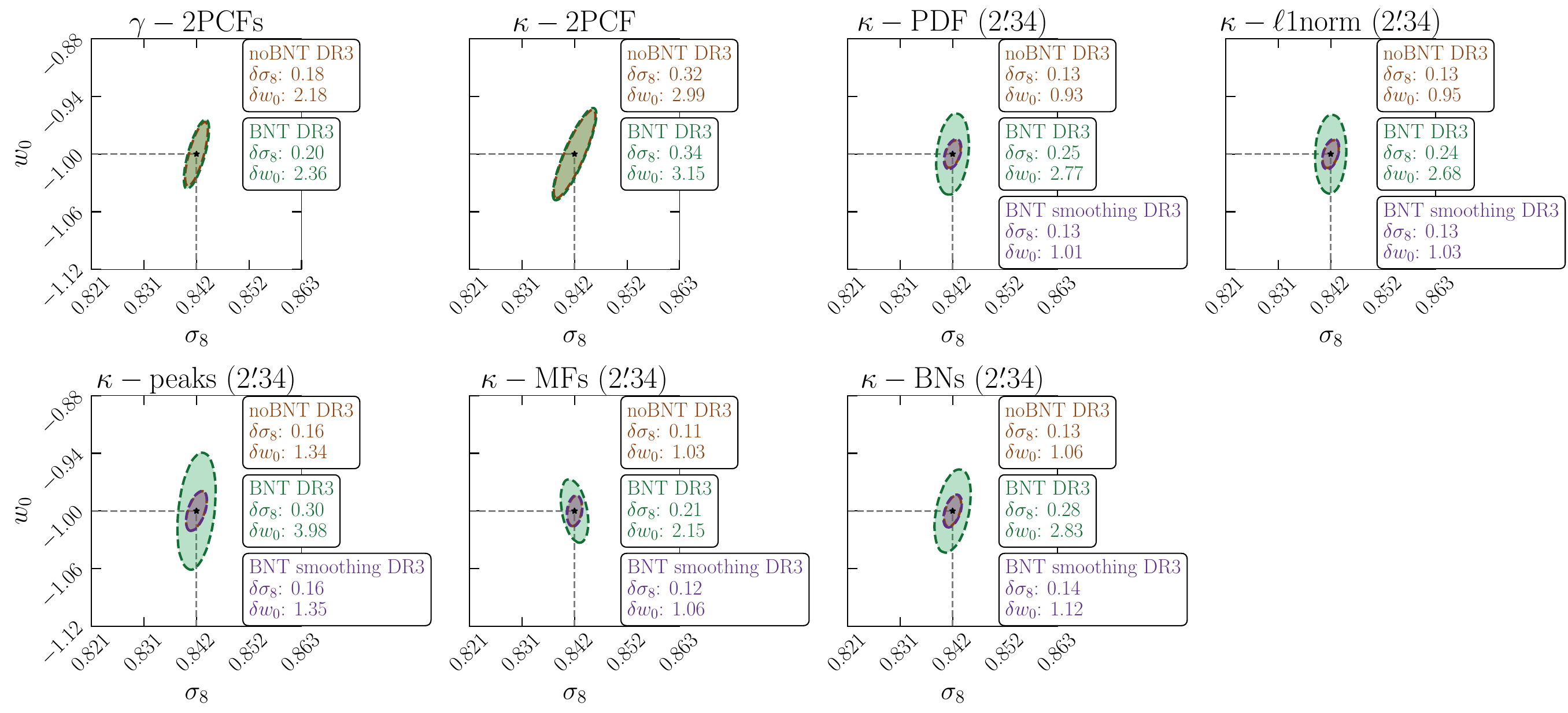}
    \caption{Impact of BNT. From left to right, DR3-like unmasked Fisher contours inferred from $\gamma$-2PCFs, $\kappa$-2PCF, and HOS (smoothing scale $\ang{;2.34;}$) measured on KS-reconstructed mass maps of all tomographic slices excluding the non-tomographic configuration in no-BNT (orange), BNT (green) and BNT smoothing (purple) setups.}
    \label{fig:no-BNT vs. BNT vs. BNT smoothing}
\end{figure*}

\subsection{BNT results}
\label{subsec: BNT-results}

Results are presented in Fig. \ref{fig:no-BNT vs. BNT vs. BNT smoothing} where we report a comparison of the DR3 unmasked Fisher contours inferred from the $\gamma$-2PCFs, $\kappa$-2PCF and $\kappa$-HOS using no-BNT (orange), standard BNT (green) and the alternative BNT smoothing (purple) method. Mass maps have been reconstructed using the KS algorithm. Within each panel, and consistently with all Fisher contours reported in the rest of the paper, the central black dot and corresponding dashed lines identify the fiducial DUSTGRAIN cosmology.
Each probe is the result of the concatenation of the full set of tomographic slices excluding the non-tomographic one on which the BNT has no effect. 
To ease the interpretation, we selected the lowest smoothing scale, $\ang{;2.34;}$, expressing the highest level of shape-noise which maximally affects the applicability of BNT for HOS.
It is worth noting that, even though a redshift-dependent smoothing radius would be more appropriate to consistently target the same physical scale across tomographic bins, in our current analysis the smoothing scale is held fixed as we are interested in testing the BNT algorithm when varying both the shape-noise level and the impact of non-linearities in our maps.
As a first observation, we report an almost perfect agreement between no-BNT and BNT ellipse size and orientation inferred by the $\gamma$-2PCFs and $\kappa$-2PCF, confirming for the latter no interplay due to the mass mapping reconstruction. 
This outcome is expected, as the BNT transformation redistributes the signal across tomographic bins \citep{Bernardeau:2013rda}: while individual bins carry less information due to the narrowing of the lensing kernels, they become less correlated.
We find that this is not the case for the $\kappa$-HOS for which the BNT-ellipses are both tilted and clearly inflated compared to the no-BNT ones. 
This behavior is explained by the effect of the shape-noise, which adds-up in the BNT linear combination of single and combined bins.
As a direct consequence, the BNT-maps become gaussianized and thus do not preserve the non-Gaussian information probed by the HOS. 
Reducing the shape-noise at the mass map level by increasing the smoothing scale, only slightly mitigates this caveat without fixing the issue.
These results are consistent with the findings of \cite{Barthelemy:2020yva} investigating the traditional BNT technique through the PDF estimator.
Finally, we highlight the agreement between the Fisher forecasts of the BNT smoothing and the no-BNT ones, which proves that the novel strategy minimizes the shape-noise diffusion.
This result is explained by the commutation between the BNT and the smoothing operators, which enables us to retain the advantages of the BNT physical scale cuts while canceling out the noise propagation prior to the mass mapping.
It is worth noticing that the shape-noise mitigation is also achieved when using combined bins.
Moreover, the entangled behavior of the lensing kernels associated with the combined bins -- noticeable in Fig. \ref{fig:lensing_kernels} -- does not prevent the ability to suppress the scale mixing.
This is because the BNT smoothing method applies filtering directly to the BNT-transformed shear maps of the single bins before combining them to construct the combined bins, as described in Eq. \eqref{BNT_combined_bins}. Consequently, the smoothing process remains unaffected by the broad profiles of the combined bin lensing kernels.

Concluding, the loss of constraining power detected in the BNT analysis based on HOS highlights the propagation of the shape-noise in the single and combined bin BNT linear combination compromising the precision gain from non-Gaussian estimators. 
However, filtering the BNT-shear maps which are then inverted back to the standard space before mass mapping circumvents this caveat. 
This strategy should enable realistic future application of the BNT to observational analyses and is further detailed in Taylor et al. (in prep.) in the context of mitigating the impact of baryonic feedback through efficient physical scale cuts.


\begin{figure*}
    \centering
    \includegraphics[scale=0.45]{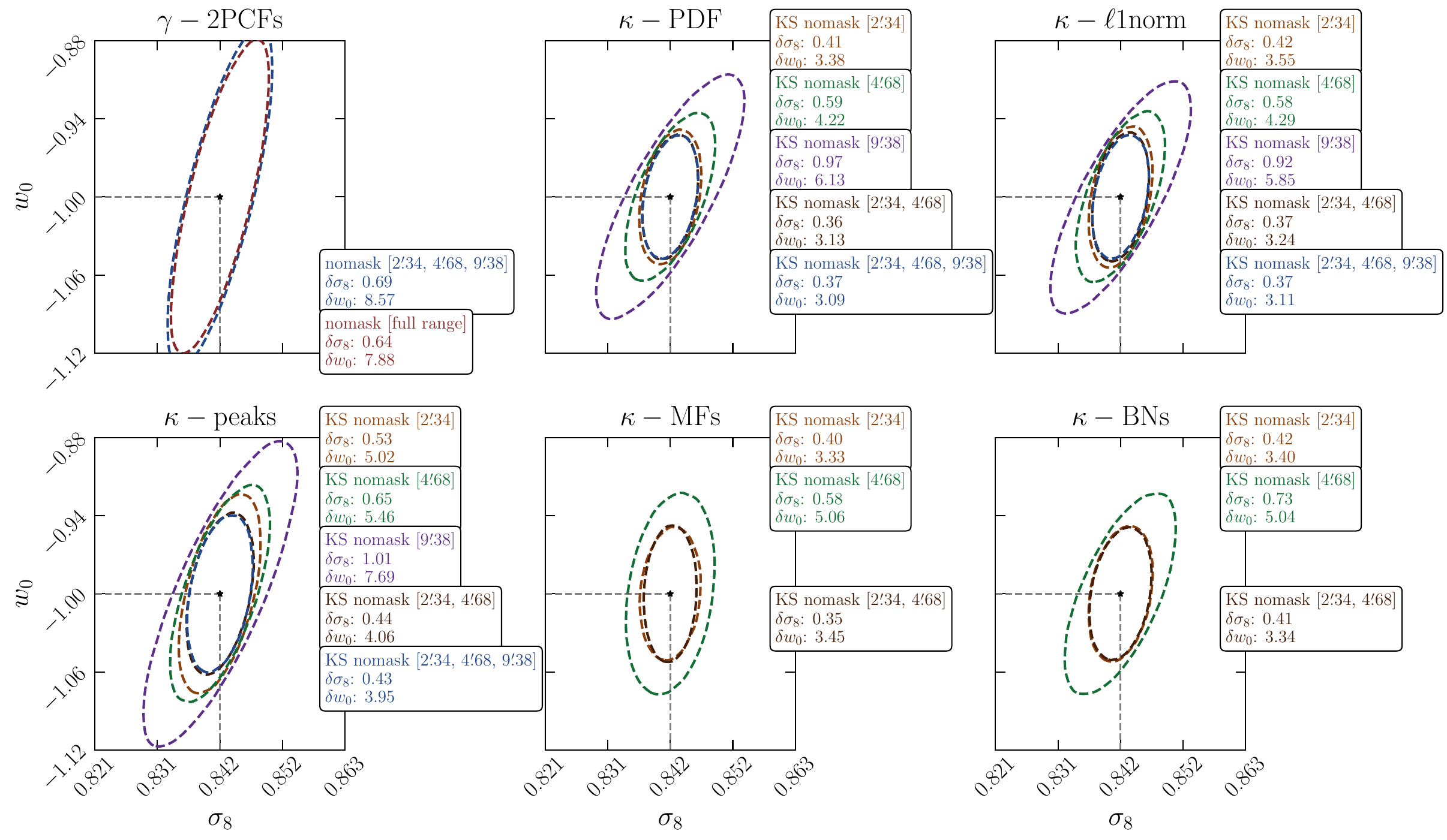}
    \caption{Multi-scale contour plots of the reference setup (KS, top-hat filter) without masks for $\gamma$-2PCFs and HOS. Fisher forecasts are computed for each individual scale of $\ang{;2.34;}$ (orange), $\ang{;4.69;}$ (green), $\ang{;9.38;}$ (purple); concatenation of $\ang{;2.34;}$ and $\ang{;4.69;}$ (brown); concatenation of $\ang{;2.34;}$, $\ang{;4.69;}$, and $\ang{;9.38;}$ (cyan); and the full-range of scales for $\gamma$-2PCFs (red). MFs and BNs do not exhibit the highest smoothing scale ($\ang{;9.38;}$) since it does not satisfy the SMAPE criterion.
    \label{fig:multiscale}}
\end{figure*}

\section{Multi-scale analysis}
\label{sec:multiscale}
After showing our ability to cut problematic scales in the last section, we now move on with extracting the combined HOS information from a multi-scale analysis.
Specifically, our multi-scale approach consists of concatenating estimators measured from mass maps filtered at different smoothing scales.
Following the dyadic decomposition, we investigate three different smoothing scales: $\ang{;2.34;}$, $\ang{;4.69;}$, and $\ang{;9.38;}$, which correspond, in our nominal pixel resolution of $\ang{;1.17;}$, to smoothing radii of two, four, and eight pixels, respectively. Larger radii have been excluded due to the limited number of smoothing apertures that could be drawn within the 25 deg$^2$ simulated field-of-view (FoV), which would result in significantly noisier measurements. For each statistic, we independently test the multi-scale smoothing using both a top-hat filter (i.e., non-compensated low-pass filter) and a Starlet wavelet filter (i.e., compensated band-pass filter, \citealp{starck:sta06,Leonard+12}). It is worth noting that the same smoothing size of a top-hat and Starlet kernel corresponds to a different physical scale given the distinct behavior of their filter functions for varying aperture radii. 

After analysing each scale individually, we combine those that exhibit Gaussian behavior according to the SMAPE test described in Sect. \ref{subsec:fisher_SMAPE_test}. This combination is carried out using the iterative scheme described in Sect. \ref{subsec:tomography_tomo-strategy}. We begin by computing the optimal tomographic DV for each scale individually, selecting $N_1$, $N_2$, and $N_3$ redshift bins for the three respective scales. We then construct the optimal multi-scale DV by applying the greedy algorithm to the combined set of $N_1 + N_2 + N_3$ bins.

The results of our multi-scale analysis are presented in Fig. \ref{fig:multiscale}. In the first panel, we show the Fisher contours for the $\gamma$-2PCFs, computed for both the full range of scales allowed by the SMAPE test and the three angular scales that closely match the smoothing radii used for the HOS, $\ang{;2.34;}$, $\ang{;4.69;}$, and $\ang{;9.38;}$. The bulk of the Gaussian information is already captured by the $\gamma$-2PCFs at these three scales, with parameter constraints degrading by less than $10\%$ compared to the full-scale analysis. Notably, this degradation remains within the uncertainty introduced by ill-conditioning.

The remaining panels of Fig. \ref{fig:multiscale} present the Fisher forecasts for each smoothing radius individually, along with their optimal combination. The multi-scale DV selection is always presented for the two smaller smoothing radii. The full combination of all three scales is included only for the one-point statistics and peak counts, as these maintain a valid Gaussian approximation across all smoothing scales. In contrast, the other topological estimators fail the SMAPE Gaussianity test at larger smoothing scales, due to the reduced number of topological features detected.

Focusing on the single-scale forecasts, we observe that the constraining power of the HOS decreases as we move from smaller to larger smoothing radii, despite the associated suppression of shape noise. This behavior is expected, as most of the non-Gaussian information probed by HOS resides in the small, highly non-linear scales, which are progressively smoothed out at larger radii.

Examining the Fisher ellipses derived from the optimal multi-scale DVs, we find that the constraining power is slightly enhanced when combining multiple scales. This is because they carry partially independent information, leading to a $[10, 20]\%$ improvement in the constraints on CPs. Our findings are consistent with those of \citet{Martinet+18}, although \citet{Ajani+20} report nearly double the improvement. However, the latter study is based on fixed redshift planes and assumes a higher level of shape noise. It is also worth noting that the level of improvement we find is comparable to the ill-conditioning uncertainty affecting our analysis, which prevents any robust conclusions regarding the benefit of the multi-scale combination. Nevertheless, the multi-scale approach enables a more efficient extraction of the signal and results in compressed DVs. For one-point statistics and peak counts, the optimal multi-scale selection across three scales rapidly reaches the same FoM as the optimal tomographic DV at the single scale of $\ang{;2.34;}$, using only about one third of the elements. For the remaining topological probes, combining the two smallest scales results in a $15\%$ compression. 

Comparing the optimal multi-scale forecasts obtained from the first two smoothing scales with those from the full set, we also note that the largest scale does not significantly enhance the constraining power. This observation is consistent with previous findings by \citet{Liu:2014fzc} and \citet{Martinet+18}, both based on the peak counts statistic, which emphasize that adding more than two smoothing scales yields negligible gains. These results are independent of the specific 2-dimensional parameter space considered. In particular, we verified that the interpretation of the results also holds in the ($\Omm$, $\sigma_8$) plane.


\section{Comparison with theory}
\label{sec:theory_validation}

In this section, we cross-validate our DUSTGRAIN $N$-body mocks with analytical predictions. We choose to focus on two summary statistics for which we have robust theory in the range of scales we have explored: $\gamma$-2PCFs (Sect. \ref{subsec:theory_validation_shear_2pcfs}) and $\kappa$-PDF (Sect. \ref{subsec:theory_validation_kappa_pdf}). We quantify the agreement both at DV level and through Fisher forecasts in Sect.~\ref{subsec:theory_validation_comparison_with_sims}.
%


\subsection{Modeling shear two-point correlation functions}
\label{subsec:theory_validation_shear_2pcfs}

The $\gamma$-2PCFs can be estimated by combining the tangential and cross components of the shear from pairs of galaxies as a function of angular scale $\theta$, yielding $\xi_\pm(\theta)$ in real space. This quantity can then be written in terms of the $E$ and $B$ modes of the convergence power spectrum in harmonic space \citep[e.g.,][]{KilbingerReview}. Neglecting higher-order effects and assuming no systematics, lensing is not expected to produce $B$-modes, so we can simply write the $\gamma$-2PCFs as

\begin{equation}
\xi^{ij}_{\pm}(\theta) = \frac{1}{2\pi} \int{C_{ij}(\ell) \, J_{0/4}(\ell \theta) \, \ell \, \text{d} \ell} \, ,
\label{eq: xipmcij}
\end{equation}
where we follow Euclid Collaboration: Cardone et al. (in prep.) in computing the angular power spectrum as

\begin{equation}
C_{ij}(\ell) = \int{\text{d}z \,
\frac{c \, q_i(z) \, q_j(z)}
{H(z) \, \chi^{2}(z)} \, P_{\rm mm}\left [\frac{\ell + 1/2}{\chi(z)}, \, z \right ]} \, ,
\label{eq: cijvsell}
\end{equation}
with $J_{0/4}$ as the 0th\,/\,4th order Bessel function, $q_{i,j}(z)$ the lensing kernels, and $P_{\rm mm}(k, z)$ the matter power spectrum evaluated at $k_{\ell} = (\ell \, + \, 1/2) \, / \, \chi(z)$ in the Limber approximation. We computed the matter power spectrum, $P_{\rm mm}(k,z)$, with \texttt{CAMB} \citep{Lewis+1999}, coupled with \texttt{HMcode2020} \citep{Mead+2020} for the non-linear part, and used two different codes (namely, {\tt LiFE}\footnote{{\tt LiFE} is a set of {\tt Mathematica} notebooks developed within the \textit{Euclid} forecasting group to compute the $3 \times 2$-pt observable in harmonic and configuration space. It implements exactly the same recipe as the one in {\tt CLOE}, the official \textit{Euclid} likelihood code (Euclid Collaboration: Cardone et al., in prep.), and computes derivatives of the observables as described below. It has been validated against Inter-SWG Taskforces (IST) Forecast codes \citep{ISTF}, and later used in the validation of the {\tt CLOE} (Euclid Collaboration: Martinelli et al., in prep.) together with {\tt PyCCL}. {\tt LiFE} is not public at the moment, but available on request.} and {\tt PyCCL}\footnote{\texttt{https://github.com/LSSTDESC/CCL}}, \citealt{Chisari+19}) to compare the results and validate the estimate of the Hankel transform in Eq. (\ref{eq: xipmcij}). To this end, we truncated the upper limit of the integration range after having checked that the results converged for $\xi_{\pm}^{ij}(\theta)$ over $\ell_{\rm max} = 10^5$ in the interval $\ang{;0.5;} \leq \theta \leq \ang{;500;}$.

Derivatives with respect to a model parameter, $p_{\mu}$, were computed as 

\begin{equation}
\frac{d \xi_{\pm}^{ij}(\theta)}{\text{d}p_{\mu}} = 
\frac{1}{2 \pi} \int{\frac{\text{d}C_{ij}(\ell)}{\text{d}p_{\mu}} \, J_{0/4}(\ell \theta) \, \ell \, \text{d} \ell} \, ,
\label{eq: dxiijdpmu}
\end{equation}
where we followed the semi--analytical method in \citet{ISTF} to compute the derivatives of the angular power spectrum as
\begin{align}
\frac{\text{d}C_{ij}(\ell)}{\text{d}p_{\mu}} & = \int{\frac{\text{d}}{\text{d}p_{\mu}}\left [ \frac{c \, q_{i}(z) \, q_{j}(z)}{H(z) \, \chi^2(z)} \right ] P_{\rm mm}\left [ \frac{\ell + 1/2}{\chi(z)}, z \right ] \, \text{d}z} \nonumber \\ 
& + \int{\frac{c \, q_{i}(z) \, q_{j}(z)}{H(z) \, \chi^2(z)} \frac{\partial P_{\rm mm}}{\partial p_{\mu}} \left [ \frac{\ell + 1/2}{\chi(z)}, z \right ] \, \text{d}z} \\ 
& - 
\int{\frac{c \, q_{i}(z) \, q_{j}(z)}{H(z) \, \chi^2(z)}
\frac{\ell + 1/2}{\chi^2(z)} \frac{\text{d}\chi(z)}{\text{d}p_{\mu}} \frac{\partial P_{\rm mm}}{\partial k}\left [ \frac{\ell + 1/2}{\chi(z)}, z \right ] \, \text{d}z} \nonumber \, ,
\label{eq: dcijdpmu}
\end{align}
where in the second (third) row $\partial P_{\rm mm}/\partial p_{\mu}$ ($\partial P_{\rm mm}/\partial k$) is the derivative of $P_{\rm mm}$ with respect to the parameter $p_{\mu}$ (the scale $k$) holding fixed the comoving distance $\chi(z)$. Note that this method helps avoid numerical derivatives since most of them can be replaced by integral of analytical quantities. As a consequence, the final derivatives of the $\gamma$-2PCFs are quite stable with respect to the details of the method used to compute the numerical derivative of the matter power spectrum.

\subsection{Modeling $\kappa$-PDF}
\label{subsec:theory_validation_kappa_pdf}

On mildly non-linear scales of the order of $\theta=10'$, large deviation theory (LDT) accurately predicts the WL $\kappa$-PDF \citep{Barthelemy20a} and its dependence on CPs \citep{Boyle2020} across different redshifts \citep{Castiblanco:2024xnd}. Moreover, the theoretical framework for the $\kappa$-PDF can be extended to account for WL systematics such as shape noise, the impact of masks in mass mapping, baryonic effects, and intrinsic alignments \citep{Barthelemy2024}. 
In a nutshell LDT describes the exponential decay of large fluctuations in series of random variables \citep{Touchette+09}. In the present case we use it to model the asymptotic exponential behavior of the matter density field PDF, as the variance goes to zero \citep{Bernardeau:2013rda}. By applying the contraction principle, it identifies the most likely mapping between initial and final densities. For projected fields, as the convergence, this relation is given by the cylindrical collapse as the dominant mapping \citep{2000A&A...364....1B}. Those principles are implemented on the level of the cumulant generating function (CGF), from which the $\kappa$-PDF can be obtained through an inverse Laplace transform. 
The convergence CGF is given in terms of the CGF of the matter density contrast  averaged over cylinders of length $L\rightarrow \infty$ and transverse sizes $\chi(z) \, \theta$. 
By focusing on the central (roughly $2 \, \sigma$) region around the peak, one can ensure accuracy of the theoretical model and Gaussianity of the DV. We computed the WL $\kappa$-PDF theoretical predictions using the \texttt{CosMomentum}\footnote{\url{https://github.com/OliverFHD/CosMomentum}} public code \citep{Friedrich_2025}. 

While $\kappa$-PDF predictions can be corrected for the impact of zero-padding in a KS reconstruction \citep{Barthelemy2024}, this is not required for our FFT-based KS reconstruction of the simulations, in which simulated maps are effectively repeated to reduce discontinuities at the patch borders. 
Finally, in contrast with the $\gamma$-2PCFs where the mean signal is insensitive to shape noise, the noisy $\kappa$-PDF is a convolution of the signal PDF and the shape noise PDF. We model the latter as a zero-mean Gaussian with a variance measured on smoothed convergence maps reconstructed from randomly rotated noisy shear catalogs probing intrinsic ellipticities only. 

\subsection{Comparison with simulations}
\label{subsec:theory_validation_comparison_with_sims}
We now compare $\gamma$-2PCFs and (KS) $\kappa$-PDF theoretical DVs with the one computed from the DUSTGRAIN simulation mocks, focusing on our reference configuration mask-free and with full tomography. Each DV is processed according to the SMAPE criterion described in Sect. \ref{subsec:fisher_SMAPE_test}, and the tomographic configurations are optimally selected following the iterative scheme reported in Sect. \ref{sec:tomography}. For $\kappa$-PDF, we explore the $\ang{;4.69;}$ and $\ang{;9.38;}$ smoothing scales, dropping the $\ang{;2.34;}$ which is not well described by the theory. Although we recommend a conservative lower bound of \ang{;10;} to ensure the robustness of the PDF theory model, we also adopt a smaller smoothing scale of \ang{;4.69;}. This choice is enabled by calibrating the noise variance using our simulations, which significantly improves the model–simulation agreement at small angular scales.

\begin{figure}
    \centering
    \includegraphics[scale=0.32]{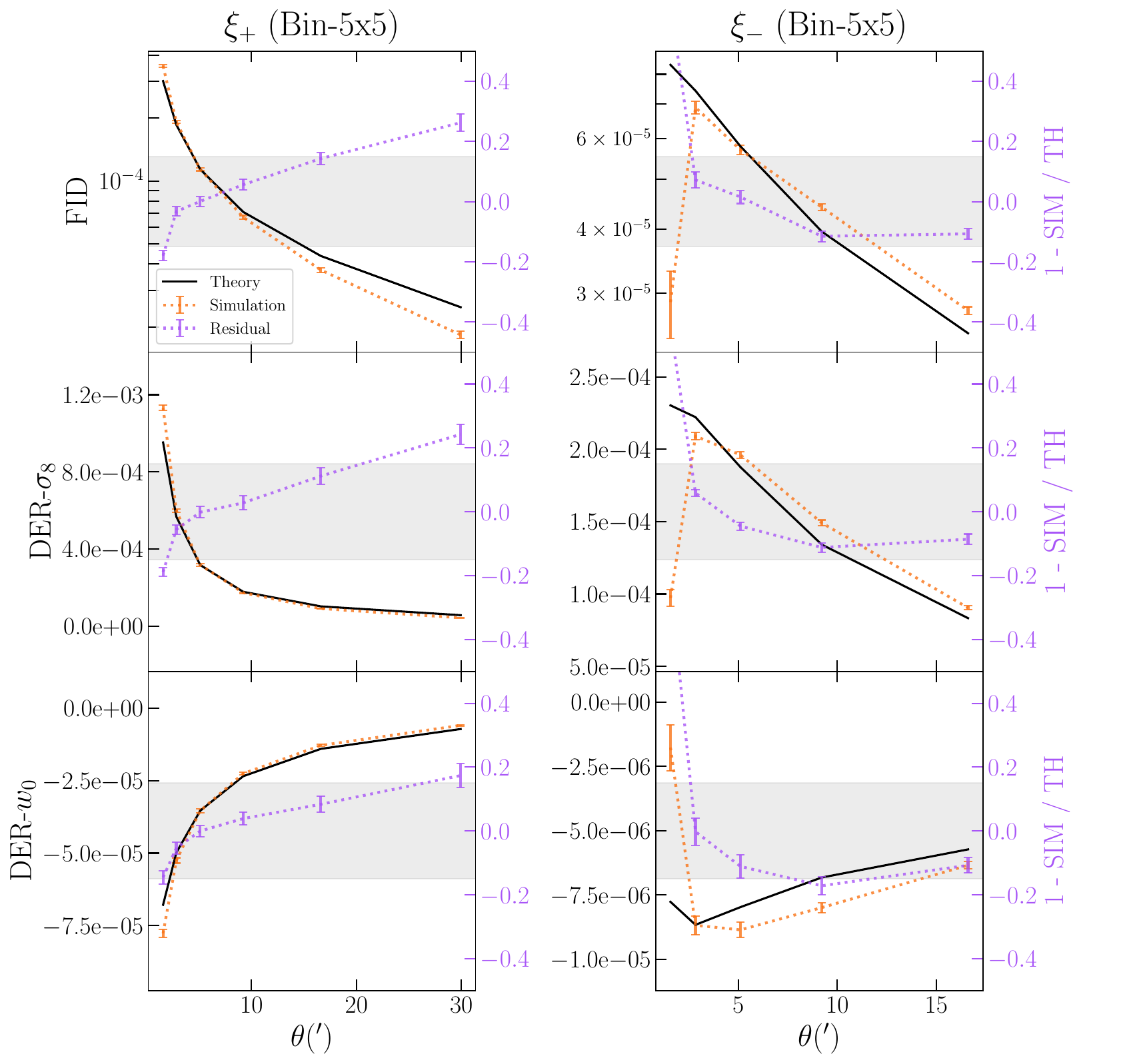}
    \caption{Shear two-point correlation functions (upper panel) and derivatives (lower panels) DV comparison for bin five between the mean computed from DUSTGRAIN mocks (orange) and theory predictions (black) for the fiducial cosmology (top), $\sigma_8$ (middle), and $w_0$ (bottom) derivatives. Fractional residuals are reported in purple on the secondary $y$-axis, emphasizing a 15\% threshold as a gray-shaded horizontal band.}
    \label{fig:shear_2pcfs_theory_validation:DV_comparison_bin4}
\end{figure}
\begin{figure}
    \centering
    \includegraphics[scale=0.35]{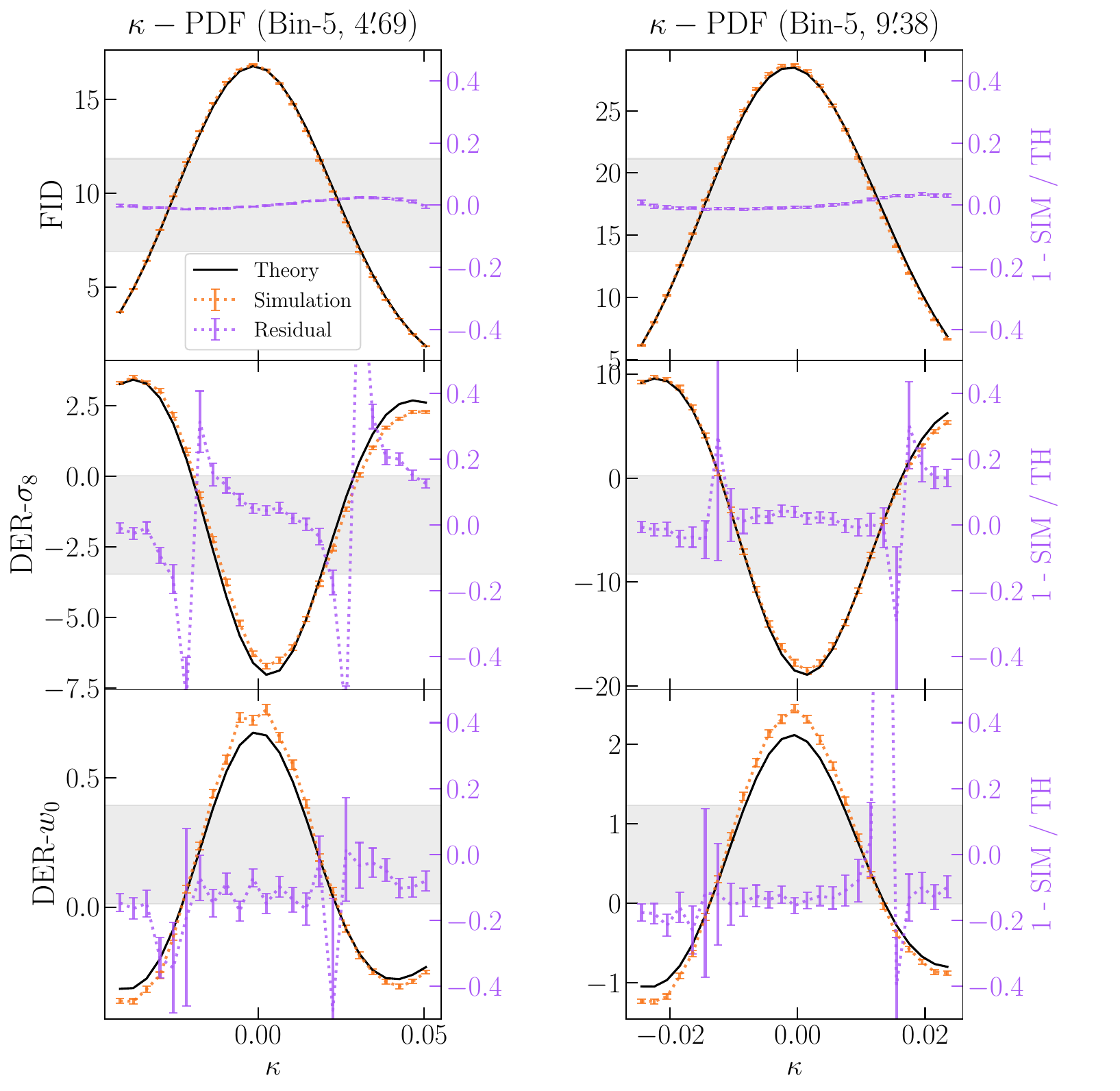}
    \caption{Same as Fig. \ref{fig:shear_2pcfs_theory_validation:DV_comparison_bin4}, but for $\kappa$-PDF of smoothing size $\ang{;4.69;}$ (left) and $\ang{;9.38;}$ (right).}
    \label{fig:kappa_pdf_theory_validation:DV_comparison_bin4}
\end{figure}

We first show a comparison at the DV level, focusing on the fifth redshift bin to ease the interpretation. This corresponds to the highest S/N bin as the cosmic shear signal is higher at higher redshift. Also, non-linear effects reduce the reliability of analytical predictions at lower redshifts, hindering good agreement between theory and simulations in the first bins.

For the $\gamma$-2PCFs (Fig.~\ref{fig:shear_2pcfs_theory_validation:DV_comparison_bin4}), we note an average disagreement between simulations and theory of $\approx 15\%$, mostly falling in the $2 \, \sigma$ range. The disagreement is higher for the lowest and highest scales respectively due to the effects of pixelization and finite FoV which are not accounted for in the theory modeling. Probably linked to these, the $\gamma$-2PCFs shows a consistent trend of the fractional residuals among the fiducial DV and its derivatives. This is supported by the better agreement we obtain between theory and simulations when rescaling the theoretical derivatives by the ratio of the simulated over analytical DVs at fiducial cosmology.

For $\kappa$-PDF (Fig.~\ref{fig:kappa_pdf_theory_validation:DV_comparison_bin4}), the residuals of the fiducial cosmology are flat and well comprised within $[2, 3]$\%. These are, however, larger for the derivatives, albeit still within the $\approx 15\%$ range and improving at larger smoothing scale. We also note that $\kappa$-PDF is better modeled in the region around the peak.

\begin{figure}
    \centering
    \includegraphics[scale=0.37]{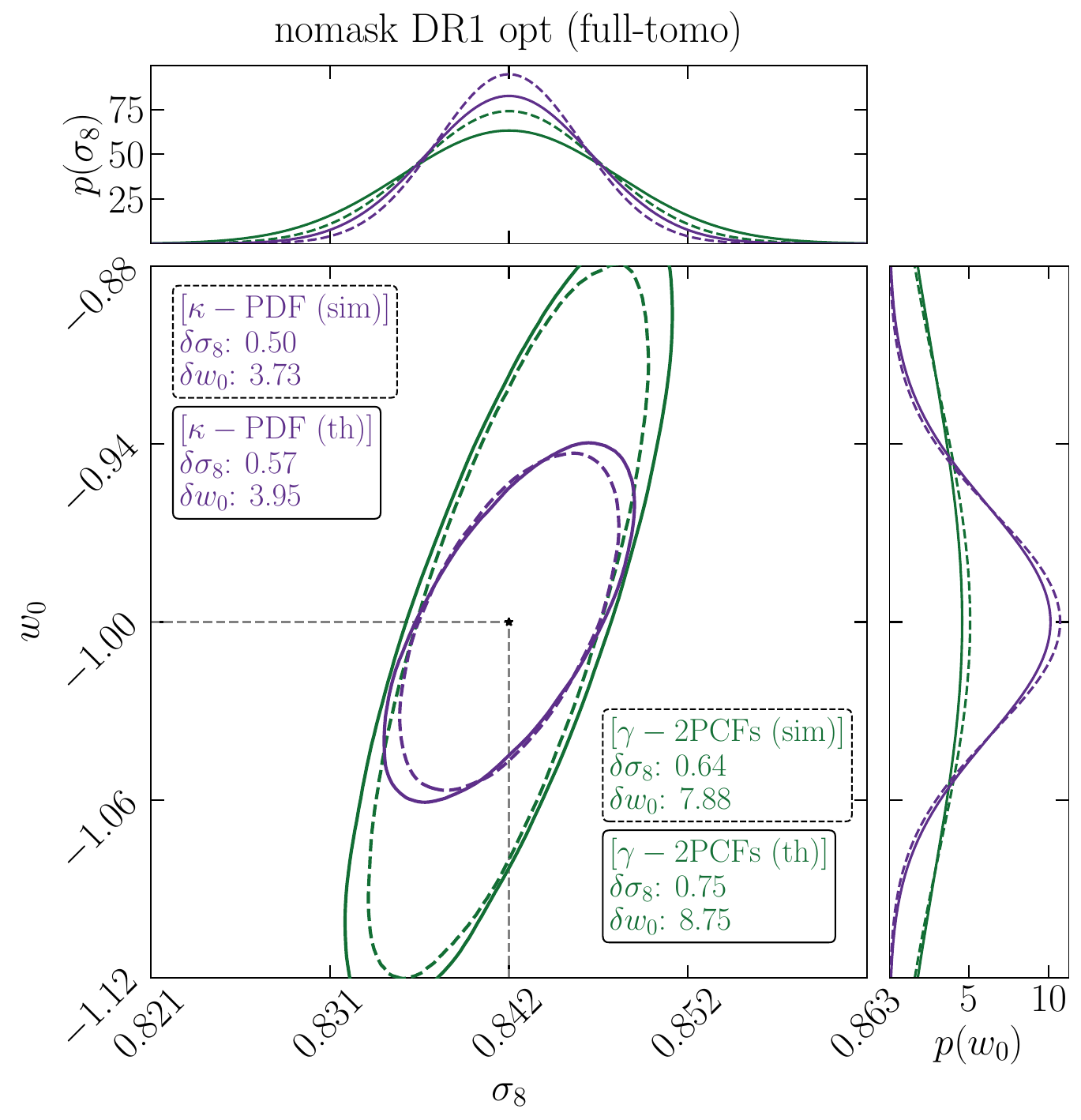}
    \caption{Fisher forecasts comparison for $\gamma$-2PCFs (green) and $\kappa$-PDF of combined smoothing size $\ang{;4.69;}, \,\ang{;9.38;}$ (purple) relying on the optimally selected full-tomographic configurations. We implement numerical derivatives of DUSTGRAIN mocks (dashed) and theory derivatives (solid). The data covariance matrix is numerically estimated from the SLICS mocks.}
    \label{fig:shear_2pcfs_kappa_pdf_theory_validation:fisher_comparison_fulltomo}
\end{figure}

In Fig. \ref{fig:shear_2pcfs_kappa_pdf_theory_validation:fisher_comparison_fulltomo}, we demonstrate how the trends reported above at the DV level translate in the Fisher forecasts of $\gamma$-2PCFs (green) and $\kappa$-PDF (purple) computed from numerical derivatives (dashed) and theoretical derivatives (solid). For both cases, we use the same SLICS numerical covariance matrix to focus on the differences in modeling the signal. We first notice that the parameter errors agree within $\approx [10, 15]\%$, with a marginal tilt of the ellipse orientations probably explained by the numerical-noise propagation. Such degree of agreement is in line with the differences observed at DV level for bin five (Figs. \ref{fig:shear_2pcfs_theory_validation:DV_comparison_bin4}--\ref{fig:kappa_pdf_theory_validation:DV_comparison_bin4}) although we concatenate multiple tomographic slices. This is because the signal is dominated by the high-$z$ bins and their cross-combinations such that the low redshift slices, where the agreement between simulation and theory is poorer, are subdominant and often even removed by our optimal tomographic selection strategy. These lower redshift bins might, however, prove useful when investigating the impact of systematics, for instance by creating a low-signal anchor point to account for intrinsic alignments \citep{Harnois-Deraps+22} and would necessitate a refined modeling for a more realistic theoretical approach. Finally, it is interesting to note that the improvement of the $\kappa$-PDF forecasts with respect to the $\gamma$-2PCFs remains the same for both theory and simulation, in agreement with figure 43 in \cite{EuclidOverview}. This validates our approach to use our simulated mocks to infer the gain brought about by HOS on cosmological forecasts.

\section{Mass mapping and masking effect}
\label{sec:mass_mapping_and_masks}

The mass inversion problem in WL refers to the challenge of reconstructing the projected mass distribution from observed subtle distortions in background galaxy shapes. 
This process is inherently non-unique because of the mass-sheet degeneracy, requiring additional assumptions or data to constrain solutions.
The performance of a mass reconstruction technique depends on several aspects: the level of noise imprinted in the shear components, assumed priors, presence of mask and missing data, and the type of filtering function (i.e., its shape in the frequency domain) used to smooth out the noise. 
Furthermore, the optimal choice of the method may vary from probe to probe.

The literature offers an extensive collection of algorithms spanning from the most popular versions, the KS and APM, into alternative implementations which make use of in-painting to better deal with masks and border effects, as the KS+, which is based on sparsity (see also \citealp{tersenov2025impactweaklensingmass}), or other techniques relying on null $B$-mode priors (e.g. \citealp{Jeffrey_2021}). In the era of machine learning, novel approaches based on neural networks can also be found (e.g. \citealp{boruah2025diffusionbasedmassmapreconstruction}).

In this analysis, we choose to compare the three methods implemented in the official \textit{Euclid} mass mapping pipeline: KS (Sect.~\ref{subsec:KS_formal_intro}), APM (Sect.~\ref{subsec:APM_formal_intro}), and KS+ (Sect.~\ref{subsec:KSPP_formal_intro}). We present the specific details of our implementation in Sect.~\ref{subsec:mass_mapping_implementation} and compare the statistical power from the same HOS applied to KS, APM, and KS+ in Sects.~\ref{subsec:mass_mapping_implementation:KS_vs_APM} and \ref{subsec:mass_mapping_implementation:KS_vs_KS+}, accounting for realistic star masks introduced in Sect.~\ref{subsec:gaia_masks_strategy_and_production}. 

\subsection{KS}
\label{subsec:KS_formal_intro}
The KS algorithm reconstructs the convergence $\kappa$ from the shear field $\gamma$ under the assumption that the observed reduced shear $g = \gamma / (1 - \kappa)$ is equal to the shear $\gamma$ in the WL regime ($\kappa, |\gamma| \ll 1$).
This process first requires to map the shear catalog into a pixelized grid.
Then, using complex notation, $\gamma = \gamma_1 + \rm i \gamma_2$ and $\kappa = \kappa_E + \rm i \kappa_B$, where $(\kappa_E, \kappa_B)$ correspond to the curl-free and gradient-free components of the shear field, they relate in the Fourier angular space as
\begin{equation}
    \hat{\gamma} = \hat{P} \, \hat{\kappa} \, ,
\end{equation}
where $\hat{P} = \hat{P}_1 + \rm i \hat{P}_2$ has components
\begin{equation}
\hat{P}_{1}(\boldsymbol{\ell}) = \frac{\ell_{1}^{2} - \ell_{2}^{2}}{\ell^{2}} \, , \quad
\hat{P}_{2}(\boldsymbol{\ell}) = \frac{2 \, \ell_{1} \ell_{2}}{\ell^{2}} \, ,
\end{equation}
where $\boldsymbol{\ell}=(\ell_1, \,\ell_2)$ and $\ell^{2} = \ell_{1}^{2} + \ell_{2}^{2}$.
The inverse relation is
\begin{equation}
    \hat{\kappa} = \hat{P}^{*} \hat{\gamma} \, ,
\label{shear-to-convergence_unityoperator}
\end{equation}
where the asterisk $(^*)$ denotes the complex conjugate.
Since a degeneracy arises for $\ell_1 = \ell_2 = 0$ (mass-sheet degeneracy), we forced the mean-convergence to be zero across the field by setting the reconstructed $\ell = 0$ mode to zero.
The above equation reads as
\begin{equation}
\hat{\kappa}_{E} = \hat{P}_{1} \, \hat{\gamma}_{1} + \hat{P}_{2} \, \hat{\gamma}_{2} \, , \quad
\hat{\kappa}_{B} = - \hat{P}_{2} \, \hat{\gamma}_{1} + \hat{P}_{1} \, \hat{\gamma}_{2} \, .
\label{convergence-shear-fuorier_domain}
\end{equation}
The KS reconstruction ends by converting back $\hat{\kappa}_{E}, \, \hat{\kappa}_{B}$ to the direct space and using the $E$-modes to infer the cosmological information.

\subsection{APM}
\label{subsec:APM_formal_intro}
The APM technique yields a local reconstruction of the projected density field within a circular aperture of radius $r$, weighted by a compensated filter function $U_{r}(\theta)$ set in this work to a Starlet wavelet kernel (Eq. 14 of \citealp{Leonard+12})
\begin{equation}
    M_{\text{ap}}(\boldsymbol{\theta}) = \int \text{d}^2\boldsymbol{\theta}' \, \kappa(\boldsymbol{\theta}') \, U_{r}(|\boldsymbol{\theta} - \boldsymbol{\theta}'|) \, .
\end{equation}
As $U_{r}$ is compensated, it satisfies $\int \, \text{d}\theta \, \theta \, U_{r}(\theta) = 0$. 
One can also express $M_{\text{ap}}$ directly in terms of the tangential shear component $\gamma_{\rm t}$
\begin{equation}
    M_{\text{ap}}(\boldsymbol{\theta}) = \int \, \text{d}^2\boldsymbol{\theta}' \, \gamma_t(\boldsymbol{\theta'}) \, Q_{r}(|\boldsymbol{\theta} - \boldsymbol{\theta}'|) \, ,
\end{equation}
where $Q_{r}(\theta)$ is related to $U_{r}(\theta)$ via
\begin{equation}
    Q_{r}(\theta) = \frac{2}{\theta^2} \int_0^\theta \, \text{d}\theta' \, \theta' \, U_{r}(\theta') \, ,
\end{equation}
and the tangential component, $\gamma_{\rm t}$ (and cross component $\gamma_\times$), are defined as
\begin{equation}
\begin{aligned}
\label{tangential_cross_shears}
\gamma_{\rm t}(\vec{\theta}^{\,\prime}, \vec{\theta}) &= -\mathrm{Re}\left[\gamma(\vec{\theta}^{\,\prime}) \, \text{e}^{-2 \rm i \phi} \right] \\
\gamma_\times(\vec{\theta}^{\,\prime}, 
\vec{\theta}) &= -\mathrm{Im}\left[\gamma(\vec{\theta}^{\,\prime}) \, \text{e}^{-2 \rm i \phi} \right]
\end{aligned}
\end{equation}
where $\mathrm{Re}$ and $\mathrm{Im}$ represent the real and imaginary parts respectively, and $\phi$ is the angle (in polar coordinates) between the reference position $\vec{\theta}$ and the galaxy position $\vec{\theta}^{\,\prime}$. Besides accounting for mass sheet degeneracy, a key advantage of APM resides in its local estimation of the impact of shape noise through the amplitude of the observed ellipticities, $|\epsilon(\boldsymbol{\theta'})|$, as
\begin{equation}
    \sigma_{M_{\text{ap}}} = \frac{1}{\sqrt{2}} \left( \int \, \text{d}^2\boldsymbol{\theta}' \, |\epsilon(\boldsymbol{\theta'})|^2 \, Q_{r}^2(|\boldsymbol{\theta} - \boldsymbol{\theta}'|) \right)^{1/2} \, .
\end{equation}

\subsection{KS+}
\label{subsec:KSPP_formal_intro}
Both KS and APM suffer from systematic effects when performed in the Fourier domain. First, we have the border effect defined by the discontinuity of the patch edges because of its finite angular size, which propagates into the reconstruction. Second, in the observational scenario, the presence of masked regions, as well as empty pixels due to the low galaxy density with respect to the chosen pixel size, introduces a leakage from $E$-modes into $B$-modes when these pixels are set to zero before the Fourier transform.

These border, empty and masked pixels effects increase the residual error between the reconstructed and the true field. The KS+ algorithm (\citealp{Pires+20}) proposes to mitigate these systematic effects by inpainting borders and empty pixels before applying the KS reconstruction. The pixel leakage is alleviated through the implementation of a sparsity prior to reconstruct iteratively a missing signal from the surrounding known regions. This helps in regularizing the borders and to replace masked/empty pixels with a definite value different from zero, thus preserving the continuity of the field. We point the reader to the Appendix A of \cite{Pires+20} for a detailed description of this procedure.\footnote{We underline here that in the implementation of the inpainting algorithm reviewed in \citet{Pires+20}, the $B$-modes were set to zero within the gaps to speed up computations, with negligible impact on the results in the non-tomographic case. This approximation can no longer be used in the tomographic set up as it would significantly increase the variance of the $E$-modes convergence maps due to the large amount of empty pixels.}

Dealing with these systematics has a cost in computational time. While KS and APM are performed in the order of $10^{-2} \, \rm s$ per map for flat patches of $5\times5 \, \, {\rm deg}^2$, KS+ takes order of $10^{2} \, \rm s$ per map mostly due to the inpainting operations which require hundreds of Fourier transforms and wavelet decompositions. Even though we reduce both CPU time and storage by not saving the KS and APM reconstructed maps given the low CPU time of such methods, we remind that I/O operations takes order of $10^{-3} \, \rm s$ per map to write the \texttt{FITS} array into the disk on our local cluster and up to $10 \, \rm s$ per map on large distributed facilities.

\subsection{Masking strategy}
\label{subsec:gaia_masks_strategy_and_production}

The collateral effect of masking propagates first into the mass mapping reconstruction and consequently through each estimator, taking particular significance in topological descriptors.
Hence, observational data analysis of WL statistics needs to account for the presence of masked pixels. 
Masked pixels arise from several unavoidable circumstances such as: pixel saturation, charged particles, CCD defects, and many others.
The most relevant among them is the pixel saturation induced by bright stars.
This produces diffraction patterns that contaminate the surrounding regions, thus affecting the photometry and morphology measurements.
We quantify their impact on the cosmological forecasts across proposed mass mapping algorithms.
However, we did not compute any bias induced by the mask due to the limitation imposed by the Fisher formalism.

We build a set of 32 independent \Euclid-like masks each covering $25 \, \rm deg^2$, matching our individual mocks's area. These masks are populated with stars from the (DR2) \texttt{Gaia star catalog} (\citealp{gaiamission_2018}) in non-overlapping patches randomly drawn from the expected \Euclid DR1 footprint. As the future DR1 analysis will make use of ground-based photometry, the radius of each star mask aperture depends on its magnitude following Eq. (2) of \citet{Coupon_2017} to mimic the typical mask size in ground-based Hyper Suprime-Cam data (\citealp{10.1093/pasj/psx066}). We only consider stars with magnitude brighter than $g \leq 17.5$ and reassign random positions to stars within a given mask.

These choices allow us to simulate $800 \, \rm deg^2$ of observed area with realistic star masks. Since the total coverage of our mocks significantly exceeds the one of the masked footprint, we batch-wise repeat the \textit{Gaia}-mask patches on our WL maps in order to guarantee a homogeneous application. 
Making use of 7616 (256) realizations for our fiducial (varied) cosmologies, we applied 238 (8) times the batch of 32 \textit{Gaia} masks. The masked area is of about 1.25 $\rm deg^{2}$ out of 25 $\rm deg^2$ ($\approx 5 \%$) on average.

\subsection{Reconstruction of mass maps from shear catalogs}
\label{subsec:mass_mapping_implementation}
Our reconstruction algorithms are applied to pixelized maps of noisy reduced shear $\gamma$. This means we approximate the shear as the reduced shear in the equations of KS and APM above, which can introduce an extra source of error at small-scales. Although there exists techniques to account for such effect in the reconstruction \citep[e.g.,][]{Seitz+96}, we neglect its impact here as this approximation was shown to be negligible for HOS analyses (\citealp[e.g.,][]{Martinet+18,ReimbergBernardeau2018,Pires+20}).

We first grid the two components of the shear $\gamma_{1/2}$ on flat sky maps of $256 \times 256$ pixels and angular side length of $5$ deg, corresponding to a pixel resolution of $\ang{;1.17;}$ . This is twice the $\ang{;0.59;}$ pixel scale used in \citetalias{Ajani-EP29}, a choice made to mitigate the impact of empty pixels. Indeed, these pixels with no galaxies induce a leakage in the Fourier domain reconstruction and become more numerous with the lower galaxy densities of the tomographic slices. It would be even more pronounced in case of realistic galaxy positions tracing the matter distribution. Each pixel of the shear maps is defined as the average (resp. sum) of the lensing signal of the galaxies falling in the pixel for KS and KS+ (resp. APM). Pixels with missing signal due to the star mask or the absence of galaxies at that position are set to zero (they will however be inpainted in the case of KS+) before applying the mass inversion algorithm in Fourier space. We mitigate the border effect due to the small simulated patches by repeating (KS) or zero-padding (APM) the image in any direction outside the FoV when performing the Fourier transform. For KS+ we also inpaint the boundary region.

As mentioned in Sect. \ref{bns_extraction} and extensively discussed in Appendix \ref{app:noremasking_vs_remasking}, there is no evidence in the literature whether reapplying or not the mask footprint after the mass reconstruction is more appropriate.
As a conservative choice, for all HOS but BNs (which cannot be computed in direct presence of masks), the reconstruction ends by reapplying the mask footprint to the computed mass map in order to neglect all regions of the sky discarded during the observations. 
This ambiguity, however, does not apply for the empty pixel effect linked to observed regions not traced by any galaxy. In contrast to KS, APM can have extra masked pixels where the aperture is more than half-empty. In our mask setup (5$\%$ masked area), this condition is rarely met around the \textit{Gaia} holes thus bringing negligible differences in regards of the fraction of masked pixels.

The impact of shape-noise imprinted into the shear field $\boldsymbol{\gamma}$ is mitigated by applying a multi-scale filtering in Fourier space.
Even with the treatment of the border effect in Fourier space, there might be some residual leakage at the edge of the maps when convolving with the filter. We further mitigate this by removing a stripe as large as the filter scale on each map side. 

\subsection{KS versus APM}
\label{subsec:mass_mapping_implementation:KS_vs_APM}

\begin{figure*}
    \centering
    \includegraphics[scale=0.5]{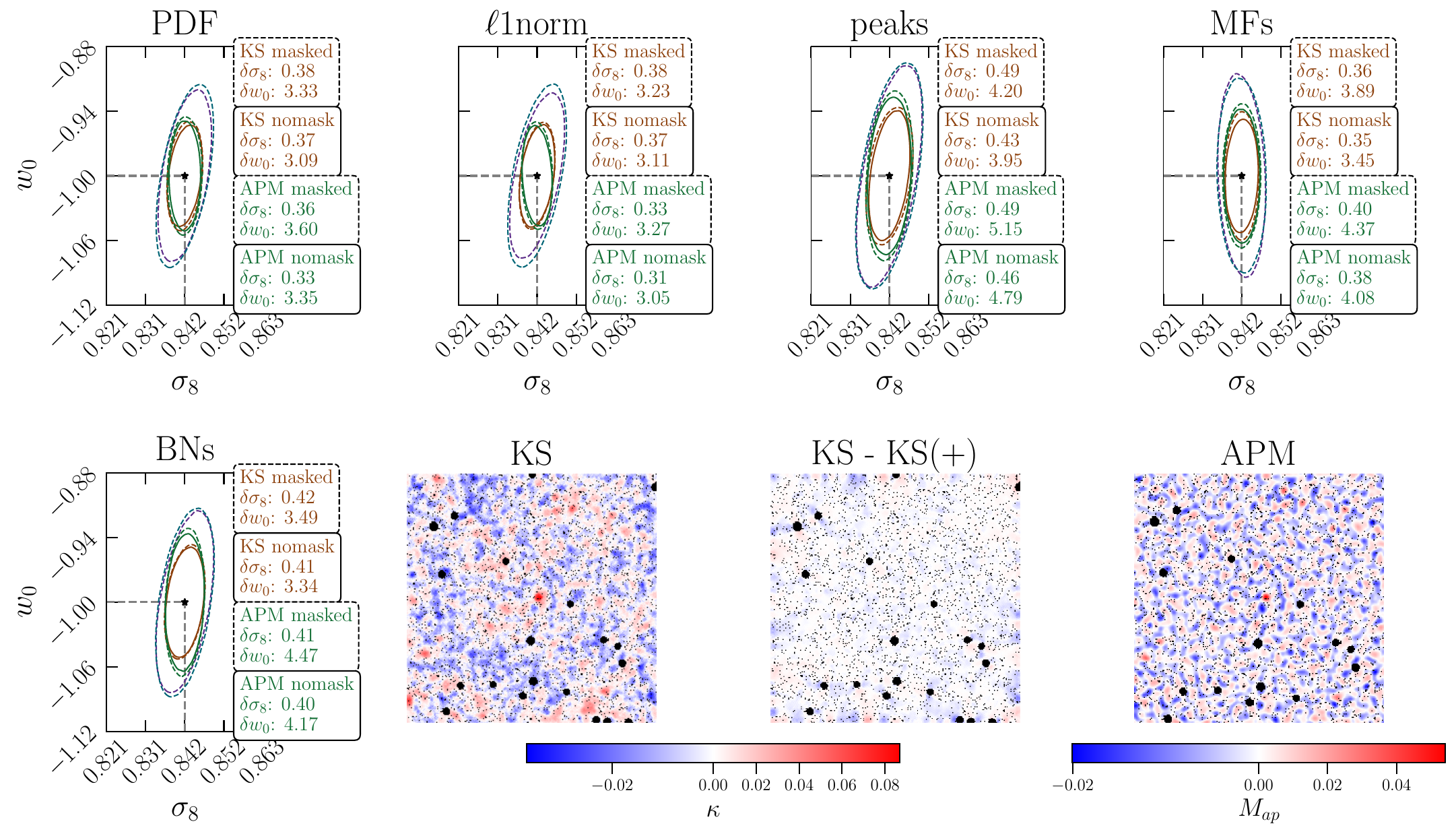}
    \caption{Fisher ellipses obtained from KS (orange) and APM (green) in unmasked (solid) and masked (dashed) configurations relying on the optimally selected full-tomographic and multi-scale approach in our DR1 fiducial setup. We also report ellipses of multi-scale KS (purple) and KS+ (cyan) inferred from the concatenation of five redshift single bins only. For a visual inspection, we show masked mass maps reconstructed from one single non-tomographic DUSTGRAIN LCDM realization: KS (left), KS - KS+ (middle) and APM (right). Maps are smoothed with $ \ang{;4.68;}$ (KS, KS+) and $\ang{;9.38;}$ (APM). \textit{Gaia} masked pixels are reported in black, while the KS/KS+ color bar values are fixed by the KS map.}
    \label{KS_vs_APM_vs_KS+}
\end{figure*}

In Fig. \ref{KS_vs_APM_vs_KS+}, we show the Fisher forecasts inferred using all probes in the DR1-opt observational setup both unmasked (solid) and masked (dashed) using KS (orange) and APM (green). Following the iterative scheme of Sect. \ref{subsec:tomography_tomo-strategy}, we concatenate the optimally selected tomographic slices and filter scales passing the SMAPE test. 

First, comparing the dashed and solid lines, for both KS and APM, the effect of masking produces a similar loss of power among each estimator increasing the 1-sigma errors accordingly to the average masked area. No tilt of the ellipses is detected in the masked case. 

Second, KS and APM show comparable performance within $10\%$, our confidence limit due to ill-conditioning. A slight trend in favor of the former algorithm is seen, particularly in the case of the peaks. We detected some differences when comparing KS and APM contours for individual tomographic slices and smoothing scales. These originate from the combination of the ill-conditioning and the mismatch of the physical scales among the filter kernels, the latter explaining the similar resolution of KS/KS+ and APM map patterns observed in Fig. \ref{KS_vs_APM_vs_KS+} for two very different smoothing radii: $ \ang{;4.68;}$ and \ang{;9.38}, respectively. On the contrary, the impact on the precision forecasts is almost negligible when recollecting the whole information through the tomographic and multi-scale approach.

\subsection{KS versus KS+}
\label{subsec:mass_mapping_implementation:KS_vs_KS+}

We now investigate how the different treatment of map borders and realistic masked patterns in the KS, KS+ algorithms affect their cosmological constraining power.
We focus on the reference setup with masks, concatenating all filter scales of only the five redshift single bins due to the high CPU time needed for KS+. The corresponding Fisher ellipses are reported in Fig. \ref{KS_vs_APM_vs_KS+}, in dashed purple (cyan) for KS (KS+).

First, it is worth noting how the KS ellipses inferred from single bins only (dashed purple) perform poorly against the full-tomographic approach (dashed orange), further motivating our tomographic strategy.
Secondly, the comparison between the two mass mapping pipelines shows very similar map patterns with noticeable differences at the borders, and a consistent agreement across all probe forecasts within $\approx 10\%$.
In this tomographic multi-scale approach, the agreement is within the ill-conditioning uncertainty.
However, we find higher departures among the individual redshift slices and smoothing scales. 
We study them in more detail in Appendix \ref{app:KS_vs_KS+_separate_effects}, by probing separately each mass mapping systematic: borders, empty pixels, and masks.

From these tests, we conclude that the tomographic multi-scale behavior is driven by two different effects: the increase of data covariance noise due to the lengthier concatenated DVs and the transfer of the KS leaked signal to small and large scales, as seen for the power spectrum.
The former partially hides any difference between the forecasts computed from the two mass mapping techniques. 
Conversely, the latter introduces higher discrepancies at fixed smoothing radii. 
These reduce when combining the scales together as the transferred leaked signal is restored among the scales.

As a final remark, the KS reconstruction appears to be more suited for the inference based on simulations being computationally cheap and maximizing the constraining power across each estimator. 
On the other hand, KS+ could be preferable for a cosmological inference based on analytical DVs. 
Indeed, KS+ considerably reduces the leakage and is able to better recover the true convergence field at the cost of a more correlated signal.



\section{Comparing probes}
\label{sec:comparing_probes}

\begin{figure*}
    \centering
    \includegraphics[scale=0.45]{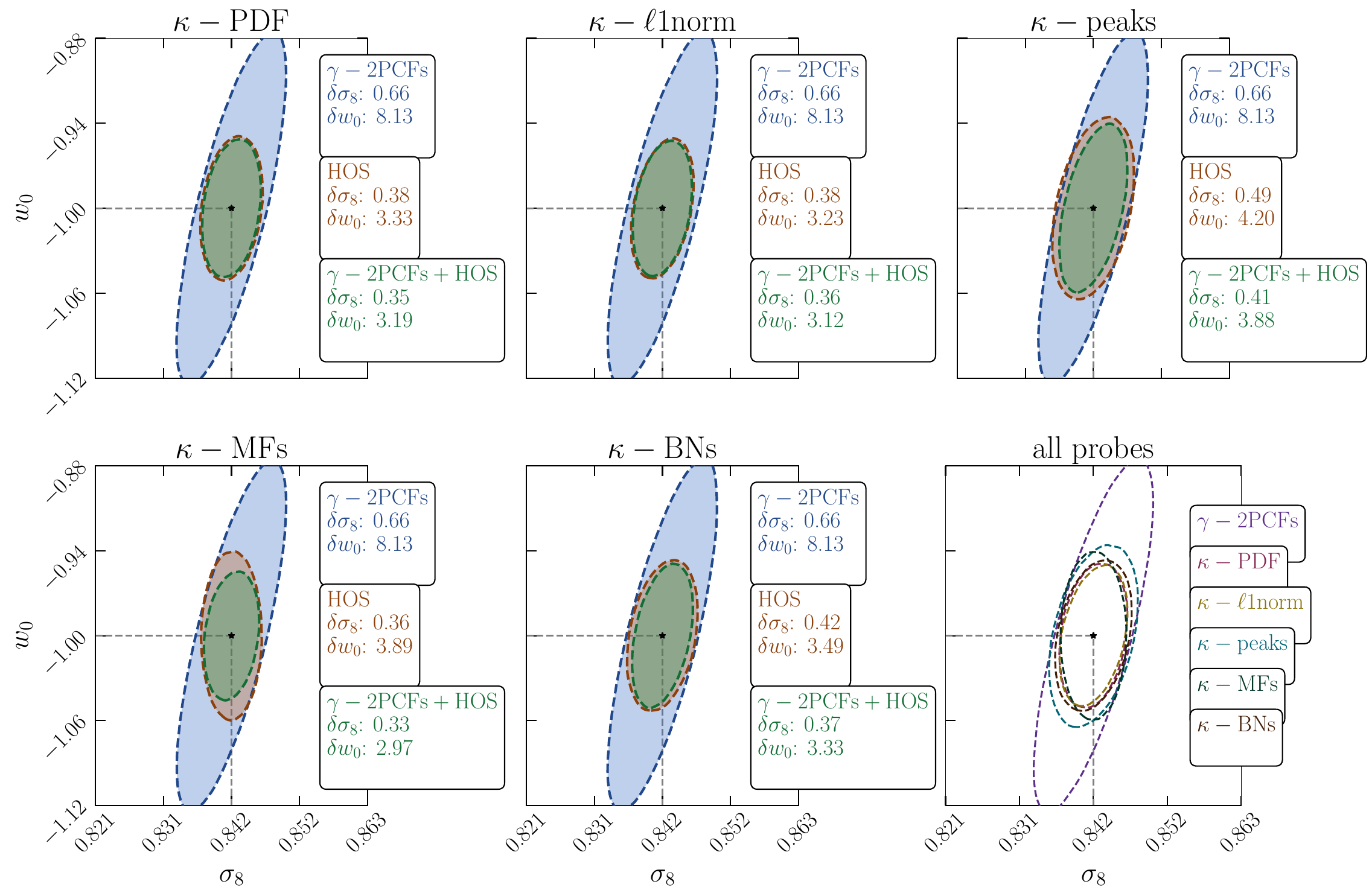}
    \caption{Fisher forecasts of reference setup with masks, full-tomography and multiscale, for $\gamma$-2PCFs (cyan) and HOS (orange), as well as for their combination (green). For a qualitative inspection, contours of all statistics are overplotted in the last panel for varying colors.}
    \label{fig:money_plot}
\end{figure*}
Figure \ref{fig:money_plot} shows the constraining power of the optimal tomographic, multi-scale DV for each probe, evaluated in our reference setup with masks. The interpretation of the results, however, remains consistent across all DR configurations and also holds for the APM reconstruction. HOS can be safely combined with the $\gamma$-2PCFs, although signs of ill-conditioning begin to appear when the $\gamma$-2PCFs is combined with either peaks or MFs. For all HOS, the multi-scale approach effectively captures the full Gaussian information content, in agreement with recent findings for Stage IV surveys (e.g. \citealp{giblin_multiscale_pdf}). As a result, combining HOS with the $\gamma$-2PCFs becomes essentially redundant, contrary to our findings of \citetalias{Ajani-EP29}, where a more substantial gain was observed. In our previous work, however, we relied on a single-scale analysis for the HOS and computed the $\gamma$-2PCFs from galaxy catalogs rather than from shear maps. Both our current results and those of \citetalias{Ajani-EP29} are consistent with \citet{Ajani+20}, which showed that combining the power spectrum with single-scale peak counts leads to significant improvement, whereas no additional gain is achieved when using multi-scale peak counts. According to these observations, the artificial tightening of the combined ellipses in Fig. \ref{fig:money_plot} serves as a red flag of emerging ill-conditioning.

All HOS outperform the $\gamma$-2PCFs, yielding improvements ranging within $[33, 50]\%$ on $\sigma_8$, and $[50, 64]\%$ on $w_0$. These gains are not expected to vanish when exploring a larger parameter space, as HOS should be particularly effective at breaking degeneracies among cosmological and nuisance parameters (\citealp{Martinet+21a}; \citetalias{Ajani-EP29}; \citealp{DES:2024jgw}). 
It is worth noting, however, that this comparison is limited to the $\gamma$-2PCFs, as we did not perform a full $3 \times 2$-pt analysis. Substantial improvement is also expected from this combination, as the $3 \times 2$-pt analysis only targets the Gaussian information (e.g. \citealp{Burger+22}; \citealp{Friedrich_2025}).

We also explored the combination of multiple HOS. However, the interplay of strong correlations and high-dimensional DVs -- which leads to invertibility issues in the data covariance matrix -- with ill-conditioning prevents us from obtaining reliable Fisher forecasts for joint probes. This issue is reflected in the combined HOS Fisher ellipses (omitted from Fig. \ref{fig:money_plot}), which exhibit a spurious tightening much more pronounced than in the case of the HOS and $\gamma$-2PCFs combination, despite the fact that the individual ellipses do not intersect and thus do not break any degeneracy. In the last panel of Fig. \ref{fig:money_plot} we simultaneously display the optimal tomographic, multi-scale forecasts from each HOS. Since they exhibit the same degeneracy direction in the ($\sigma_{8}$, $w_{0}$) plane, we observe nested ellipses, except for MFs that show a slightly tilted contour with respect to the others. This suggests, at least qualitatively, that no significant improvement is expected in this two-dimensional parameter space when combining multiple HOS. We found consistent results in the ($\Omm$, $\sigma_8$) plane. This result stands in contrast to the findings reported in \citetalias{Ajani-EP29}. In our previous work, however, we did not adopt a consistent choice of filter types and scales across the different statistics. As a result, the improvement observed from their combination was primarily driven by multi-scale information rather than by the combination of multiple statistics per se. This behavior observed in the present work, however, is likely specific to our particular analysis setup. In a full MCMC framework, where the non-Gaussian tails of the statistics can be included and a broader set of cosmological and nuisance parameters is explored, the different HOS may carry partially uncorrelated information. As a result, their combination may yield improvements both in the constraints on CPs and systematic biases (see \citealp{DES:2024jgw} as an illustrative example). 

Figure \ref{fig:money_plot} also enables a direct comparison between the probes that we have investigated in this work. It is important to keep in mind that our analysis includes an additional $\approx10\%$ uncertainty budget due to ill-conditioning, and the differences among statistics should be interpreted in light of this. It is also worth noting that both MFs and BNs required very aggressive tail cuts to achieve a valid Gaussian approximation, and hence they are slightly penalized in comparison with the other probes. 
Among the topological estimators, the MFs yield tighter parameter constraints on $\sigma_8$, with improvements of approximately $14\%$ and $27\%$ compared to BNs and peaks, respectively. In the case of $w_0$, BNs slightly outperform MFs by about $8\%$, while both estimators show $17\%$ and $7\%$ improvements over peaks, respectively. It is worth noting that MFs and BNs are specifically designed to capture partially uncorrelated information, as hinted by the small degeneracy difference in their Fisher ellipses. Their combination is therefore expected to yield additional gains in a more comprehensive analysis. Furthermore, the peaks analysis could be enhanced by exploring the redshift evolution of WL peaks, which is also largely insensitive to baryonic effects (\citealp{Broxterman:2024cam}).

The one-point statistics, namely the PDF and the $\ell1$-norm, exhibit comparable constraining power, as both probe similar features of the convergence field. However, the PDF proves slightly more efficient in extracting this information, requiring shorter DVs. 

When comparing the topological estimators with the one-point statistics, we find that the latter provide tighter constraints on the dark energy equation of state parameter $w_0$, with improvements of $5\%$, $15\%$, and $22\%$ over BNs, MFs, and peaks, respectively. In the case of $\sigma_8$, the MFs yield the best performance, outperforming the one-point statistics by $5\%$. However, one-point probes exhibit better results than BNs and peaks, achieving a gain of $10\%$ and $22\%$, respectively. Different results emerge from our APM analysis, where the Starlet filter targets smaller physical scales. As a consequence, the HOS measurements are more affected by shape noise and the one-point probes show better performance, being more robust to noise. It is also important to recall that all improvements below $10\%$ are of limited significance when considering the additional uncertainty budget introduced by ill-conditioning. Nevertheless, all HOS offer an unquestionable improvement over the $\gamma$-2PCFs, with enhancement factors ranging in $[1.35, 1.85]$ for $\sigma_8$ and $[1.95, 2.45]$ for $w_0$. 

Our results are clearly specific to the present analysis framework, as they are affected by Fisher matrix ill-conditioning, the varying tail cuts needed to ensure Gaussianity, and the absence of systematics beyond shape noise. Such limitations also prevented us from investigating additional HOS such as higher-order moments, persistent homology, and scattering transforms. However, despite the shortcomings, our findings remain informative for guiding a full MCMC analysis: combining the best-performing topological probe (MFs) with the most efficient one-point statistic (PDF) -- which are expected to be only partially correlated, as shown in \citet{Novaes:2024dyh} for the ($\Omm$, $S_8$) plane -- would indeed constitute a solid starting point for a comprehensive HOS-based investigation.


\section{Conclusion}
\label{sec:conclusions}

We aim to unveil the potential of HOS as cosmological tools leveraging \textit{Euclid} WL data. 
To prepare for DR1, we improve the Fisher analysis of \citetalias{Ajani-EP29} by applying various tomographic configurations, increasing the realism of our mocks through realistic photometric redshifts, and exploring new treatment of the data: BNT, multiscale, and three mass mapping techniques.

Specifically, we developed a unique pipeline involving Gaussian and non-Gaussian statistics, for a total of two two-point correlation functions ($\gamma$-2PCFs, $\kappa$-2PCF) and five HOS (PDF, $\ell1$-norm, peaks, MFs and BNs).
Our dataset is now composed of 7616 pseudo-independent mock catalogs built from the SLICS simulations to numerically estimate the data covariance matrix and $256 \times 7$ DUSTGRAIN-\textit{pathfinder} simulations to determine Fisher derivatives with respect to ($\Omm$, $\sigma_8$, $w_0$).
These $\approx 10 \, 000$ mocks now mimic a realistic photometric redshift galaxy distribution sampled from the \textit{Euclid} Flagship-2 simulation.
By applying photometric cuts, we constructed DR1- and DR3-like configurations to investigate the HOS performance in different tomographic scenarios.
We relied on five bins, also implementing combined bins by merging shear catalogs in ten pairs, ten triplets, and five quadruplets built from the individual slices. 
This approach included the mass reconstruction of 31 tomographic configurations (comprising the non-tomographic case) through three algorithms (KS, APM, KS+)  in two DR-like setups, for a total of $\approx 2 \, 600 \, 000$ mass maps on which HOS have been measured in homogeneous settings.

The wide ensemble of simulated data has been fully exploited to cover multiple aspects of the HOS analysis.
As a first result, we obtained a $2.5$-fold improvement on the dark energy equation of state parameter $w_0$ from each individual non-Gaussian estimator over the standard Gaussian two-point statistics, in line with \cite{Martinet+21a} and as suggested by the similar gain noted in \citetalias{Ajani-EP29} for $\Omm$ and $\sigma_{8}$.
Such improvement is further confirmed when comparing the forecasts of $\gamma$-2PCFs and $\kappa$-PDF computed using the theoretical derivatives.
This gain has been obtained in a DR1-like full-tomographic and multi-scale analysis with realistic 
\textit{Gaia} masks neglecting $\approx 5\%$ of the map area. Although the result is reported in the $(\sigma_{8}, w_{0})$ plane, we observe a similarly improvement in the $(\Omega_{\rm m}, \sigma_{8})$ configuration, notwithstanding the numerical instability associated with the latter parameter pair. 
The concatenation of multiple smoothing scales revealed beneficial in breaking parameter degeneracies, enabling HOS to almost fully capture the Gaussian content decoded by classical two-point correlation functions.
In more detail, HOS showed comparable performance, with a slight hint in favor of PDF/$\ell1$-norm and MFs.
Among the explored mass mapping techniques, none of them seemed to be preferred, while the impact of masking in the reconstruction follows the effective probed area.
Interestingly, the KS+ inpainting, based on a sparsity scheme, proved to better reconstruct the true field by significantly easing the leakage induced by empty  and masked pixels in Fourier space.
On the other hand, by introducing noisy correlated pixels, the sparsity interpolation degrades slightly the final constraints, thus making the standard KS more suited for simulation-based analyses, especially since KS+ is significantly slower.
Finally, even though we obtain similar performance across the mass mapping techniques involved in this study, more advanced reconstruction algorithms may reveal non negligible differences in the final parameter constraints (e.g. \citealp{tersenov2025impactweaklensingmass}).
The numerous set of explored tomographic configurations limited the reliability of our numerical data covariance. We therefore implemented a tomographic selection to reduce the overall DV length by neglecting a small amount of cosmological information. 
Moreover, it allowed us to spot any optimal sub-sample of redshift bins among the HOS. 
In particular, merged bins proposed in \cite{Castiblanco:2024xnd}, i.e., combined bins composed by contiguous redshift slices, are consistently found within the most constraining configurations defined by the concatenation scheme for one-point estimators.
Differently, the constraining power of topological descriptors tends to be maximized when relying on a two-point-like tomographic setup.
Finally, in order to open up the possibility of efficiently removing scales which we cannot accurately model in the presence of known systematic effects such as baryonic feedback, we applied the BNT transformation and its alternative method, the BNT smoothing (Taylor et al., in prep.).
We confirmed the non-Gaussian information to be compromised in the standard BNT because of the propagation of the shape-noise (\citealp{Barthelemy20a}). Lastly, we probed the BNT smoothing as a way to ease this caveat and thus enable future implementations.

We note several caveats in our analysis.
The first one we address is related to the Gaussian likelihood assumption on which the Fisher formalism is structurally based.
In line with \citetalias{Ajani-EP29}, a non negligible number of HOS (four out of nine) did not pass the tomographic SMAPE test we designed, giving strong symptoms of non-Gaussian likelihood behavior. 
Out of the discarded probes, most of the statistics, including the two-point correlation functions, needed tailored cuts to remove elements which were not meeting this requirement.
The following result suggests that several estimators would be rather described by a non-Gaussian likelihood.
Secondly, we detected signatures of ill-conditioning both in our data covariance and Fisher matrices. 
While the ill-conditioning on the former revealed not to impact our forecasts, the latter induced a substantial scattering in the inverse Fisher matrix introducing an average uncertainty on the final forecasts within $\approx 10 \%$.
Although this caveat appears to be highly relevant for the Fisher formalism only, we will carefully check its effect through the data covariance matrices used in future MCMC analyses.
Finally, likely linked to ill-conditioning, we noticed unreliable forecasts when jointly combining the HOS, thus preventing us from probing eventual extra gains brought by their combination. 
This behavior is attributed to the numerical noise, which propagates into the Fisher forecasts computation.

Concluding, the results presented in this work are limited to the Fisher formalism and do not consider most of the systematic biases which could alter our conclusions for the DR1 analysis.
However, within this challenging perspective, we clearly demonstrate the substantial precision gain enabled by non-Gaussian statistics over the standard two-point correlation functions in the future \textit{Euclid} observational analyses.


\begin{acknowledgements}

\AckEC 
\AckCosmoHub

This work has made use of data from the European Space Agency (ESA) mission {\it Gaia} (\url{https://www.cosmos.esa.int/gaia}), processed by the {\it Gaia} Data Processing and Analysis Consortium (DPAC, \url{https://www.cosmos.esa.int/web/gaia/dpac/consortium}). Funding for the DPAC has been provided by national institutions, in particular the institutions participating in the {\it Gaia} Multilateral Agreement. Simone Vinciguerra \& Nicolas Martinet acknowledge the funding of the French Agence Nationale de la Recherche for the PISCO project (grant ANR-22-CE31-0004).  Andreas Tersenov and Jean-Starck   acknowledge the funding of the TITAN ERA Chair project
(contract no. 101086741) within the Horizon Europe Framework Program of
the European Commission and of the French Agence Nationale de la Recherche
(ANR-22-CE31-0014-01 TOSCA).
\\~\\

\end{acknowledgements}

\bibliographystyle{aa}

\bibliography{HOWLS_KPpaper2}

\appendix

\section{Measuring statistics on masked mass maps: No remasking versus remasking}
\label{app:noremasking_vs_remasking}
The algorithms of BNs, persistent BNs and persistent heatmaps allow for reliable measurements as long as the fraction of masked pixels in the input map is negligible.
This is due to their intrinsic nature: while BNs describe the connectivity of minima, maxima and saddle points in a field, persistent BNs track their birth and death across filtration levels (persistence diagrams). Persistent heatmaps further summarize this by smoothing persistence diagrams with a Gaussian kernel, thus enhancing robustness in the statistical analysis. As a consequence, such topological features are no longer properly defined when a significant number of masked pixels are present.
We face this caveat in our analysis as we perform the mass reconstruction in Fourier space, inducing a leakage of the signal into the masked coordinates, the latter now containing a well-defined value after the inversion that is finally masked to recover the original mask footprint. 

In the literature, there is no clear evidence whether the leaked masked coordinates should be neglected in the final reconstructed map (i.e., remasking it) or not (i.e., without remasking it). 
Indeed, not reapplying the mask-footprint to the final mass map significantly reduces the presence of masked pixels, consequently allowing for the standard BNs and homology measurements.
This is the case of \citet{Heydenreich:2021} and \citet{Heydenreich2022}, where persistent BNs and persistent heatmaps are investigated using observational data. 
Following on these results, further investigation is needed to understand the impact on precision and bias forecasts when performing measurements from mass maps with and without final remasking applied. 
Nevertheless, the CP errors inferred in our observational setup from the two approaches agree within $\approx [2, 3] \%$ across all others $\kappa$-based statistics, validating our choice not to remask in case of BNs.


\section{Single scale comparison between KS and KS+}
\label{app:KS_vs_KS+_separate_effects}

\begin{figure*}
    \centering
    \includegraphics[scale=0.55]{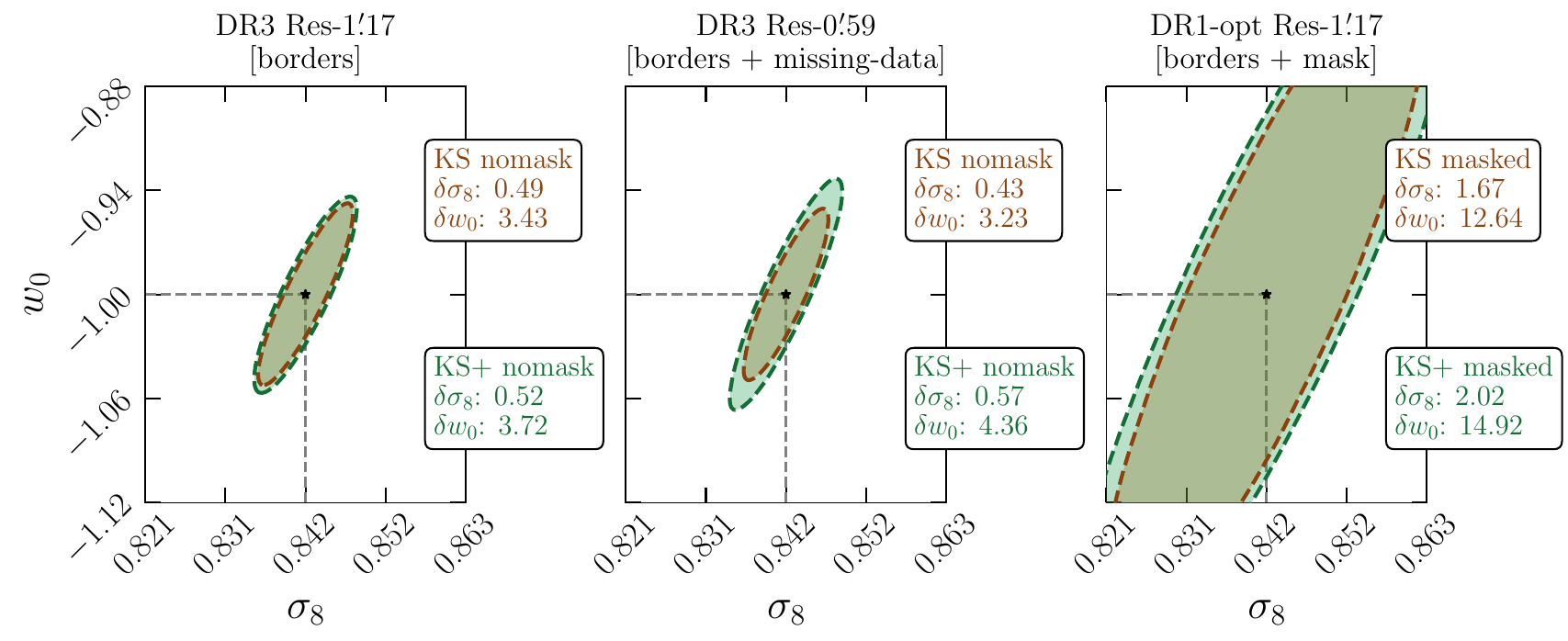}
    \caption{Fisher contours of the PDF measured from the KS (orange) and  KS+ (green) reconstructed redshift bin five top-hat smoothed with $\ang{;9.38;}$ scale. In the left, middle and right panel, we explore the DR3 (\ang{;1.17;}), DR3 (\ang{;0.59;}), DR1-opt (\ang{;1.17;}) DR-configuration (pixel scale) respectively.}
    \label{fig:KS_vs_KS+_bin4_onescale}
\end{figure*}

In Fig. \ref{fig:KS_vs_KS+_bin4_onescale}, we show the individual impact of borders, masking and tomographic missing data observed in the KS and KS+ techniques at Fisher contour level for $\kappa$-PDF in a single redshift bin (5) and smoothing scale (\ang{;9.38;}).

Specifically, in the left panel, map borders are addressed in the unmasked DR3-like configuration using $\ang{;1.17;}$ as pixel resolution in which no empty pixels are found.
In the middle panel, tomographic missing data are investigated again through the unmasked DR3-like configuration but with $\ang{;0.59;}$ pixel resolution, defining $\approx 20\%$ of empty pixels in the single redshift bins.
Finally, in the last plot, the impact of realistic masks is quantified making use of the DR1-opt setting with $\ang{;1.17;}$ pixel resolution fed with \textit{Gaia} masks ($\approx 5\%$ of masked area) and in absence of empty pixels beside the star mask.
It is worth noting that the effect of map borders is unavoidably present in all cases, but it is expected to be sub-dominant when compared to the one arising from masks or empty pixels.

In the presence of map borders only, the two methods show consistent parameter constraints within the 10\% of the ill-conditioning uncertainty.
This result suggests the effect is not prominent, in particular thanks to the removal of patch stripes from the smoothed maps as explained in Sect. \ref{subsec:mass_mapping_implementation}.

\begin{figure}
    \centering
    \includegraphics[scale=0.6]{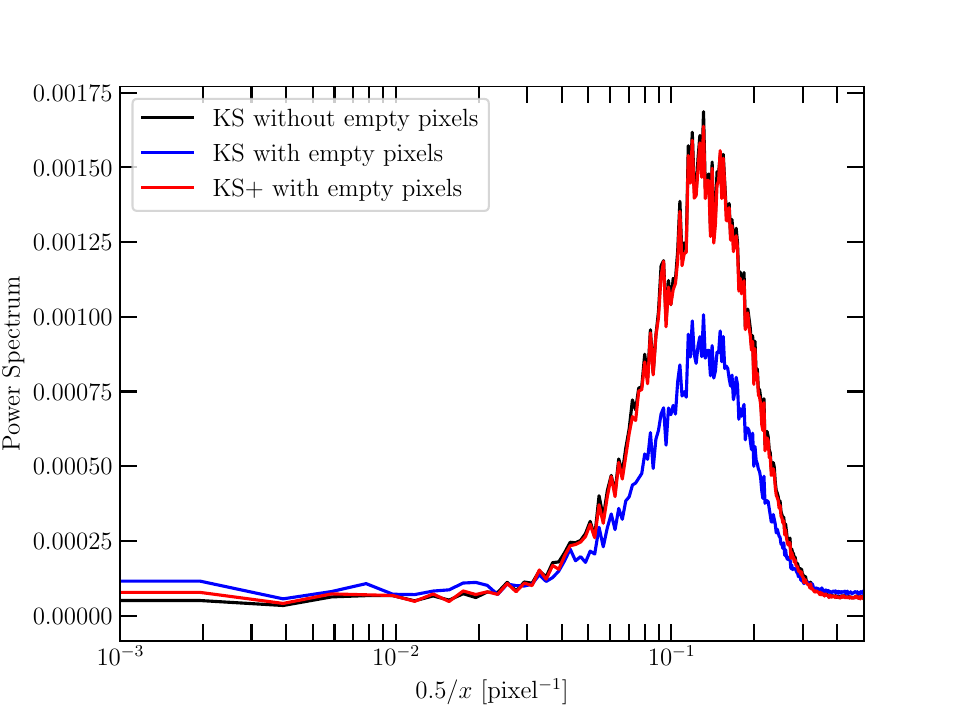}
    \caption{Power spectra of a noisy $\kappa$-map of $512 \times 512$ pixels (DR3-like configuration with pixel resolution $\ang{;0.59;}$) reconstructed from complete shear maps (black using KS) and from incomplete shear maps with  $\approx 20\%$ of empty pixels (blue using KS and red using KS+) randomly set to zero before the mass inversion. In order to focus on a specific scale, a $\ang{;2.34;}$ wavelet filter kernel is applied to the complete shear maps before applying random empty pixels. The signal leakage into the empty pixels is thus only induced by the FFT mass reconstruction.}
    \label{fig:KS_vs_KS+_powerspectrum_leakage}
\end{figure}

When introducing empty pixels randomly distributed across the maps, the agreement deteriorates, with discrepancies rising to 30\% in specific tomographic and smoothing configurations. 
This is explained by the distinct handling of missing data: KS sets empty pixels to zero, leading to signal leakage into those regions, while KS+ inpaints them mitigating the leakage.
Specifically, the KS leakage transfers to smaller and wider scales which explains why the constraints between KS and KS+ become more consistent when combining all the scales.
This is shown in Fig. \ref{fig:KS_vs_KS+_powerspectrum_leakage}, where the KS and KS+ power spectra are measured in presence of empty pixels.
We note that the KS power spectrum loses power around the filter scale, reporting an excess on the other modes as compared to the KS+.

In the setup including \textit{Gaia} masks but no missing data, a moderate 20\% increase of the disagreement is observed in the worst case.
This is consistent with the inpainted masked area.

In a last instance, the KS constraining power is found to be slightly higher than the one of KS+.
While the numerical derivatives are consistent, data covariance matrix terms are lower-valued in the KS method.
We thus explain this second trend as the direct effect of the inpainting which introduces noisy correlated pixels in the shear maps. 
In fact, the extra correlation propagates across all pixels of the KS+ reconstructed mass maps, increasing its data covariance values and finally bringing to degraded constraints.


\end{document}